\documentclass{book}
\usepackage{natbib}
\usepackage{graphicx}       
\usepackage{amsmath}

\usepackage[english]{babel}

\begin{document}
\vspace*{-1.5cm}
\begin{center}
  \large
  \begin{figure}[!h]
  \centering
  \includegraphics[height = 3 cm,width = 2.5 cm]{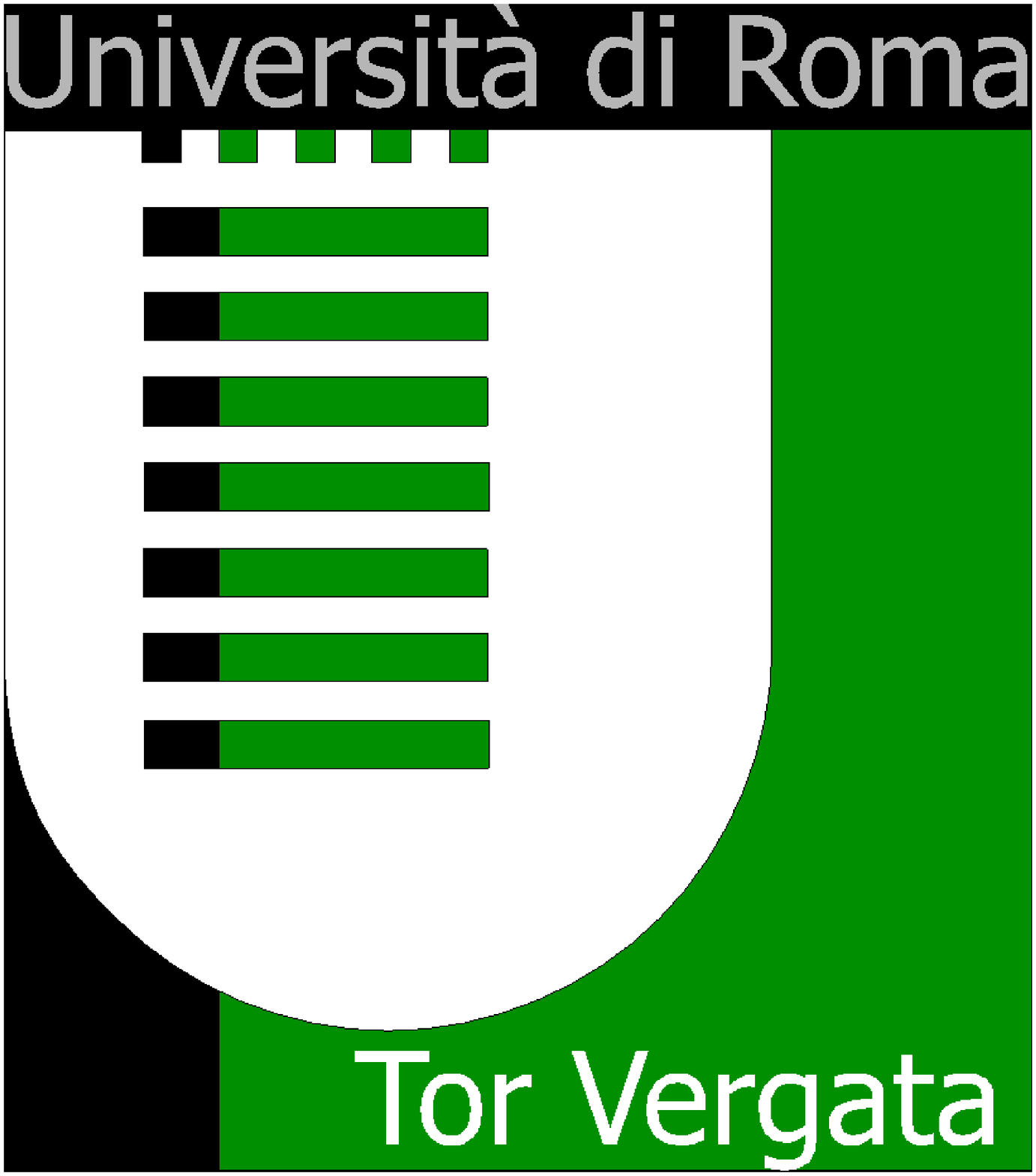}
  \end{figure}
  UNIVERSIT\`A DEGLI STUDI DI ROMA\\``TOR VERGATA"\\
 \vspace*{1.3cm}

 \vspace*{0.3cm}
  \normalsize

  \vspace*{0.3cm} \large

 DOTTORATO DI RICERCA IN ASTRONOMIA\\
 XXVII CICLO\\
 \vspace*{.75truecm} \LARGE
\bf{Spectrophotometric analysis of cometary nuclei from in situ observations\\
}
\end{center}
\vspace*{.5cm} 
\large
\begin{center}
\bf{Dott. Andrea Raponi}
\end{center}

\vspace*{1.5cm}

\begin{flushleft}

Tutor: \\

Prof. Francesco Berrilli\\

 $ $

 Relatore esterno: \\
 Dott. Fabrizio Capaccioni\\
 
 INAF-IAPS, Roma

 $ $

Coordinatore del corso di dottorato: \\

Prof. Pasquale Mazzotta\\
 
\end{flushleft}
\vspace*{0.5cm}
\begin{center}

 A. A. 2013-2014
\end{center} 
\clearpage

\tableofcontents
\listoffigures
\listoftables

\chapter*{Introduction}
Space science has certainly brought important breakthroughs in recent science history. It allowed to overcome the shield of terrestrial atmosphere to observe photons emitted by of planets, stars, galaxy and beyond, across the entire electromagnetic spectrum. In particular for Solar System studies, it represents something very novel for astrophysics: the possibility to enter directly in contact, or very close, to the object under investigation, thanks to missions which are intergenerational in most cases, but whose travel time certainly worth the potential of discoveries. 

Not much time has passed since the first scientific space mission has sent to Earth the first data. We can say that space exploration is still at the beginning, and every new data from in situ observations can bring a new comprehension of the Solar System, and thus of the part of the universe closer to us.

Topic of this work are comets, small and elusive objects that may hold great secrets about the origin of the Solar System and life on Earth, being among the most primitive objects.

The method of investigation addressed in this work is the visible-infrared spectrophotometry by remote sensing instruments. 

Spectrophotometry has already had very successful results in planetary exploration, in particular thanks to a new class of instruments: imaging spectrometers, designed for the observation of remote planetary atmospheres and surfaces, capable to acquire hyperspectral data with high spatial and spectral resolution. 

This work has to be intended as a preparation for the huge amount of data coming from VIRTIS (Visible and InfraRed Thermal Imaging Spectrometer) onboard Rosetta spacecraft, which will orbit and follows comet 67P/Churyumov-Gerasimenko (CG) during its journey toward the Sun, and will deliver a lander on the surface for the first time in space exploration history. 

The context under which this mission moves its steps is described in the first chapter. 

In the second chapter the performances of the VIRTS instrument are analyzed in detail. In particular the modeling of the signal to noise ratio is the main argument of this chapter. 

The third chapter try to answer the question: how does VIRTIS manage cometary spectra? Here simulations of possible spectra of the comet's nucleus are performed, which are useful for both a comparison with real spectra, and for a planning of the observations. In this chapter it is also introduced the Hapke's radiative transfer model we intend to use to invert acquired data to infer physical properties.

The fourth chapter introduces a method for spectral modeling. It takes advantage of fast algorithms for optimization of least square problems to perform accurate retrieval of the investigated parameters. Moreover, the method includes the information on the instrumental noise, permitting the analysis of the goodness of the models, and an estimation of the error of the retrieved parameters.

In order to be prepared to manage VIRTIS data of the comet, we followed two different approaches: analysis of comet's data acquired by another spectrometer and of VIRTIS data of another target (fifth and sixth chapters). 

In the fifth chapter data coming from previous missions to comets are analyzed: Deep Impact to Tempel 1 and its extended investigation to Hartley 2.  Among the scientific payload of Deep Imapct, the probe carries a spectrometer similar to VIRTIS (HRI). Its data allowed us to test the reliability of the method developed and to produce further investigations on those targets.  

In the sixth chapter analysis of another small body is presented: Lutetia asteroid. Its data are coming from VIRTIS itself thanks to a flyby of Rosetta during its cruise phase.  

This work have paved the way to the analysis of the final target. The tools presented are currently used by the VIRTIS-Team to produce works on the comet, that are recommended to the reader. Since a complete analysis on the comet is outside the scope of this work, just preliminary results are shown here.

\chapter{Comets and space missions}       
\section{Cometary science history}       
\label{sec:history}

Comets are impressive celestial phenomena: bright blotches of light with long, beautiful tails, suddenly appearing across the sky. They travel on elliptical orbits that swing around our Sun and return back, in most cases beyond the orbit of Neptune. These incredibly elongated orbits mean that it can take years, from a few to hundreds or even thousands, for a comet to reappear to an observer on the Earth.

The seemingly transient and unpredictable behavior of comets rendered them completely mysterious to ancient observer. Rather than being viewed as beautiful and intriguing, comets were instead regarded as omens of impending doom, or meteorological phenomena in terrestrial atmosphere and were not yet clearly established as astronomical bodies.

\begin{figure}
\centerline{\includegraphics[width=0.4\textwidth,clip=]{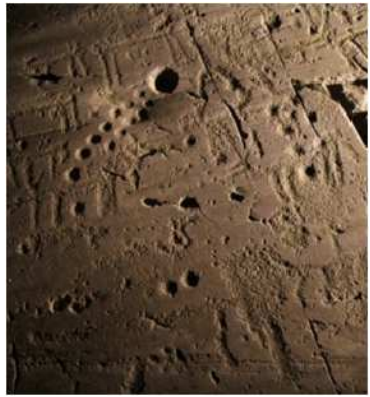}}
\caption{Comet symbol on Rock 35 from Area di Foppe, Nadro di Ceto (Brescia, Italy). Riserva naturale Incisioni rupestri di Ceto, Cimbergo e Paspardo. Copyright: Centro Camuno di Studi Preistorici, Dipartimento Valcamonica e Lombardia}
\label{Fig:Valcamonica}
\end{figure}

Humans have been captivated by comets for much of history. Some archaeologists suggest that prehistoric rock paintings, found in several sites across the globe, may portray comets. Early rock carvings resembling comets have been found in Scotland dating back to the second millennium BC, and a comet-like shape found on rock carvings in Val Camonica, Italy, dates back to the late Iron Age (see Fig. \ref{Fig:Valcamonica}).

The word ``comet'' comes from the Greek ``kometes'', which literally means ``long-haired'', but the earliest extant records of cometary observations date from around 1000 BC in China \citep{Ho1962} and probably from about the same time in Chaldea. 

Ancient Greek philosophers put forward a number of different ideas to explain the physical nature of comets. To some, comets were celestial bodies just like planets. To others, they were linked to celestial mechanics such as planetary conjunctions, and to yet others they were burning clouds or optical phenomena in the Earth's atmosphere. Although far from correct, the latter view was adopted by Aristotle in the fourth century BC.

Aristotle suggested that they were ``windy exhalations'' from the Earth that reached out into our atmosphere. The Aristotelian ideas about planets and comets were upheld for an entire millennium during which there was little scientific advancement in the field of astronomy.

The only voice that dared contradict this view was that of Seneca, in the first century BC. According to Seneca's \textit{Naturales Quaestiones}, comets were more like planets. He also recognized that their sporadic behavior would make studying them quite complex, but hoped that future observations and investigations would shed light on their nature.

Aristotle's view of the heavens dominated almost undisputed throughout the Middle Ages. This situation started to change, however, with the dawn of the Renaissance in Europe. New ideas emerged as astronomers assembled an increasingly larger body of data. 

In the fifteenth century, Italian scientist Paolo dal Pozzo Toscanelli conducted the first systematic observations of several comets. He recognized the importance of measuring both the orientation of the comet's coma, the nebulous envelope surrounding the comet's nucleus, and the position of the nucleus. 

In the sixteenth century, German astronomer Peter Apian and the Italian scientist Girolamo Fracastoro realized that a comet's tail always points away from the Sun, a revelation that would be key to correctly interpreting these objects and their physics.

The exceptionally accurate observations of the Danish astronomer Tycho Brahe (1578) initiated a new era for observational astronomy, as he demonstrated that the horizontal parallax of the bright Comet C/1577 V1 was certainly smaller than 15 arcmin, corresponding to a distance in excess of 230 Earth radii, consequently farther away than the Moon. 

The German astronomer Johannes Kepler observed several comets throughout his life, and tried to study their motion. However, he believed comets to be interstellar objects moving along straight lines, and so never fully grasped their dynamics. 

Some progress on understanding cometary orbits came from Polish astronomer Johannes Hevelius, who suggested in his \textit{Cometographia} treatise, published in 1668, that comets moved on open-ended parabolic orbits, meaning that they would be seen once and never again. 

\begin{figure}
\centerline{\includegraphics[width=0.6\textwidth,clip=]{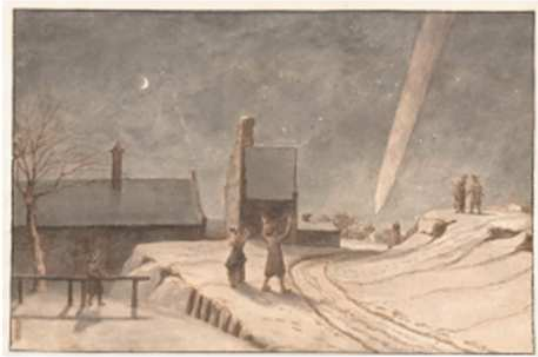}}
\caption{The Great Comet of 1680 over the Dutch town of Alkmaar, in a painting by Lambert Doomer. Rijksmuseum, Amsterdam. In public domain.}
\label{Fig:gratcomet}
\end{figure}

The English physicist and mathematician Isaac Newton, in his \textit{Philosophiae Naturalis Principia Mathematica}, developed the brilliant tool that could link all these observations, together with the studies of Galileo Galilei. With his new theory, Newton calculated the orbit of the Great Comet of 1680 (Fig.  \ref{Fig:gratcomet}). His results were in excellent agreement with observations. Shortly after, the English astronomer Edmond Halley applied Newton's theory to the study of comets. This work would produce two major results: a striking confirmation of Newton's theory of universal gravitation, and a revolutionary change in our understanding of comets. With this, Halley was the first to suggest that comets may be periodic, moving along very elongated ellipses rather than parabolic paths. With an estimated period of about 76 years, Halley expected the comet to be visible again between the years of 1758 and 1759. Halley died before he could view it with his own eyes, but others were able to test and confirm his conjecture and, with it, Newton's theory of gravity, demonstrating that it can be applied not only to planets but to other celestial bodies too. It also marked one of the greatest triumphs of science.

Although it seemed like our understanding of comets was becoming more complete, one vital piece was missing from this puzzle: where do they come from? The debate shifted from nature to origin thanks to the German philosopher Immanuel Kant, who wrote his \textit{General History of Nature and Theory of the Heavens} in 1755. In this early work, Kant suggested that the Sun and its planets formed from an extended diffuse nebula. However, the idea of an interstellar origin of comets prevailed until the second half of the nineteenth century, when the Italian astronomer Giovanni Schiaparelli suggested that comets belong to the Solar System and surround the Sun in an almost uniform cloud. He also established the link between comets and meteors, observing that the Perseid and Leonid meteor streams coincided with the orbits of Comets 109P/Swift-Tuttle (1862 III) and 55P/Tempel-Tuttle (1866 I) respectively. This was the proof that comets were indeed losing solid particles. 

After Halley, the German astronomer Johann Franz Encke (1820) was the second to successfully predict the return of a comet (in 1822). Comet 2P/ Encke has the shortest period of all known comets, 3.3 years, which provides similar Sun-Earth-comet configurations every 10 years. The comet was subsequently found to arrive systematically at perihelion about 0.1 days earlier than predicted, even when taking planetary perturbations into account. 

Inspired by his observations of an asymmetric distribution of luminous matter in the head of Comet Halley in 1835, the German astronomer Friedrich Bessel (1836) interpreted this as a Sun-oriented asymmetric outflow and suggested that a non-gravitational effect might arise due to the rocket-type impulse imparted by such an outflow, possibly explaining perihelion shifts such as those observed for Comet 2P/Encke. It would take over a century for this idea to be fully accepted by the scientific community because the existence of a solid body at the center of the coma was far from being unanimously accepted. In fact, the theory that comets were a swarm of solid particles was the most favored by scientists at that time.

While astronomers were making great progress using telescopes, the arrival of photography in the mid-nineteenth century opened up a new way to study our skies. The first comet to be photographed was the Great Comet of 1858, also known as Donati' s Comet after Italian astronomer Giovanni Donati who discovered it. A few years later, Donati was also the first astronomer to use spectroscopy to study the composition of a comet. By splitting the light from celestial bodies into its constituent colours through a prism, spectroscopy allows astronomers to investigate the chemical composition of distant and otherwise inaccessible objects. Donati recorded the spectrum of a comet, now known as Comet C/1864 N1, that had been discovered by German astronomer Ernst Wilhelm Tempel in 1864. The spectrum contained three features that are now known to be produced by molecules of diatomic carbon ($C_{2}$) in the comet's coma. Further observations around the turn of the twentieth century uncovered more about the chemical makeup of cometary comas, identifying sodium ions and a variety of carbon-, oxygen- and nitrogen-based molecules. 

In the first half of the twentieth century astronomers were collecting and studying more high-quality astronomical data than ever before, building up an impressive database. This allowed them to delve into the physical nature and origin of comets in great detail.

In 1950, American astronomer Fred Whipple proposed a new model to describe comets. Rather than a loose collection of dust and debris kept together by ice, he suggested that comets have an icy nucleus, consisting primarily of frozen volatiles like water, carbon dioxide, methane, and ammonia, and containing only traces of dust and rock. 

Then, the German astronomer Ludwig Biermann (1951) gave the correct explanation for the motions of features in cometary plasma tails caused by their interactions with a flow of charged particles emanating from the Sun's surface (i.e., the solar wind). 

Finally, from dynamical studies of the distribution of semimajor axes of comets, came the identification by the Dutch astronomer Jan Oort (1950) of a distant population of comets now known as the Oort cloud, a huge cloud of ``dormant'' comets extending over a thousand times farther than the orbits of Neptune and Pluto. This cloud is not the only reservoir of comets in the Solar System; as suggested by Dutch astronomer Gerard Kuiper in 1951, most comets with a relatively short period are located in a flattened, ring-like distribution that begins just outside Neptune's orbit, now known as the Kuiper belt. 

None of these ideas resulted directly from new observational evidence, but this was the first time that the known facts were effectively combined, leading to a comprehensive description. A new picture of comets, and the existence of a family of celestial bodies, were suddenly revealed at the same time.

Advances in technology towards the end of the twentieth century provided exciting new opportunities for comet scientists. Whereas before scientists had relied solely on ground-based telescopes to observe and study comets, with the advent of the space age  there was the novel opportunity to approach these icy bodies as they journeyed towards the inner regions of the Solar System, passing relatively close to Earth.

In 1985, NASA's International Cometary Explorer (ICE) became the first space mission to approach a comet passing approximately 7800 kilometres from the nucleus of Comet 21P/Giacobini-Zinner. 

Just one year later, when Halley's Comet returned to the vicinity of the Sun, an armada of spacecraft was sent to study the comet up close: these included two probes from Russia (Vega-1 and Vega-2), two from Japan (Sakigake and Suisei), and ESA's Giotto spacecraft.

\begin{figure}
\centerline{\includegraphics[width=0.6\textwidth,clip=]{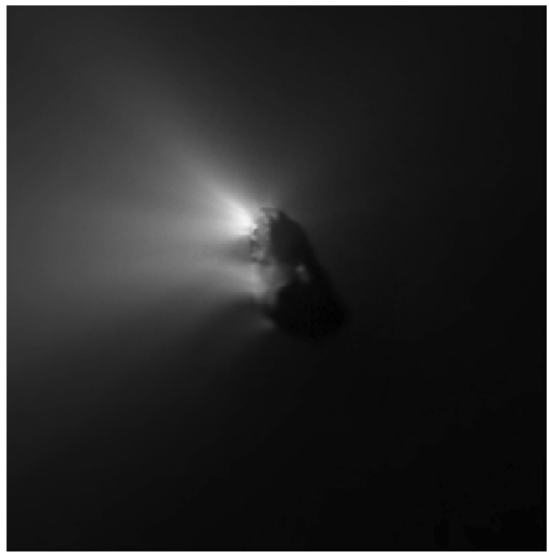}}
\caption{The nucleus of Comet Halley by ESA's Giotto in 1986, from a distance of about 2000 km. Credit: ESA/MPS}
\label{Fig:halley}
\end{figure}

Named after the Italian painter who, in 1303, depicted the star of Bethlehem as a comet, the Giotto mission obtained the closest pictures ever taken of a comet (Fig. \ref{Fig:halley}). After Vega-1 and Vega-2 came within 8900 and 8000 kilometres respectively from the nucleus of Halley's Comet in early March 1986, ESA scientists and engineers used the telemetry data from these probes to guide the Giotto spacecraft even closer (reported from \textit{http://sci.esa.int/rosetta/}).

On 13 March 1986, Giotto flew past the comet's nucleus at a distance of less than 600 kilometres, relaying unprecedented images. It is only with this first spacecraft flyby that the idea of a single solid body at the center of the comet activity was widely accepted by scientific community  \citep{Keller,Reinhard,Sagdeeva,Sagdeevb}.

The nucleus  was found to be larger (equivalent radius 5.5 km) and darker (albedo $\approx 4\%$ )  than expected. In the Giotto images, surface features (craters, ridges, mountains, etc.) and source regions were observed \citep{Keller2}. 

The coma was found to be highly structured on all scales (jets, shells, ion streamers, etc.) and the gaseous component was analyzed in situ by mass spectroscopy. Signals at atomic masses of 1 and from 12 to $\approx55$ amu were detected. $H_{2}O$ was confirmed to represent 85\% by weight of the gas phase, and the likely presence of large organic polymeric molecules was indicated. The dust was analyzed by size and composition and there was an unexpectedly high fraction of very small grains, down to the sensitivity limit of $\approx10^{-19}$ kg. Particles rich in metals and silicates were found as expected, but particles rich in H, O, C, and N (``CHONs'') were seen for the first time and were thought to be related to the smallest grains mentioned above \citep{Kissel}. The integrated mass loss experienced by the nucleus at this passage, was $\approx 0.5\%$ of the total mass of the nucleus.  The various predicted plasma effects were confirmed, including the existence of a bow shock, and the adjacent interplanetary medium was found to be kinematically and magnetically extremely turbulent \citep{Festou}. 

Since the turn of the new millennium, four more comets have been visited by various NASA spacecrafts. 

Comet 19P/Borrelly was visited by the Deep Space 1 in 2001 \citep{Soderblom1} and the first ever disk-resolved photometric study was performed on a cometary surface \citep{Buratti,Li1}. However, since all images were obtained with a single filter, color information for Borrelly was not available. The infrared spectrometer onboard Deep Space 1 measured the temperature with a single one-dimensional scan \citep{Soderblom2}.

A few years later in 2004, the Stardust mission flew by Comet 81P/Wild 2 \citep{Brownlee}, collecting samples from the comet's coma using an aerogel collector and returning them to Earth via a sample-return capsule in 2006. The comet samples suggested that high-temperature inner Solar System material formed and was subsequently transferred to the Kuiper Belt \citep{Matzel}. Moreover, in 2009, glycine, an amino acid, was detected in the gathered material \citep{Elsila}. 

Comet 9P/Tempel 1 was visited in 2005 and 2011 respectively by Deep Impact \citep{AHearn} and Stardust extended mission \citep{Veverka,Li2}. Deep Impact mission was also extended, and in 2010 visited the very active comet 103P/Hartley 2 \citep{Li3,Groussin1}. Main goal of the mission was achieved with the detection of water ice ejected from the interior of Tempel 1 as a result of the collision of the impactor dropped by the probe \citep{Sunshine1}. Deep Impact, allowed a direct observation of the inner coma and thus of the parent species, directly linked to the nucleus composition \citep{Feaga}. Moreover, Deep Impact permitted a first map of the surface temperature of a comet \citep{Groussin2}, and the first spectral detection of exposed water ice on the nucleus surface \citep{Sunshine06}. For further details see Chapter \ref{DI}.

\section{Comets: general view and open questions}
\label{sec:questions}
Comets are the most primitive observable objects remaining from the era of planetary formation. As such, they provide information on the thermophysical conditions of the protoplanetary disk and on the formation mechanism for the icy planetesimals from which the cores of the outer planets were built. Furthermore, the physical evolution of cometary nuclei over the past 4.6 Gy must be explained within the context of any unified theory of the solar system, and comparative studies of cometary nuclei and dynamically related bodies should provide insights into the physical and collisional histories of these objects.
Through impacts over the age of the solar system, cometary nuclei have significantly affected the formation and evolution of planetary atmospheres and have provided an important source of volatiles, including water and organic material, to the terrestrial planets. Another important motivation for studying cometary nuclei is that their bulk properties may dictate what steps should be taken for hazard mitigation in the event of a predicted collision \citep{Lamy,Festou}.

Although the long history of cometary science we have still not or few information of basic physical parameters like distribution of size and mass of the cometary nuclei. 
Because of they are small bodies enveloped in their coma it is possible to have resolved images only with the most powerful telescopes on Earth, or with is situ missions. The principal problem with the ground observations is that estimates of the size distribution from optical data require an assumed albedo. Many of the data, furthermore, consist of only single observations rather than complete rotational lightcurves, which adds scatter to the distribution. The cumulative size distribution (CSD) of the ECs obeys a single power law with an exponent $1.9 \pm 0.3$ down to a radius of ~1.6 km. Below this value there is an apparent deficiency of nuclei, possibly owing to observational bias and/or mass loss \citep{Lamy}. 
Regarding the mass we must emphasize that we still do not have a single measured mass for a cometary nucleus, and thus not a single, measured density. A lower limit of $\sim 0.6 \: g/cm^{3}$ for the nucleus bulk density could be found by combining rotational periods (from 5 to 70 h) and shape data when available to ensure stability against centrifugal disruption \citep{Lamy}.

Determination of the interior structure is limited to what have been learned from the Deep Impact
mission until the results of Rosetta  rendezvous and soft lander mission (see section \ref{sec:rosetta}). Deep Impact have been excavate a large crater with a probe that impacted with the nucleus of Comet 9P/Tempel 1 in order to study the outermost tens of meters. The expansion rate of the ejecta plume imaged during the look-back phase of observations leads to an estimate of comet mass ($2.3 - 12.0 \times 10^{13}$ kg) and a bulk density of  $0.2 - 1.0$ $g/cm^{3}$ with large error due to uncertainties in the magnitude of coma gas pressure effects on the ejecta particles in flight. Medium-to-high porosity of the material was also established from ejecta plume behavior and from the size of the crater \citep{Richardson}. Moreover, implications for internal stratigraphy of devolatilized material and water ice of Tempel 1 were also determined \citep{Sunshine1}.

Our knowledge of composition is limited almost entirely to the coma, the spectra of nuclei being almost featureless, with the exception of the water ice features seen on the surface of Tempel 1 and Hartley 2. Other than that, the surfaces of comets are known only to be very dark, presumably from a combination of particle shadowing due to porosity and the inclusion of very dark, carbonaceous material as one of the abundant components at the surface. We are thus faced with the problem of deciding the extent to which the composition in the coma is representative of the composition in the nucleus \citep{AHearn2}. 
Interpretation of the chemistry of the coma is further limited by the fact that most of the species observed, including virtually all the easily observed species at optical and ultraviolet wavelengths, are clearly fragments of larger molecules that existed in the nucleus.
Furthermore, the relative abundances of species vary dramatically with heliocentric distance even in a single comet. A few cases of variation can be explained in terms of processes in the coma, but most are not explained at all and in virtually no case is there consensus that we can correct for the variation with
heliocentric distance adequately to say something definitive about nuclear abundances \citep{AHearn2}. Our knowledge of the composition of the gaseous coma is quite extensive, coming from remote sensing at wavelengths from the X-ray to the radio and from in situ measurements. Water is the most abundant constituent of cometary ices and its production rate is used for quantifying cometary activity and for abundance determinations. Other molecules whose production rate have been observed to overcome 1\% relative to water ice are: $CO, CO_{2}, CH_{4}, CH_{3}OH, H_{2}CO, NH_{3}, H_{2}S$. Other sulfur, nitrogen, and CHO - bearing molecules are frequently observed \citep{Bock}. 

The refractory species are much less well known. Remote sensing has brought us primarily the silicate feature, including identification of crystalline olivine, and pyroxene. In situ measurements brought us CHON particles, whose presence filled a major gap in our understanding of the overall abundances, since combining these particles with the volatiles leads to more or less solar abundances \citep{AHearn2}.

Comets contain some of the least-altered material surviving from the early solar nebula. Cometary dust may contain both presolar particulates and solar nebula condensates. Their structure and mineralogy may hold important clues about the chemical and physical processes in the early solar system. The key questions in this area are (1) whether interstellar ices survived the accretion shock and were incorporated directly into comets, (2) whether any chemical reactions (either in the gas phase or on grain surfaces) were important in that part of the accretion disk in which comets formed, and therefore, (3) whether or not the details of the abundances of ices in comets are good constraints on the conditions in the early solar system  \citep{AHearn2}.

Finally, among the main issues, there is the role of comets at the terrestrial planets. Did comets deliver most of Earth's water and organics? The difference
in D/H ratios between comets and terrestrial ocean water has provided clues in this direction. Measurements of this ratio in several Oort cloud comets resulted in a value twice as high as that in the Earth oceans, leading to the generally accepted conclusion that comets are unlikely to be the primary source of ocean water. But measurement of the D/H ratio of 103P/Hartley 2 resulted to be consistent with terrestrial value \citep{Hartogh}. 

\section{Rosetta space mission}
\label{sec:rosetta}
\begin{figure}[!h]
\centerline{\includegraphics[width=0.7\textwidth,clip=]{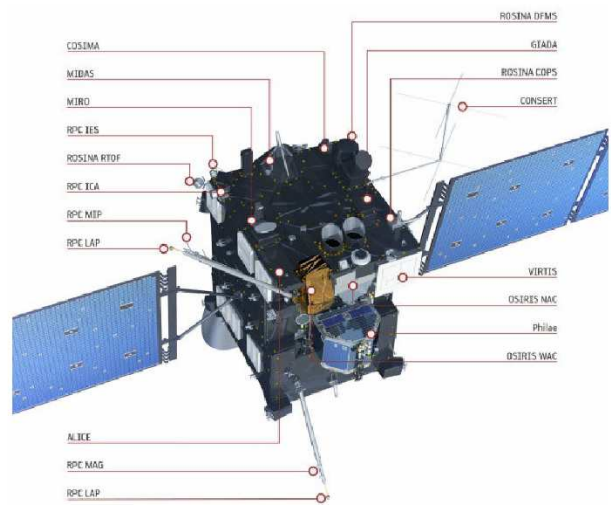}}
\caption{The Rosetta orbiter and the scientific instruments onboard.}
\label{fig:orbiter}
\end{figure}

\begin{figure}[!h]
\centerline{\includegraphics[width=0.7\textwidth,clip=]{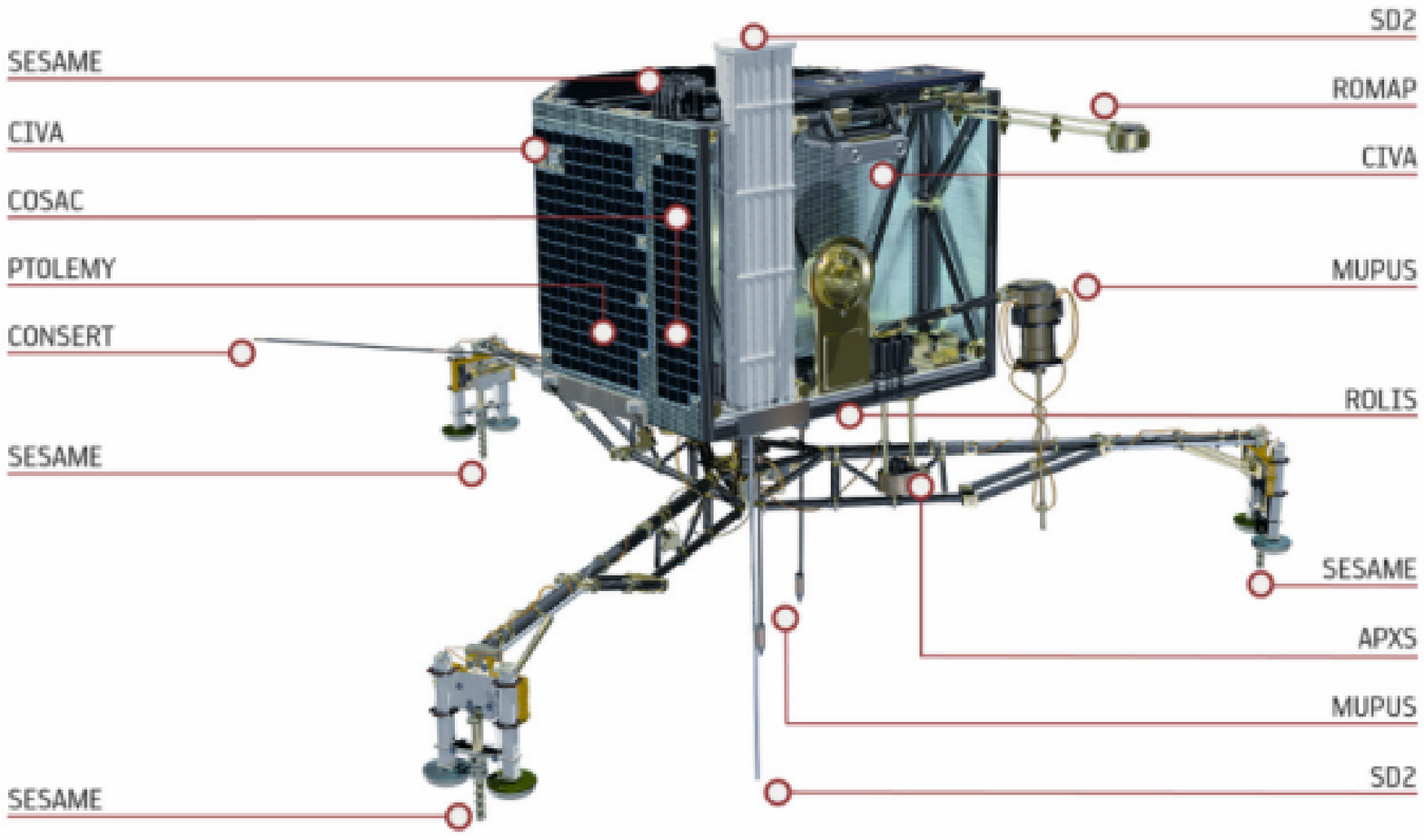}}
\caption{The Rosetta lander and the scientific instruments onboard.}
\label{fig:lander}
\end{figure}

The International Rosetta Mission, approved in November 1993 by the Science Programme Committee of the European Space Agency (ESA), is a logical follow-up mission of ESA's very successful first mission, Giotto, to comet 1P/Halley \citep{Reinhard}. The prime scientific objectives of the Rosetta mission are to investigate the origin of our solar system by studying the origin of comets (see section \ref{sec:questions}). It consists of two mission elements, the Rosetta orbiter (see Table \ref{tab:orbiter} and Figure \ref{fig:orbiter}) and the Rosetta lander Philae (see Table \ref{tab:lander} and Figure \ref{fig:lander}). The mission is named after a plate of volcanic basalt currently in the British Museum in London, called the Rosetta Stone. This plate was the key to unravel the civilization of ancient Egypt. Just as the Rosetta Stone provided the key to an ancient civilization, the scientific instruments onboard the Rosetta spacecraft are designed to unlock the mysteries of the oldest building blocks of our solar system, the comets \citep{Glassmeier}.

Rosetta will be the first spacecraft to orbit a cometary nucleus and the first spacecraft to fly alongside a comet as it heads towards the inner Solar System. This allows for an unprecedented opportunity to examine from close proximity how a comet, coming from deep space, is transformed while heated by the Sun. The lander Philae will perform the first controlled touchdown of a robotic lander on a comet nucleus, and therefore the mission will provide the first images taken from the cometary surface and the first in situ analysis results of cometary material. The Rosetta mission also comprises the first flybys of two main-belt asteroids.
Rosetta is also the first spacecraft to fly close to the Jovian orbit using solar cells as its main power source. This is due to its large solar panels and the use of specially optimized low intensity low temperature (LILT) technology \citep{Strobl}. 

\begin{table}[!h]
\caption{The Rosetta Orbiter (11 science instrument packages)}
\begin{center}
\begin{tabular}{||c|c||}
\hline
\hline
ALICE & Ultraviolet Imaging Spectrometer\\
\hline
CONSERT & Comet Nucleus Sounding experiment \\
\hline
COSIMA & Cometary Secondary Ion Mass Analyser \\
\hline
GIADA & Grain Impact Analyser and Dust Accumulator \\
\hline
MIDAS & Micro-Imaging Analysis System \\
\hline
MIRO & Microwave Instrument for the Rosetta Orbiter \\
\hline
OSIRIS & Rosetta Orbiter Imaging System \\
\hline
ROSINA & Rosetta Orbiter Spectrometer for Ion and Neutral Analysis \\
\hline
RPC & Rosetta Plasma Consortium \\
\hline
RSI & Radio Science Investigation \\
\hline
VIRTIS & Visible and Infrared Thermal Imaging Spectrometer \\
\hline
\end{tabular}
\end{center}
\label{tab:orbiter}
\end{table}

\begin{table}[!h]
\caption{The Rosetta Lander (10 science instrument packages)}
\begin{center}
\begin{tabular}{||c|c||}
\hline
\hline
APXS & Alpha Proton X-ray Spectrometer \\
\hline
CIVA and ROLIS & Rosetta Lander Imaging Systems \\
\hline
CONSERT & Comet Nucleus Sounding experiment \\
\hline
COSAC & Cometary Sampling and Composition experiment \\
\hline
MODULUS & PTOLEMY	Evolved Gas Analyser \\
\hline
MUPUS & Multi-Purpose Sensor for Surface \\&and Subsurface Science \\
\hline
ROMAP & Rosetta lander Magnetometer and Plasma Monitor \\
\hline
SD2 & Sample and Distribution Device \\
\hline
SESAME & Surface Electric Sounding \\&and Acoustic Monitoring Experiment \\
\hline
RSI & Radio Science Investigation \\
\hline
\end{tabular}
\end{center}
\label{tab:lander}
\end{table}

The idea that became Rosetta was conceived in the early 1980's even before ESA's Giotto mission flew by comet 1P/Halley, returning the first detailed picture of a comet's nucleus ever seen. The success of Giotto meant that plans for a follow-on mission were enthusiastically considered. Almost 20 years later, Rosetta was built, tested and ready to launch. But disaster struck just one month before the planned liftoff. In December 2002, an Ariane 5, similar to the one designated to launch Rosetta, failed while lifting a communications satellite into orbit. The difficult decision was made to postpone the attempt until the launcher failure was understood. This robbed the mission of its original target, comet 46P/Wirtanen. While the engineers worked to understand and prevent the loss of another Ariane 5, scientists and engineers searched for a replacement target. Eventually they settled on comet 67P/Churyumov-Gerasimenko, a somewhat more massive comet than Wirtanen. This led to the strengthening of the legs on Rosetta's lander Philae, to cope with the slightly faster landing speed now expected. The launch took place on 2 March 2004. 

Rosetta could not head straight for the comet. Instead it began a series of looping orbits around the Sun that brought it back for three Earth fly-bys and one Mars fly-by. Each time, the spacecraft changed its velocity and trajectory as it extracted energy from the gravitational field of Earth or Mars. During these planetary fly-bys, the science teams checked out their instruments and, in some cases, took the opportunity to carry out science observations coordinated with other ESA spacecraft such as Mars Express, ENVISAT and Cluster (\textit{http://sci.esa.int/rosetta/}). 
From 8 June 2011 to 20 January 2014, because too far from the Sun, there would not be enough power to keep all the spacecraft's systems operating. For this reason the spacecraft has been placed in hibernation, when it was in deep space cruise \citep{Glassmeier}. Between wake-up and rendezvous with the comet, all twenty-one instruments had to be brought online and checked out. Software was updated and the spacecraft had to perform a series of 10 manoeuvres to reduce its speed sufficiently to rendezvous with the comet rather than fly by it (\textit{http://sci.esa.int/rosetta/}).

The Rosetta mission is comprised of twenty-five experiments, making it unprecedented in scale. 
The measurement goals of the Rosetta mission include \citep{SS}:
\begin{itemize}
\item a global characterization of the nucleus,
\item the determination of its dynamic properties,
\item the surface morphology and composition,
\item the determination of chemical, mineralogical and isotopic compositions of volatiles and refractories in the cometary nucleus,
\item the determination of the physical properties and interrelation of volatiles and refractories in the cometary nucleus,
\item studies of the development of cometary activity and the processes in the surface layer of the nucleus and inner coma, that is dust/gas interaction,
\item studies of the evolution of the interaction region of the solar wind and the outgassing comet during perihelion approach.
\end{itemize}

To achieve these goals Rosetta combines two strategies for characterizing the properties of a cometary nucleus. On one hand, the comet’s evolution along the orbit with decreasing heliocentric distance will be investigated with the orbiter instruments by monitoring the physical and chemical properties of the nucleus and in situ analysis of the near-nucleus environment. On the other hand, the Rosetta lander Philae will provide ground truth by analyzing the nucleus material directly \citep{Glassmeier}.

The Main ROSETTA Target Comet 67P/Churyumov-Gerasimenko is a short-periodic comet from the class of the Jupiter Family comets (JFC) \citep{Lamy2}. It was discovered on September 9, 1969 by Klim Ivanovic Churyumov and Svetlana Ivanovna Gerasimenko at the Alma Ata Observatory, Tadchik Republic.
JFCs are believed to originate from the Kuiper Belt. Since the size distribution in the Kuiper Belt is collision-dominated for bodies smaller than about 50 km, JFCs may actually represent the low-mass fragments from collision events in the belt. They were ejected from the Kuiper Belt region by Neptune. Repeated gravitational
scattering at the outer major planets let them cascade inward. Backward calculations have revealed several encounters with Jupiter over the past 200 years with the one on February 4, 1959 being the closest one at just 0.052 AU distance from the planet \citep{Belyaev}. This close encounter has significantly changed the orbit and the perihelion distance from 2.7 AU before to 1.2 AU after, bringing the comet within reach for the Rosetta mission \citep{Glassmeier}.

During cruise to its main target the Rosetta spacecraft has also visited two asteroids, namely 2867 Steins and 21 Lutetia. Both objects are main belt asteroids.

\begin{figure}
\centerline{\includegraphics[width=0.5\textwidth,clip=]{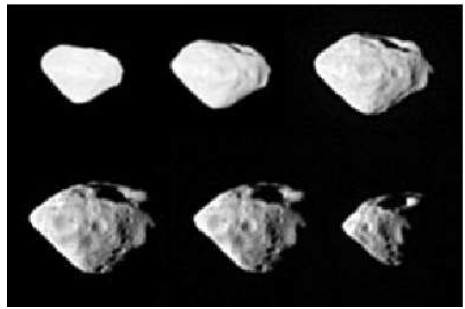}}
\caption{Steins as imaged by the camera OSIRIS onboard Rosetta during the flyby on 5 September 2008, at a distance of 800 km (\textit{http://sci.esa.int/rosetta/}).}
\label{fig:steins}
\end{figure}

Steins had been classified as a rare E-type asteroid based on its visual and near-infrared spectrum, and its high albedo. The observations by OSIRIS (Figure \ref{fig:steins}) and VIRTIS on Rosetta brought new information that could not have been gained from the ground. The dimensions of Steins were found to be $6.67 km \times 5.81 km \times 4.47 km$. Steins was probably part of a larger differentiated object that had broken up. It was later struck by other objects, creating impact craters. However, the interior is thought to be a rubble pile and the asteroid will eventually break up (\textit{http://sci.esa.int/rosetta/}).
Steins probably gained its diamond shape from the YORP effect - the Yarkovsky - O'Keefe - Radzievskii - Paddack effect - in which photons from the Sun are re-radiated as infrared emission taking momentum from the body and altering the rotation rate. In this case, the change in rotation rate resulted in some material moving towards the equator of asteroid \citep{Keller3}. Surface reshaping could be also due to the YORP effect, accounting for the lack of smaller craters on Steins. The effect would have caused landslides to fill in the smaller craters. This is the first time that the YORP effect has been seen in a main-belt asteroid (\textit{http://sci.esa.int/rosetta/}).

On 10 July 2010, ESA's Rosetta spacecraft flew past asteroid (21) Lutetia, one of the largest objects orbiting within the main asteroid belt between Mars and Jupiter.

\begin{figure}
\centerline{\includegraphics[width=0.5\textwidth,clip=]{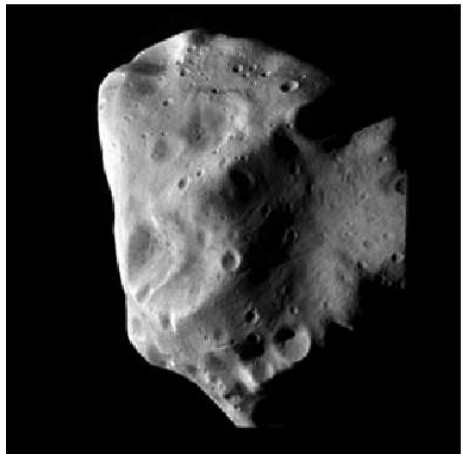}}
\caption{Lutetia as imaged by the camera OSIRIS onboard Rosetta on 10 July 2010. Rosetta flew past at a greater distance, 3162 km, in order to allow the full asteroid to appear in the field of view of the scientific cameras (\textit{http://sci.esa.int/rosetta/}).}
\label{fig:lutetia}
\end{figure}

Discovered in 1852, Lutetia was among the first objects to be classified as an M-type (metallic) asteroid, but radar observations revealed an unusually low albedo, or reflectivity, that was inconsistent with a metallic surface. Meanwhile, spectra obtained at visible and infrared wavelengths found similarities with meteorites known as enstatite chondrites and with carbonaceous chondrite meteorites, typically associated with C-type asteroids.
The flyby of Lutetia took place on 10 July 2010, when Rosetta flew past the asteroid at a distance of 3170 km.
As Rosetta closed on the rotating asteroid, the OSIRIS imaging system returned 462 pictures of the illuminated northern hemisphere, showing more than 50 per cent of the asteroid's surface. OSIRIS revealed an irregular, cratered object (Figure \ref{fig:lutetia}) whose overall dimensions were $121 km \times 101 km \times 75 km$. Its diverse surface displayed a variety of small, localised features, as well as a number of geological features that were on a far larger scale than any seen on other asteroids.
More than 350 craters were identified with diameters between 600 metres and 55 km and depths of up to 10 km. However, the crater density varied considerably, the regions with the largest number of craters were clearly older than those with fewer craters (\textit{http://sci.esa.int/rosetta/}).
Its ancient surface age (determined from crater counting) coupled with its complex geology and high density suggest that Lutetia is most likely a primordial planetesimal \citep{Sierks}. This was also confirmed by the spectroscopic observations performed by the VIRTIS instrument \citep{CoradiniL}, which have shown no absorption features, of either silicates or hydrated minerals.

\chapter{VIRTIS instrument performance analysis}       

\section{VIRTIS}       
\label{sec:virtis}
Spectrophotometry is a very powerful diagnostic tool in remote sensing to study the composition and the physical properties of the surfaces of objects under investigation. The amount of solar radiation, as a function of the wavelength, scattered from a surface towards the observer, is a function of several parameters such as the composition of the materials making up the surface, their grain size, the porosity, the surface roughness and the scattering properties of the regolith. On the other hand the thermal emission, is a function of temperature and emissivity, which give clues on the composition, thermal inertia and active spots on the surface.

The identification of the nature of the main constituents of the comets is a primary goal of the Rosetta mission, and particularly for VIRTIS \citep{Coradini} (Visual InfraRed and Thermal Imaging Spectrometer) that is part of the scientific payload of the Rosetta Orbiter.

The VIRTIS experiment has been one of the most successful experiments built in Europe for Planetary Exploration. It was developed under the supervision of Angioletta Coradini. 

The VIRTIS scientific and technical teams have taken advantage of their previous experience in the design and development of spectrometers for space applications. In fact the various groups contributing to the VIRTIS experiment from Italy France and Germany have been deeply involved in the CASSINI mission with the experiments VIMS and CIRS.

The instrument is composed by a high spectral resolution channel (-H) and two high spatial resolution channels devoted to spectral mapping (-M), covering the spectral range 0.22 - 1.04 $\mu m$ (Visible) and 0.95 - 5.06 $\mu m$ (Infrared), both with 432 spectral bands. 
VIRTIS will detect and characterize the evolution of specific spectral features, such as the typical spectral bands of minerals and ices, arising on the nucleus surface and in the coma. It will also study the surface thermal evolution during comet's approach to the Sun. Moreover, it will perform photometric analysis of the surface thanks to observations at different phase angles with resolution up to the few-meter scale along the mission duration. 
This work will focus on the M channel. Here we intend to show the performance analysis in order to characterize the Signal to Noise ratio.

\begin{figure}
\centerline{\includegraphics[width=0.8\textwidth,clip=]{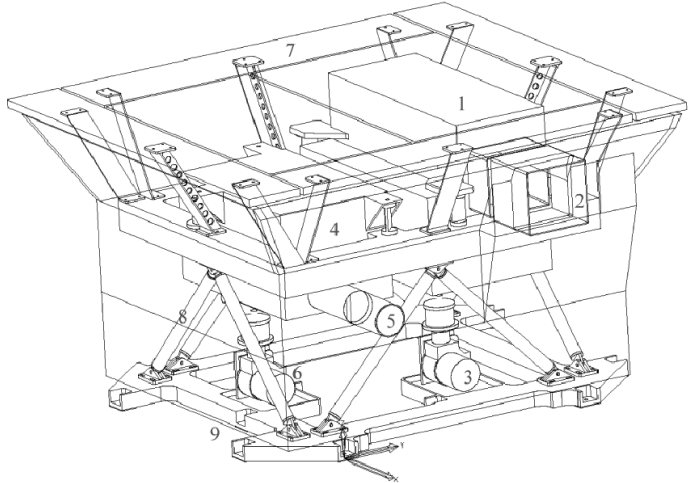}}
\caption{Scheme of VIRTIS: 1. VIRTIS-M; 2. Aperture VIRTIS-M; 3. Cryocooler VIRTIS-M; 4. VIRTIS-H; 5. Aperture VIRTIS-H; 6.Cryocooler VIRTIS-H; 7. Radiator; 8. Isostatic mount in titanium; 9. Pallets baseplate (Selex ES).}
\label{fig:virtis}
\end{figure}

\begin{figure}
\centerline{\includegraphics[width=0.5\textwidth,clip=]{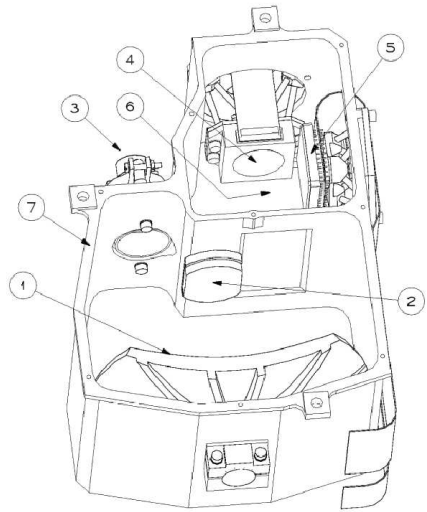}}
\caption{Offner spectrometer of VIRTIS-M: 1) mirror; 2) diffraction grating; 3) the entrance slit and electromechanical shutter; 4) infrared focal plane; 5) visible focal plane; 6) position of the mirror on visible focal plane (not shown); 7) mechanical structure (Selex ES).}
\label{fig:offner}
\end{figure}

VIRTIS-M shares the same optical system to analyze two spectral channels (VIS and IR). The angular resolution is 250  $\mu rad$ for a single pixel with a FOV of 64 $mrad$ for the whole slit (256 pixels). The spectral resolution is 1.8 nm/band for VIS and 9.8 nm/band for IR channel.
The instrument is designed as a rigid structure in order to resist to vibrations caused by the launch, and to maintain a high alignment of the two optical heads in flight conditions. Two cryocoolers (used for active cooling of the focal planes of the IR channels M and H), the boxes of electronics and eight pillars in titanium for the upper part of the instrument are mounted on the pallet baseplate (349 x 409 mm). This is located at the base of the structure and acts as a thermo-mechanical interface with the spacecraft. The optical heads are positioned on the upper part of the instrument, in contact with a radiator that keeps the entire instrument in radiative equilibrium; the outside of the whole structure is covered by a multi-layer insulation. The focus of the telescope, is placed at the entrance slit of the instrument, the components of which are listed in Figure \ref{fig:virtis} \citep{Filacchione}. VIRTIS-M is equipped with a Offner spectrometer, which does not require collimators, lenses or beam splitter, and this allows for a substantial reduction of the size of the optics (see Figure \ref{fig:offner}).  

The VIS focal plane is constituted by a Thomson CCD, whose sensor is cooled to temperatures in the range 150-180 K.
The IR focal plane is composed by a Raytheon two-dimensional sensor made up of photosensitive material (Mercury, Cadmium and Tellurium, HgCdTe), with the sensor actively cooled to temperatures 86 - 88 K. It is divided into six spectral zones, corresponding to different filters.
Since the two focal planes have different thermal requirements, their assembly has been realized on orthogonal planes, although very close one each other. The CCD receives the diffracted light from the grating by means of a mirror that deflects only visible light, while the infrared sensor receives the light in a direct way.

VIRTIS has an internal calibration mode, in order to monitor both the spectral and radiometric performance during the mission. In this mode acquisitions are performed following a sequence with different configurations of the mechanical components (shutter, cover) and two calibration sources, one optimized for the VIS channel and one for the IR channel. This provides information on some parameters that can affect the experimental data. In particular, the acquisitions with closed shutter permit to obtain the signal of the dark current added to the thermal background, the latter being relevant for the IR channel. 

The stored data are saved in a format called ``cube": this name is used as VIRTIS-M is in fact capable of recording images at various wavelengths. Two dimensions are spatial and the third is spectral. Thus, each cube gives as many monochromatic images as spectral bands. For further details refer to \citep{Coradini} and \citep{Filacchione}. 

\section{Dark current}       
\label{sec:dark}
In order to model the dark current in terms of integration time ($it$) and temperature of the focal plane ($T_{FP}$) for both VIS and IR channel, we assume the following relation:

\begin{equation}
Dark(\lambda)_{DN} = a(\lambda) + b(\lambda) \times it  + c(\lambda) \times T_{FP}
\label{eq:dark}
\end{equation}

The calculation of the coefficients $a, b, c$, together with the readout noise analysis, is performed for the two spectral channels as follows. The calculation is performed along the 432 bands ($\lambda$). The dependence on the samples (256 along the slit) is implicit.

\subsection{IR channel}
\label{sub:darkir}
\begin{figure}[!h]
\centerline{\includegraphics[width=0.7\textwidth,clip=]{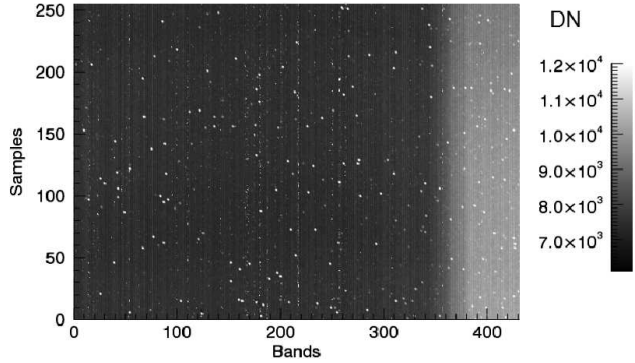}}
\caption{Example of acquisition with closed shutter with IR channel ($it =  0.4, T_{FP} = 87.5 K$). 432 bands (horizontal axis) x 256 samples (vertical axis). It includes the electronic offset, the dark current and the thermal background. The background thermal emission of the spectrometer ($T\approx 140 K$) is visible at longer wavelengths. The spots are defective pixels.}
\label{fig:darkir}
\end{figure}

\begin{figure}[!h]
\centerline{\includegraphics[width=0.7\textwidth,clip=]{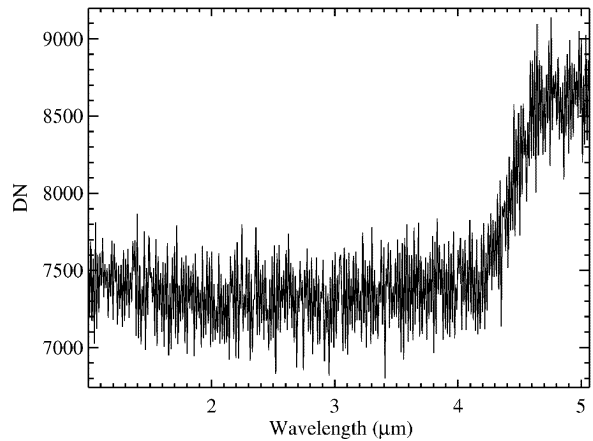}}
\caption{Equivalent to figure \ref{fig:darkir} for one single sample. The rise above $4\:\mu m$ is due to the thermal background. The different electronic offset along the bands caused by the multiplexer readout circuit can also be noted.}
\label{fig:darkir1s}
\end{figure}

The acquisitions of dark current investigated in this section have all been carried out by VIRTIS during the cruise phase and until June 2010 when the S/C entered its hibernation phase. They also include dark frames acquired during internal calibration mode. The dark frames analyzed include the electronic offset (Figure \ref{fig:darkir}, \ref{fig:darkir1s}). 

During the measurements the IR focal plane is stabilized in temperature by an active cryocooler, while the spectrometer temperature is regulated by a passive radiator and could vary according to the S/C orientation \citep{Coradini}.

\begin{figure}[!h]
\centerline{\includegraphics[width=1.0\textwidth,clip=]{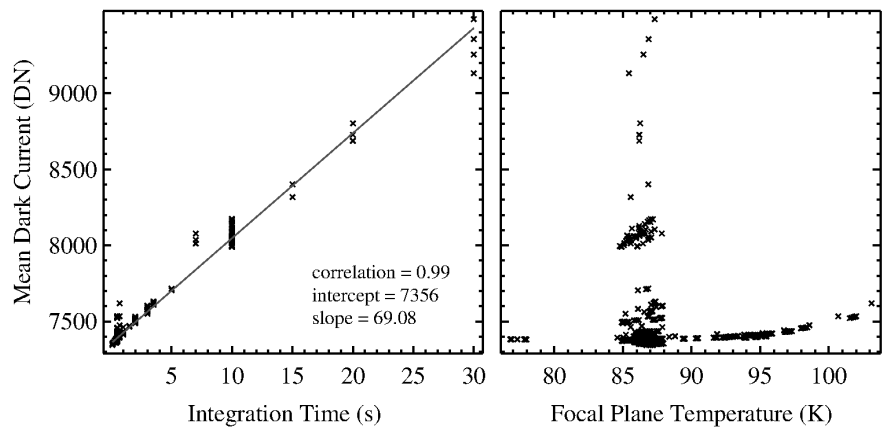}}
\caption{Left panel: Mean Dark current versus integration time. Each point is the average of all the pixels in the first 250 bands. The solid line is a linear best fit. The Pearson correlation coefficient is used for the analysis. The intercept of the line represents the mean electronic offset in DN while the slope corresponds to the dark current rate in DN/s. Right panel: same points of left panel as a function of the Focal Plane Temperature.}
\label{fig:darkirfun}
\end{figure}

In order to separate the contribution of the thermal background from the dark current, only the first 250 bands ($1.0 - 3.6\:\mu m$) are taken into account. 

The dependence of the dark current on integration time and on the focal plane temperature is calculated pixel by pixel. The general trend of these relations is represented in Fig. \ref{fig:darkirfun} where the average of the pixels of each frame is reported (256 sample x 250 bands). 

While there is a clear correlation between the dark current and the integration time, we have observed a negligible dependence of the dark current on the focal plane temperature when $T_{FP} = 85 - 88 K$. Many points in the plot are overcoming the 88 K. They correspond to acquisitions in internal calibration mode with an integration time of 0.5 s. The pattern they show is clearly exponential, although still negligible if compared to the effect of the integration time. The higher temperature of these points is due to the short time left available to the cryocooler to cool the focal plane. For larger integration times and for temperatures $> 88$ K this exponential behavior could be significant. However, the model of dark current does not take into account these higher temperatures because the planned acquisitions are supposed to operate within the range where the observed correlation is negligible.

\subsection{VIS channel}
\label{sub:darkvis}
As in the case of the IR channel also for the VIS channel the acquisitions of dark current taken into account are all those carried out by VIRTIS during the overall cruise phase.

\begin{figure}[!h]
\centerline{\includegraphics[width=0.7\textwidth,clip=]{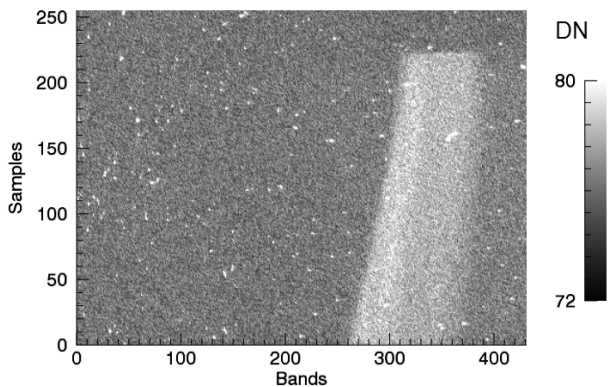}}
\caption{This dark frame of VIS channel clearly shows an external source of light at longer wavelengths. Frames like this are not taken into account in the analysis.}
\label{fig:darkvis}
\end{figure}

For VIS channel a selection of the dark frames was needed because in few cases dark current acquisitions are affected by sun light contamination passing through a narrow aperture between the radiator and the spectrometer outer chassis. This effect happen only for some specific orientation of the spacecraft. Figure \ref{fig:darkvis} shows one of the dark acquisitions to be discarded for this analysis.

The amount of dark current is analyzed as a function of the integration time and of the temperature of the focal plane.

\begin{figure}[!h]
\centerline{\includegraphics[width=1.0\textwidth,clip=]{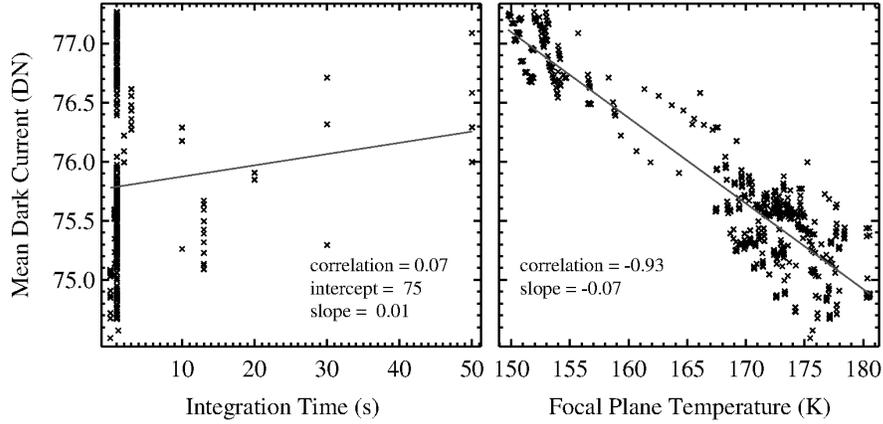}}
\caption{Left panel : Mean Dark current versus integration time. The solid line is a linear best fit. The Pearson correlation coefficient is used for the analysis. The intercept of the line represents the mean electronic offset in DN while the slope corresponds to the dark current rate in DN/s. 
Right panel: same points of left panel as a function of the Focal Plane Temperature. The negative correlation observed is due to the low temperature of the sensor in which the transfer of the charges is blocked by the freezing of the crystal lattice of the semiconductor \citep{Filacchione}.}
\label{fig:darkvisfun}
\end{figure}

Unlike the IR channel the amount of dark current is more affected by the temperature of the focal plane than the integration time in the range of possible variation of this quantities (see Figure \ref{fig:darkvisfun}).

\section{Readout noise}       
\label{sec:readout}

\subsection{IR channel}
\label{sub:readoutir}
\begin{figure}[!h]
\centerline{\includegraphics[width=0.7\textwidth,clip=]{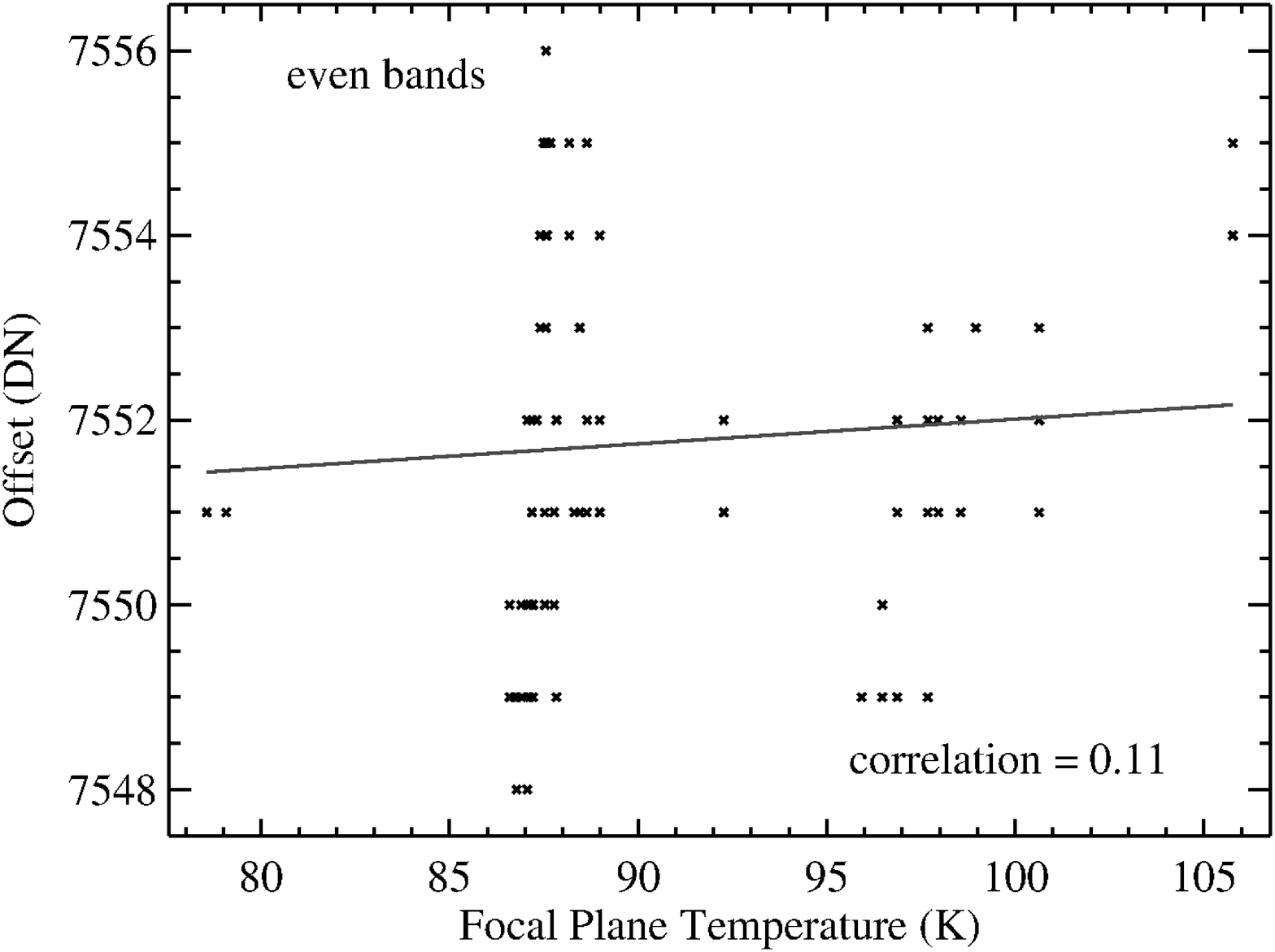}}
\caption{Electronic offset for even bands (IR channel) in function of the temperature of the focal plane. There is not significant correlation.}
\label{fig:offsetireven}
\end{figure}

\begin{figure}[!h]
\centerline{\includegraphics[width=0.7\textwidth,clip=]{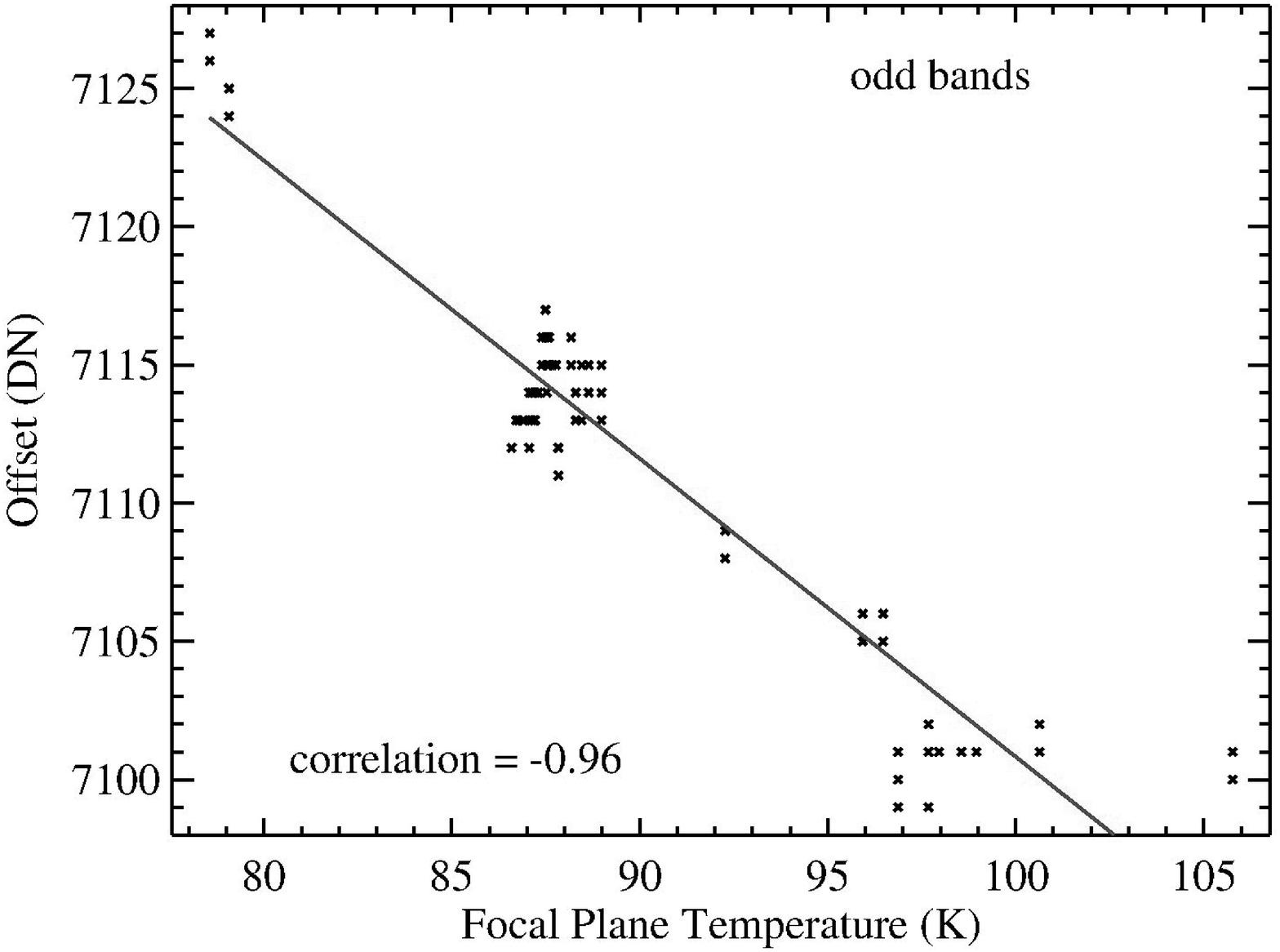}}
\caption{Electronic offset for odd bands (IR channel) in function of the temperature of the focal plane. There is a negative correlation.}
\label{fig:offsetirodd}
\end{figure}

The analysis of the readout noise is performed by investigating the acquisitions of electronic offset during the internal calibration mode. Variations that are not caused by the readout noise have to be avoided in the analysis. To this aim the electronic offset is studied as a function of the temperature of the focal plane, dividing odd and even bands. The results are shown in figure \ref{fig:offsetireven} and \ref{fig:offsetirodd}. 

While even bands do not present any significant correlation, odd bands seem to be correlated with the temperature of the focal plane. In order to evaluate the readout noise, the contribution of the temperature has to be eliminated choosing frames acquired at the same temperature. Many (50) acquisitions of the offset with a focal plane temperature of $\sim 87$ K (the most frequent) are selected for the analysis. Then the standard deviation is performed pixel by pixel. The average of the standard deviations over all pixels is a measurement of the readout noise in digital number. The result for IR channel is 2.6 DN, equivalent to 175 photoelectrons. It is assumed to be independent from instrumental parameters.

\subsection{VIS channel}
\label{sub:readoutvis}
Here the same procedure of section \ref{sub:readoutir} is followed. For the visible channel there is no difference for the electronic offset between odd and even bands. The correlation of the electronic offset with the temperature of the focal plane is negative as for the dark current (see section \ref{sub:darkvis}). In order to analyze the readout noise 50 acquisitions with the same focal plane temperature ($\sim 155$ K) are analyzed. The resulting readout noise is 1 DN, equivalent to 65 photoelectrons. 

\begin{figure}[!h]
\centerline{\includegraphics[width=0.7\textwidth,clip=]{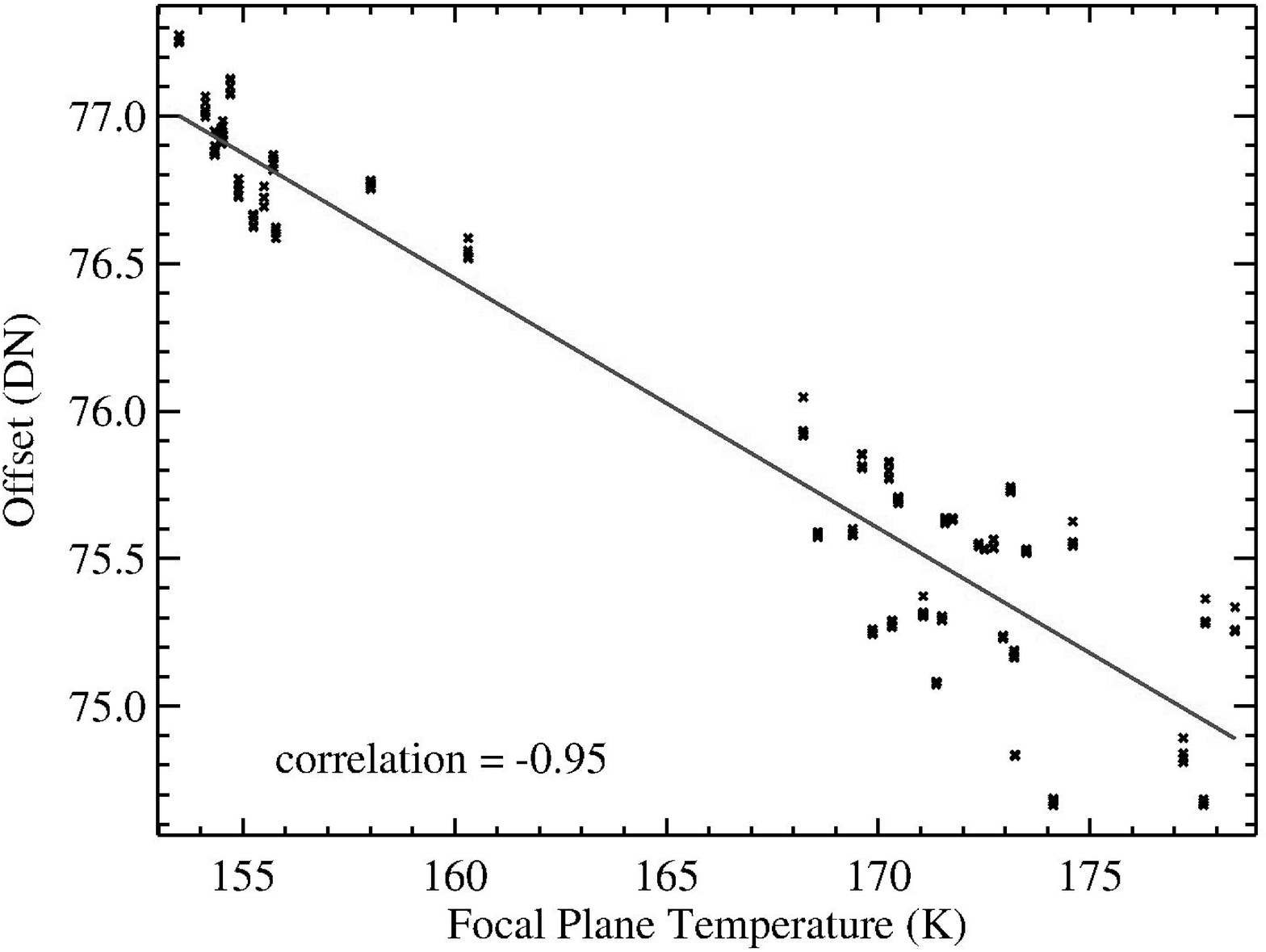}}
\caption{Mean offset of VIS channel vs Temperature of the focal plane. The correlation is negative.}
\label{fig:offsetvis}
\end{figure}

\section{Thermal background}       
\label{sec:thermalback}
The IR channel is further affected by the thermal background which has to be taken into account for the calculation of the noise and the saturation limit. 

We model the thermal background as a function of the integration time ($it$) and the temperature of the spectrometer ($T_{S}$): 

\begin{equation}
Background(\lambda)_{DN} = K(\lambda) \times B(\lambda,T_{S}) \times it \times ITF(\lambda) 
\label{eq:termalback}
\end{equation}

where:
\begin{itemize}
\item[]{$B(\lambda,T_{S})$ is the internal thermal radiance depending on wavelength and spectrometer temperature;}
\item[]{$K(\lambda)$ accounts for the filter transmittance, the effective area of the spectrometer seen by the detector and other possible contributions.}
\item[]{$ITF(\lambda)$ is the instrument transfer function in ($DN\:m^{2}\:\mu m\:sterad\:/ W\:/ s$) \citep{Filacchione, Filacchione2, Migliorini}.}
\end{itemize}

\begin{figure}[!h]
\centerline{\includegraphics[width=0.7\textwidth,clip=]{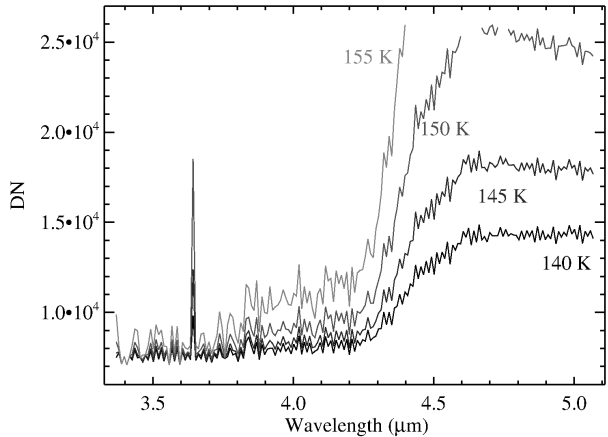}}
\caption{The simulated DN as a result of four different temperatures of the spectrometer, with an integration time of 1 s. Thermal Background affects the spectrum at decreasing wavelength for increasing temperature. The pixels are saturated where the line is missing. This simulation is performed for the central sample of the slit. The result is similar for the other samples.}
\label{fig:thermalback}
\end{figure}

The characterization of  the $K(\lambda)$ parameter is crucial for modeling the thermal background. 
$K(\lambda)$ can be derived from measured acquisitions of thermal background with known $T_{S}$ and $it$ by inverting the Eq. \ref{eq:termalback}. Acquisitions with different $T_{S}$ and $it$ are taken into account to perform this analysis. Figure \ref{fig:thermalback} shows a simulation of the Thermal Background with different spectrometer temperatures.

\section{Signal to noise ratio modeling}
\label{sec:S/N}
The S/N Simulator is a software tool able to calculate the expected instrument signal to noise ratio (S/N) for different input signals and observing conditions. Only Poissonian noise is taken into account during the calculation, as other kind of noises are assumed negligible for our purpose.

The calculation performed to obtain the S/N depends on the kind of signal to treat: measured or simulated. 

\subsection{Measured signal}
A correct characterization of the noise for measured signals is important in the retrieval phase of the surface properties as discussed in chapter \ref{chap:modelextractor}. 

Prior to the calibration, a raw signal is recorded in Digital Number (DN), directly convertible to PhotoElectrons (PE). From the latter the Poissonian noise and the resulting S/N are straightforward:

\begin{equation}
Noise(\lambda)_{PE} =  \sqrt{Total\:signal(\lambda)_{PE} + readout\:noise^{2}}
\label{eq:noise}
\end{equation}

\begin{equation}
S/N(\lambda) = R(\lambda) / Noise(\lambda) 
\label{eq:s/n}
\end{equation}

\begin{itemize}
\item[]{$Total Signal(\lambda)$ includes the signal of the target and all other sources of signal such as the dark current and thermal background. It is obtained by subtracting the electronic offset to the measured acquisition.}
\item[]{$R(\lambda)$ is the signal of the target. It is isolated from other sources of signal by subtracting an acquisition with closed shutter to the acquisition with open shutter, being the former sufficiently close in time to the latter.}
\end{itemize}
Moreover, the tool checks for saturated pixels, if  $Total Signal + electronic\;offset$ overcomes the saturation limit of the instrument.

\subsection{Simulated signal}
The analysis of the expected noise for a given simulated signal is mandatory to plan the observations. An example is presented in section \ref{sec:minor} where the study of the detectability of minor components is analyzed as a function of the integration time and other parameters which affect the instrumental noise. 

In order to calculate the S/N, for simulated signals we have to simulate the Total Signal as well, which include dark current and thermal background. 

\begin{equation}
Total\:signal(\lambda) =  R(\lambda)  + Dark(\lambda) + Background(\lambda)
\label{eq:totsignal}
\end{equation}

From Eq. \ref{eq:totsignal} we can then compute Eq. \ref{eq:noise} and \ref{eq:s/n} to derive the S/N.

The terms of Eq. \ref{eq:totsignal} depend on the integration time and the temperature of different parts of the instrument. In order to perform simulation of the dark current and thermal background we have analyzed the DN acquired by the instrument in different conditions for both Visible and Infrared channels as explained in sections \ref{sec:dark}, \ref{sec:readout} and \ref{sec:thermalback}. The simulation of the target signals $R(\lambda)$ is addressed in chapter \ref{chap:spectrasimulation}.

\subsection{Testing the S/N simulator}

\begin{figure}[!h]
\centerline{\includegraphics[width=0.7\textwidth,clip=]{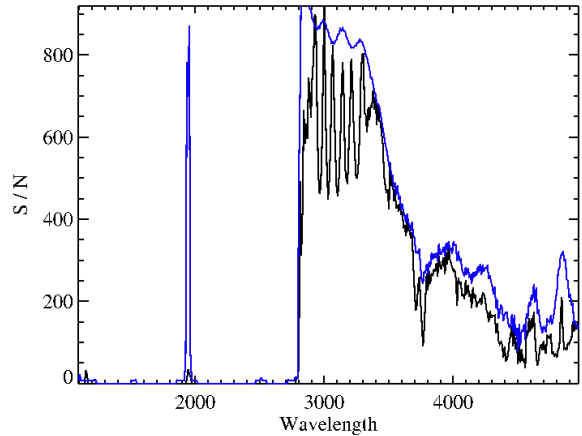}}
\caption{The simulated S/N (blue line) is over plotted to the S/N extracted from the visible lamp signal (black line) acquired during internal calibration mode with IR channel. The integration time is 0.1 s. The average overestimation of the simulator is 30\%. This comparison also represents a satisfactory test for the saturation of IR channel (where the S/N is set to 0). The band at $\sim 2800$ nm is the limit before which the signal is saturated. This is what emerges from the real data and the simulation.}
\label{fig:Itotvislampcheck}
\end{figure}

\begin{figure}[!h]
\centerline{\includegraphics[width=0.7\textwidth,clip=]{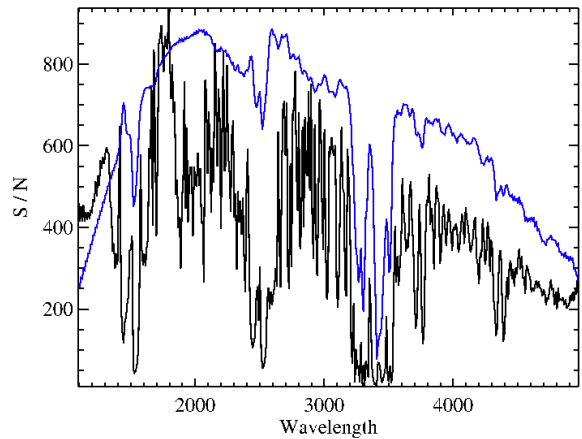}}
\caption{The simulated S/N (blue line) is over plotted to the S/N extracted from the infrared lamp signal (black line) acquired during internal calibration mode with IR channel. The average overestimation of the simulator is 40\%. Integration time = 0.5 s.}
\label{fig:Itotirlampcheck}
\end{figure}

\begin{figure}[!h]
\centerline{\includegraphics[width=0.7\textwidth,clip=]{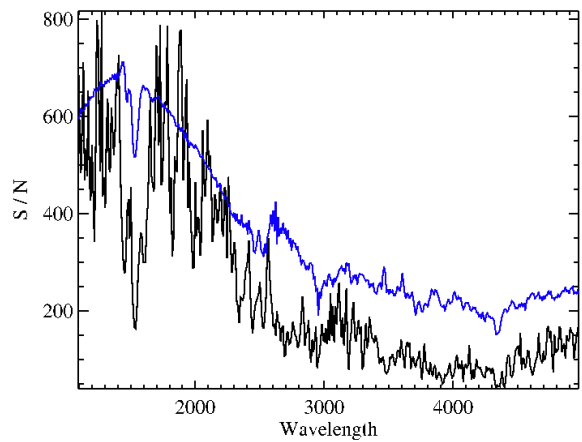}}
\caption{The simulated S/N (blue line) is overplotted to the S/N extracted from the Lutetia spectral cube (black line) in IR channel. The integration time is 0.7 s. The average overestimation of the simulator is 40\%.}
\label{fig:lutetiaircheck}
\end{figure}

\begin{figure}[!h]
\centerline{\includegraphics[width=0.7\textwidth,clip=]{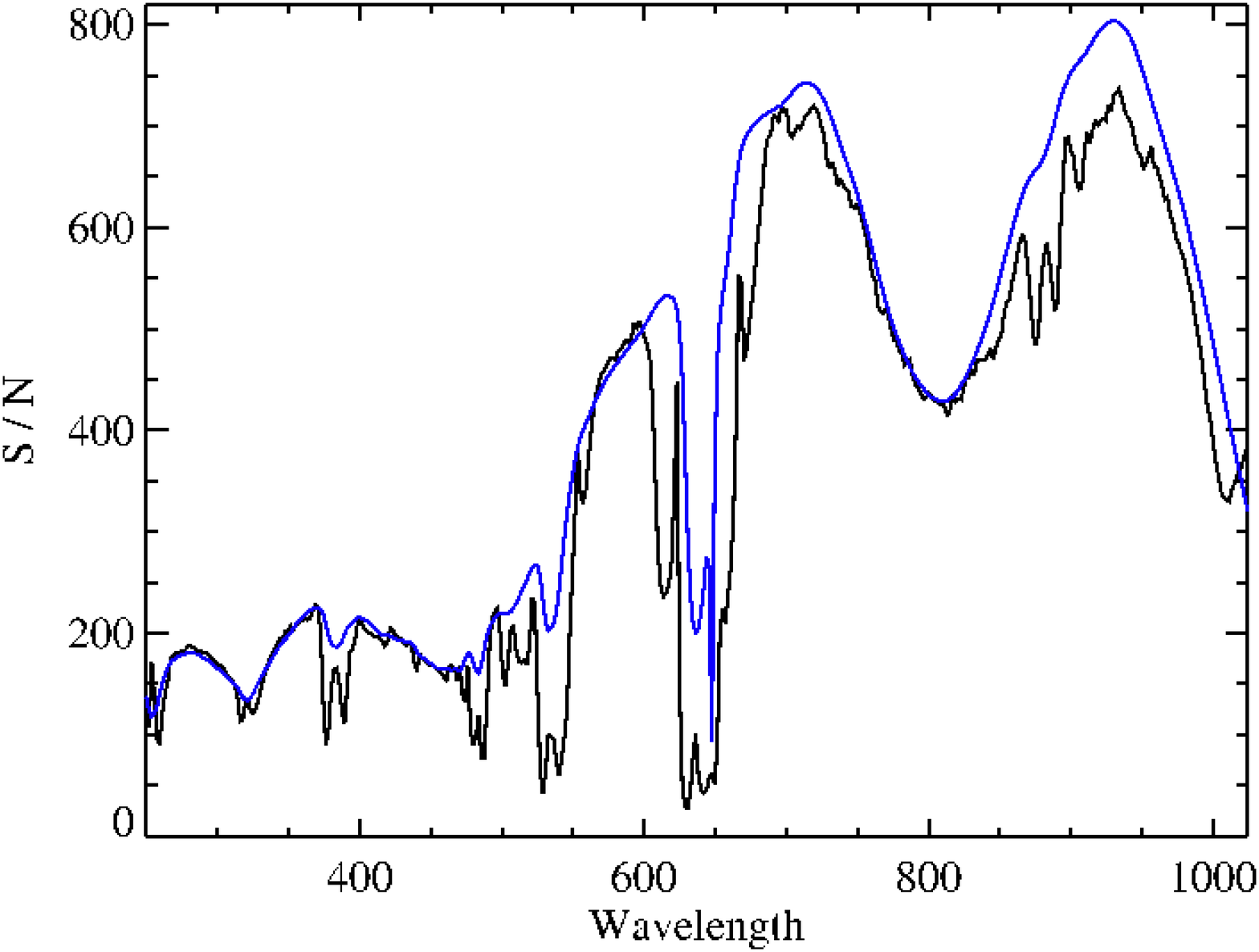}}
\caption{The simulated S/N (blue line) is over plotted to the S/N extracted from the visible lamp signal (black line) acquired during internal calibration mode with VIS channel. The integration time is 1 s. The average overestimation of the simulator is 15\%.}
\label{fig:Vtotvislampcheck}
\end{figure}

\begin{figure}[!h]
\centerline{\includegraphics[width=0.7\textwidth,clip=]{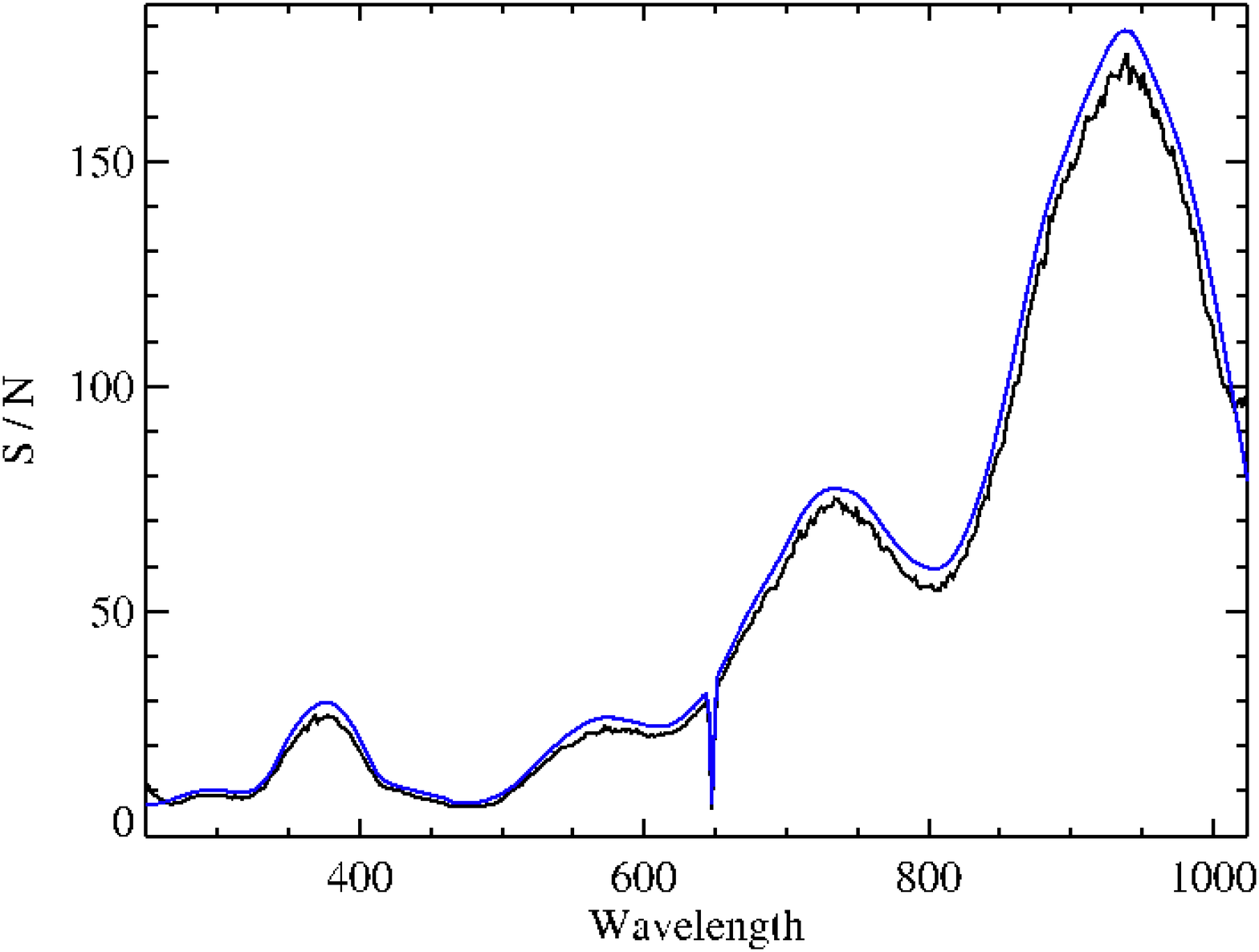}}
\caption{The simulated S/N (blue line) is over plotted to the S/N extracted from the infrared lamp signal (black line) acquired during internal calibration mode with VIS channel. The integration time is 20 s. The average overestimation of the simulator is 10\%.}
\label{fig:Vtotirlampcheck}
\end{figure}

\begin{figure}[!h]
\centerline{\includegraphics[width=0.7\textwidth,clip=]{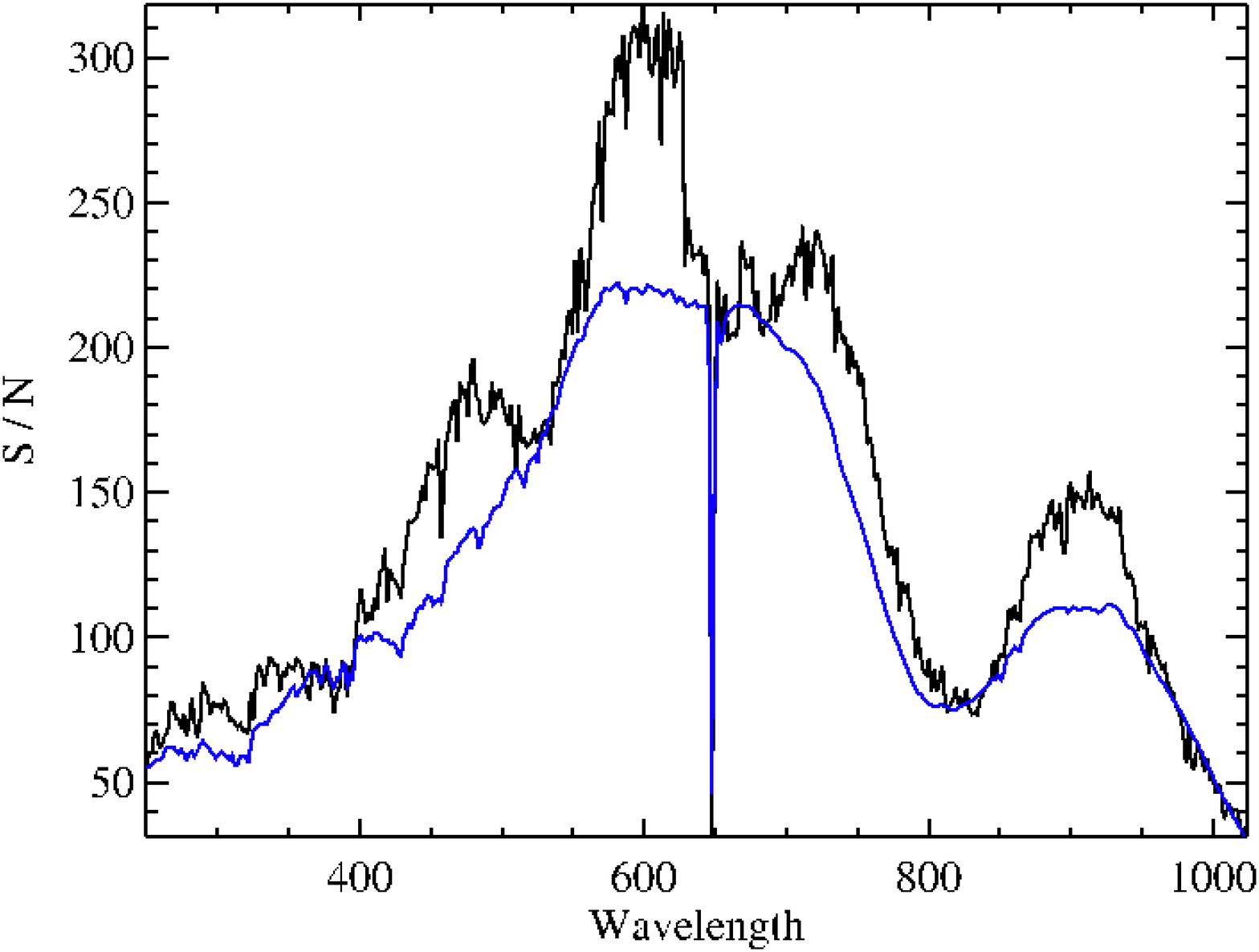}}
\caption{The simulated S/N (blue line) is over plotted to the S/N extracted from the Lutetia spectral cube (black line) in VIS channel. The integration time is 1 s. The simulator underestimates the S/N in average by a factor 15\%. This underestimation of S/N is due to the underestimation of the signal. In fact the VIS channel is affected by contamination from the high orders of the IR channel \citep{Filacchione}, and this contamination is not well characterized by the ITF. This is a further source of error that is not taken into account in this analysis.}
\label{fig:lutetiavischeck}
\end{figure}

To test the accuracy of the tool in reproducing the S/N, a method has been developed to extrapolate the S/N from real acquisitions of the VIRTIS instrument, which are to be compared with the results of the simulator.

The method is aimed to study the signal poissonian fluctuations compared to the average, excluding fluctuations due to changes in radiance.

A number N of radiances spectra, from the same source, that must remain as stable as possible, have been investigated. The mean radiance is the reference against which the fluctuation over N radiances is calculated (for each band). The fluctuation due to changes in radiance are overcome by adapting all the spectra to the mean spectrum by a running box of the size of 20 bands: the signal of each band (central band of the box) is divided by the average value relative to the box bands and multiplied by the average of the box bands of the mean radiance. This procedure is repeated for each one of the N radiance spectra. Residual fluctuations of the resulting spectra for each band are due to poissonian noise, and from that the S/N is calculated. 

As stated above, the source has to be as stable as possible. Three sequences of acquisitions have been identified that comply with this requirement: two from calibration lamps \citep{Melchiorri} whose signal is acquired during the internal calibration mode, and a sequence of acquisition of Lutetia asteroid, during its flyby.

The available acquisitions are: 80 for Lutetia's frames, that is a full sequence; 50 for internal lamps, that is a sequence maintaining (enough for our purpose) the same focal plane and spectrometer temperatures.

The quality of the result depends on the available statistics. The larger is the number of spectra the more reliable will be the result. An average of the resulting S/N along the samples could compensate the poor statistics. 

The available samples are: 12 for Lutetia’s frames, that are the samples always filled with asteroid surface during the sequence; 256 for internal lamps acquired in visible channel, that is the whole slit; 128 for internal lamps acquired in infrared channel, that is the half slit, avoiding the part affected by a defect of the order sorting filter \citep{Filacchione}.

Figures \ref{fig:Itotvislampcheck}, \ref{fig:Itotirlampcheck} and \ref{fig:lutetiaircheck} show the result of the comparison for IR channel, while VIS channel is described by figures \ref{fig:Vtotvislampcheck}, \ref{fig:Vtotirlampcheck} and \ref{fig:lutetiavischeck}. As expected, in general the simulator overestimates the S/N because it does not take into account other sources of noise than the Poissonian noise. To compensate this overestimation, in the analysis preformed in section \ref{chap:spectrasimulation}, a conservative approach is used, lowering the resulting S/N by 40 \% for IR channel. 

Looking at the results, for VIS channel a correction is not needed.

\chapter{Spectral simulation}
\label{chap:spectrasimulation}
In order to simulate how VIRTIS-M handles the signal coming from the nucleus' surface of the comet, we have to define the physical properties of the surface and apply a radiative transfer model to calculate the spectral radiance entering the instrument. The simulation of the spectra is also finalized to test the developed tools for the spectrophotometric analysis discussed in next sections.

\section{Hapke model}
\label{sec:hapke}

The Hapke model \citep{Hapke93, Hapke} is used both to simulate the spectra and to retrieve quantitative information on the parameters describing the surface physical properties. The most recent version accounts for the effects of regolith porosity on the overall reflectance. Recent works have established its greater reliability than the previous version \citep{HelfShe, Ciarni14}.

In Hapke's theory the surface bidirectional reflectance can be expressed as:

\begin{equation}
\begin{split}
r\:(i,e,g)  = &\:K(\omega/4\pi)\:[\mu_{0e} /(\mu_{e}+\mu_{0e})] \\ 
&\left\{ p(g)[1+B_{S0}B_{S}(g)]+H(\mu_{0e}/K) H(\mu_{e}/K) -1 \right\} \\
&\:[1+B_{C0}\:B_{C}(g)]\:S(i,e,g,\overline{\theta})
\label{eq:hapke}
\end{split}
\end{equation}

where: 

\begin{itemize}
\item[]{$\mu_{0e},  \mu_{e}$: effective incidence and emission angle cosine. They are the cosines of the angles corrected by the effect of the surface roughness (see section 12.4 in \citet{Hapke}.}
\item[]{$\omega$: single scattering albedo (SSA). It contains the spectral information. Once the optical constants and particle diameter are fixed it is possible, following Hapke's model, to compute single scattering albedo for a given type of particles. It can assume values in the range 0-1.} 
\item[]{$p(g)$: single scattering phase function ($g$ is the phase angle) modeled with Heyney-Greenstein function with 2 parameters $b, v$: the first one describes the angular width of both forward and back scattering lobes (0-1 range), while the second one describes their relative amplitude \citep{HeyGree,DominVer}:}

\begin{equation}
\begin{split}
p(g) = &\:\frac{(1+v)}{2}\cdot\frac{(1-b^{2})}{(1-2b cos(g) + b^{2})^{3/2}}\:+\\
+&\:\frac{(1-v)}{2}\cdot \frac{(1-b^{2})}{(1+2b cos(g) +b^{2})^{3/2}} 
\label{eq:phase}
\end{split}
\end{equation}

it must be noted that $p(g)$ cannot be negative. Thus in dependence of the $b$ values, $v$ has physical meaning within the range permitted by \citep{HapkeH}:

\begin{equation}
 \left|v\right| < \frac{(1 + 3b^{2})}{b(3+b^{2})}   
\label{eq:limitv}
\end{equation}

\item[]{$B_{S0}, B_{S}(g), B_{C0}$ and $B_{C}(g)$ describe the observed non-linear increase in reflectance towards small phase angles due to two mechanisms: the Shadow Hiding Opposition Effect (SHOE) and the Coherent Backscattering Opposition Effect (CBOE). In principle the amplitude of both ($B_{S0}, B_{C0}$) should not be greater than 1. $B_{S}(g)$, and $B_{C}(g)$ are linked to the angular amplitude $h_{S}$ and $h_{C}$. To model them we choose the approximated equations (respectively 9.22 pag. 232, and 9.43 pag. 244 in \citet{Hapke}).}
\item[]{$K$: porosity parameter. It is linked to the filling factor $\phi$, that is the total fraction of the volume occupied by the particles (Eq 7.45b, pag. 167 in \citet{Hapke}), and it is meaningful up to the critical point $K \sim 1.9$, corresponding to $\phiΦ < 52\%$ where coherent effects become important \citep{Hapke}}
\item[]{$H$: Ambartsumian-Chandrasekhar function (Eq. 8.56, p.204 in \citet{Hapke}), it is related to the multiple scattering, which involves $\omega$ in a non-linear way.}
\item[]{$S(i,e,g,\overline{\theta})$: shadowing function, is the correction due to surface roughness. Its value is always less than 1 and decreases with increasing roughness parameter $\overline{\theta}$, which is an average slope of the facets composing the surface (see section 12.4 in \citet{Hapke}).}
\end{itemize}

\noindent
We remand to \citet{Hapke} for further details.

\noindent
To obtain the equivalent radiance spectra we need to model the solar irradiance and the thermal emission:

\begin{equation}
R = J/D^{2} r + B (T, \epsilon_{eff})  
\label{eq:rad}
\end{equation}

where:
\begin{itemize}
\item[]{$r$ :	bidirectional reflectance (Eq. \ref{eq:hapke})}
\item[]{$J$ :	solar irradiance measured at 1 AU 

(\it{http://rredc.nrel.gov/solar/spectra/am0/modtran.html})}
\item[]{$D$ :	heliocentric distance in AU}
\item[]{$B$ : 	thermal emission}
\item[]{$T$ : 	temperature}
\item[]{$\epsilon_{eff}$ : 	effective integrated emissivity \citep{Davidsson}}
\end{itemize}

\section{Dark terrain definition}
\label{darkterrain}

To obtain realistic spectra of the surface we need to define a main component which for comets has typically a very low albedo (referred to hereafter as Dark Terrain). The definition of the SSA for the Dark Terrain is crucial for the final model of reflectance coming from the mixing with minor components. This in turn defines the simulated detectability for the VIRTIS instrument, and so the conditions (environmental and instrumental) required to perform abundance measurement. To define the SSA of the Dark Terrain we take advantage of the data obtained by the HRI-Deep Impact spacecraft, that encountered Comet 9P/Tempel.

Ground observations suggest that comet Tempel 1 has very similar photometric properties to that of CG 67/P, target of the Rosetta mission \citep{Lamy3}.

We analyzed publicly available data (http://pds.jpl.nasa.gov/) obtained by the HRI (High Resolution Imager) onboard Deep Impact consisting of a telescope serving both a camera and an imaging spectrometer in the infrared range \citep{Hampton}. We have selected data taken before the impact, where the nucleus is resolved and has a maximum spatial resolution of 120 m per pixel. We use the data of the spectrometer in the range 1.2 - 3.5 $\mu m$ which present an higher radiometric accuracy \citep{Klaasen1, Klaasen2}. To obtain the reflectance from the radiance, we modeled it as the sum of two contributes: the reflected sunlight ($1.2 - 2.5\:\mu m$) and the thermal emission ($2.5 - 3.5\:\mu m$), setting as free parameters the temperature, the effective integrated emissivity and the spectral reddening, similarly to \citet{Davidsson}. The data of the camera are available for seven different bands from 0.35 to 0.95 $\mu m$. These images are rescaled to the same spatial resolution of the hyper-spectral cubes obtained from the spectrometer. In this way we are able to extend the spectra of each pixel to the visible wavelengths. This task is straightforward because all the data we have used were acquired in a short time interval compared with the comet's rotation period. Therefore, the dataset we are analyzing contains only observations taken on the same nucleus hemisphere. Since the radiometric accuracy of the VIS channel is higher than the IR, we have scaled the latter to the former before bridging the spectra of each pixel. Out of statistics pixels are then removed by applying a the despiking filter discussed in section \ref{sec:artifact} on the whole dataset.

\begin{figure}[!h]
\centerline{\includegraphics[width=0.9\textwidth,clip=]{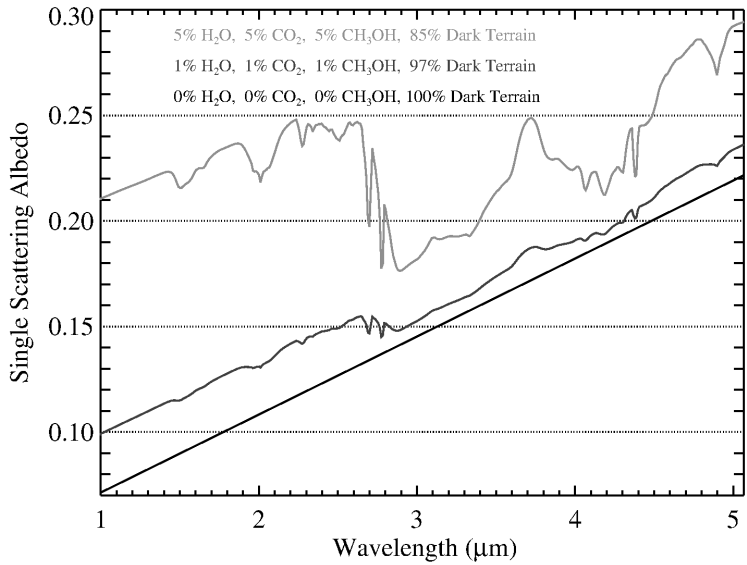}}
\caption{The defined SSA of the Dark Terrain is plotted in black. Two different mixing are over plotted. The Dark Terrain is mixed with water ice, carbon dioxide and methanol, with different abundances.}
\label{fig:ssavari}
\end{figure}

The hyperspectral cubes containing both visible and infrared data are powerful tools to recognize icy regions. After having assigned thresholds for the reflectance at visible wavelengths and for the 2.0 $\mu m$ band depth is possible to identify the ice-rich units. The results obtained \citep{Raponi} are similar to that of \citet{Sunshine06}. To define a SSA of the Dark Terrain we need to exclude the icy units from the analysis. Taking into account the Hapke model (Eq. \ref{eq:hapke}), we can derive a SSA from the average spectrum of reflectance of the non-icy units. However, because the available observations are all taken at a single phase angle ($63^{o}$) we have to assume the value of the parameters in the model. At this specific phase angle it is assumed that the opposition effect (significant at small phase angles) and the shadowing caused by the roughness (significant at high phase angles) are negligible. Moreover, the porosity parameter and the single particle phase function are fixed to 1. In this way the SSA has been derived  by inverting the Eq. \ref{eq:hapke}, and a linear interpolation of it is performed (see Fig. \ref{fig:ssavari} and \ref{fig:SSA2comets}).

\section{Synthetic spectra}
\label{sec:synthetic}
The various compositional end-members present on the surface can be mixed in different ways. Here we selected two mixing modalities: areal and intimate mixing. 

In the case of areal mixing, namely the surface covered by patches of different composition, the SSA of each depends on the spectral properties of a single component. The resulting reflectance is then a linear combination of the reflectances of the various components, weighted for the relative end-members abundances.
 
In the case of intimate mixing particles of different composition are mixed together (``salt and pepper" mixture) and the final SSA is a linear combination of the SSA of the different components.

In modeling the comet surface we take into account different minor components (and the respective optical constants) which could be found in icy form on the surface. They are mixed with the Dark Terrain as defined in previous section.

The SSA of the minor components simulated in this work comes out by the following optical constants:
\begin{itemize}
\item{water ice ($H_{2}O$) \citep{Warren, Mastrapa08, Mastrapa09, ClarkRN}}
\item{carbon dioxide ($CO_{2}$) \citep{Quirico} (GhoSST service \it{http://ghosst-prod.obs.ujf-grenoble.fr/})}
\item{methanol ($CH_{3}OH$) \citep{Trotta} (GhoSST service \it{http://ghosst-prod.obs.ujf-grenoble.fr/})}
\end{itemize}

\noindent
To simulate a real spectrum we add noise to the theoretical spectrum. 

\noindent
The base for its calculation is the S/N($\lambda$) coming from the S/N simulator (section \ref{sec:S/N}).
The S/N simulator converts the simulated radiance $R$ into digital number (and photoelectrons) as a function of the integration time: 

\begin{equation}
R(\lambda)_{DN} = R(\lambda) \times ITF(\lambda) \times it
\end{equation}

\noindent
This quantity is used in Eq. \ref{eq:totsignal} for the calculation of the S/N.

\noindent
The error bars for radiance and reflectance spectra are: 

\begin{align}
\sigma_{Rad}(\lambda) &= Radiance / (S/N)\\
\sigma_{Refl}(\lambda) &= Radiance / (S/N) / Solar Irradiance
\end{align}

\noindent
Therefore, for a given radiance spectrum in input and given instrumental and observational conditions the S/N simulator gives as output the error bars of the spectrum. They could be considered as the standard deviations of the random fluctuations of the signal around the value of each band.

\section{Minor component detectability}
\label{sec:minor}
The detection of the spectral bands is the first step in order to detect the composition of the surface, from which follows the retrieval of the surface model parameters as the abundance of the end-members and the grain size.

Since the detection of the spectral bands depends on environmental and instrument conditions (in addition to the composition of the nucleus), we need to summarize the possible observation scenarios, in order to obtain a general view, and to identify the optimal integration time as a function of the case considered.

In modeling comet's surface we have assumed three different minor components (and their respective optical constants) which could be found in icy form on the surface: water, carbon dioxide, and methanol. The presence of the ice is expected in regions cold enough to prevent sublimation. In particular in the early morning we could observe frost ice formed in night time as a result of the deposition of material ejected by active spots. The ice could also be found in regions with higher temperature if it is physically separated from non-icy components, as assumed in the case of Tempel 1 \citep{Sunshine06}. 

Four different compositions of the nucleus are simulated: areal and intimate mixing, both with high and low abundance of the components selected (see previous section):

\begin{enumerate}
\item{Intimate mixing with 1\% of water ice, 1\% carbon dioxide, 1\% methanol, 97\% dark terrain}
\item{Areal mixing with 1\% of water ice, 1\% carbon dioxide, 1\% methanol, 97\% dark terrain}
\item{Intimate mixing with 5\% of water ice, 5\% carbon dioxide, 5\% methanol, 85\% dark terrain}
\item{Areal mixing with 5\% of water ice, 5\% carbon dioxide, 5\% methanol, 85\% dark terrain}
\end{enumerate}

\noindent
In order to simulate typical VIRTIS-M observations of Churyumov-Gerasimenko comet from Rosetta's orbit we have considered the following observation scenario:

\begin{itemize}
\item{$e = 0^{o}$, because the geometry of view is supposed to be in nadir direction.}
\item{$i = 80^{o}$, the output strongly depends on this parameter and this choice represents an observation in the early morning (local time).}
\item{Grain size = 50 $\mu m$. It is an intermediate value with respect to the results obtained in our Tempel 1 data analysis \citep{Raponi}.}
\item{Single particle phase function backscattering: Heyney-Greenstein function parameters: $b=0.2, v=0.73$, being their value linked with the ``hockey stick relation" \citep{HapkeH}}
\item{Roughness parameter $\overline{\theta} = 16^{o}$, as resulting from Tempel 1 photometric data analysis \citep{Li06}.}
\item{Porosity and opposition effects assumed negligible.}
\item{Heliocentric distance in AU = 3.2, 2.5, 1.5. These values are related to different phases of Rosetta's mission.}
\end{itemize}

\noindent
The temperature of the nucleus is linked to the heliocentric distance with the simple equation representing the energetic equilibrium:

\begin{equation}
S(1-A)\:\mu / D^{2} = (1 -  \overline{\epsilon}_{h} \xi)\: \overline{\epsilon}_{h} \sigma_{SB} \:T^{4}
\label{eq:selfheating}
\end{equation}

Where:
\begin{itemize}
\item[]{S is the solar constant;}
\item[]{A is the bond albedo, calculated from spectral information of the components;}
\item[]{D is the heliocentric distance;}
\item[]{$\overline{\epsilon}_{h}$ is the integrated emissivity, calculated from spectral information of the components (see section \ref{sec:ThermalTempel});}
\item[]{$\xi$ is the self-heating parameter. It is assumed equal to 0.5 that is an intermediate value with respect to those obtained by \citet{Davidsson} for Tempel 1 surface;}
\item[]{$\sigma_{SB}$ is the Stefan-Boltzmann constant;}
\item[]{$T$ is the temperature of the nucleus surface.}
\end{itemize}

This is an approximated expression because it does not consider the contribution of the thermal inertia of the comet's nucleus, but for our purpose it is sufficient in order to provide different plausible temperatures of the nucleus' surface. 

\begin{figure}[!h]
\centerline{\includegraphics[width=0.8\textwidth,clip=]{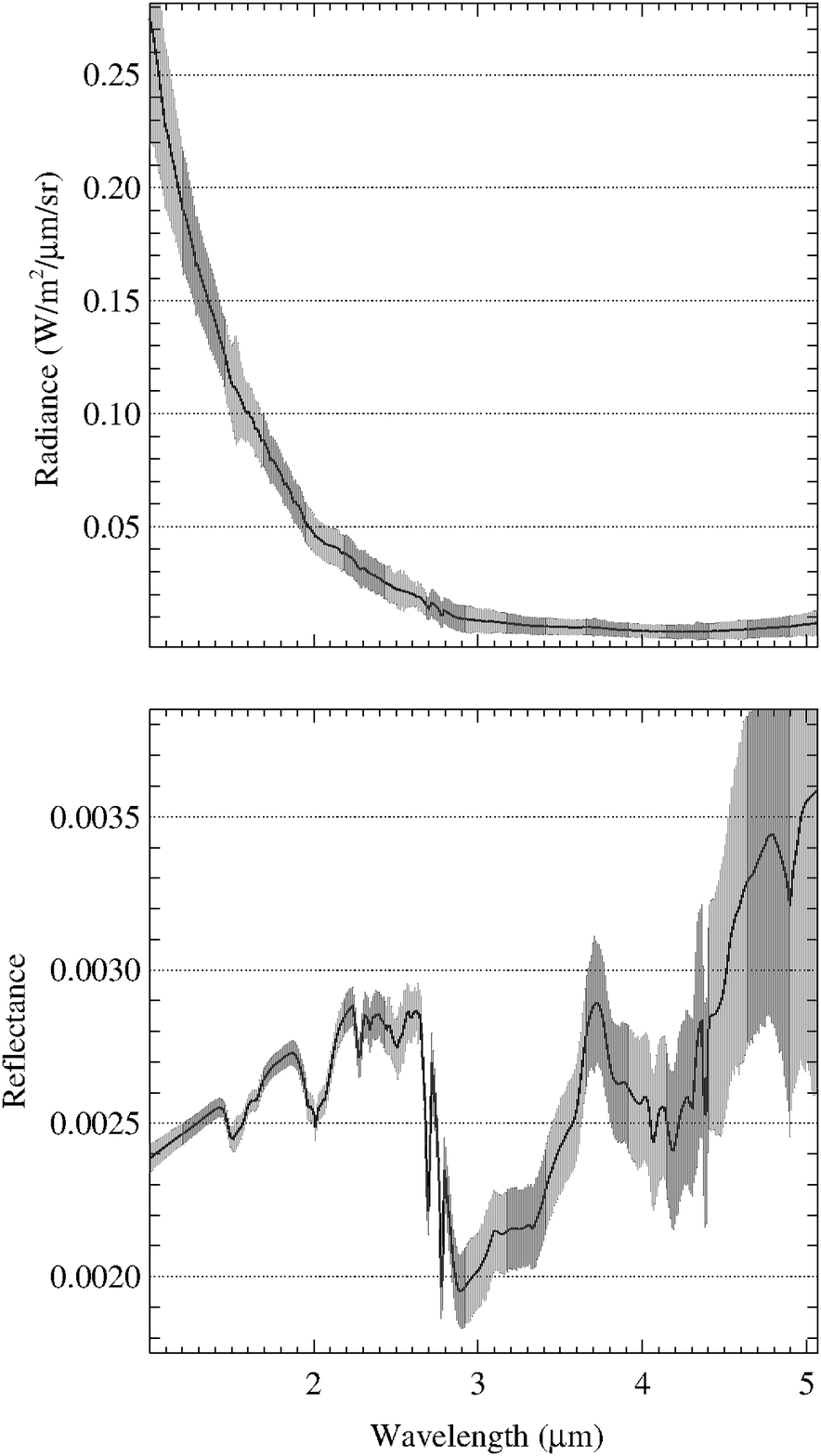}}
\caption{Retrieved radiance (up panel) and reflectance (down panel) with error bars. In this example the simulated composition is: intimate mixing with 5\% water ice, 5\% carbon dioxide, 5\% methanol, 85\% dark terrain. The heliocentric distance is fixed to 2.5 AU, from which the resulting temperature of the surface is 189 K. The integration time is fixed to 2 s. The error bars are calculated as explained in section 4.3. The error bars for radiance spectrum are enhanced by a factor 10 to make them clearly visible. Other parameters are fixed as stated in the text.}
\label{fig:radrefl}
\end{figure}

For a given heliocentric distance, and a radiance spectrum in input we have simulated the reflectance spectra and the error bars associated to each band, being the latter resulting from the calculated  S/N (Fig. \ref{fig:radrefl}). 

In order to determine the achievable accuracy in the characterization of spectral features it has been developed a tool that calculates the error associated to each band area thanks to the information coming from the S/N simulator. The band area is calculated with the formula: $\int 1 - reflectance/continuum \;d\lambda$.
 
\noindent
The used algorithm performs the calculation taking into account possible variations in the shape and wings position of the absorption bands due to different mixing methodologies. 

The following diagnostic features are considered : 

\begin{itemize}
\item{$H_{2}O$ ice (1500 nm)}
\item{$CO_{2}$ ice (4200 nm)}
\item{$CH_{3}OH$ ice (3600 nm)}
\end{itemize}

\begin{figure}[!h]
\centerline{\includegraphics[width=1.\textwidth,clip=]{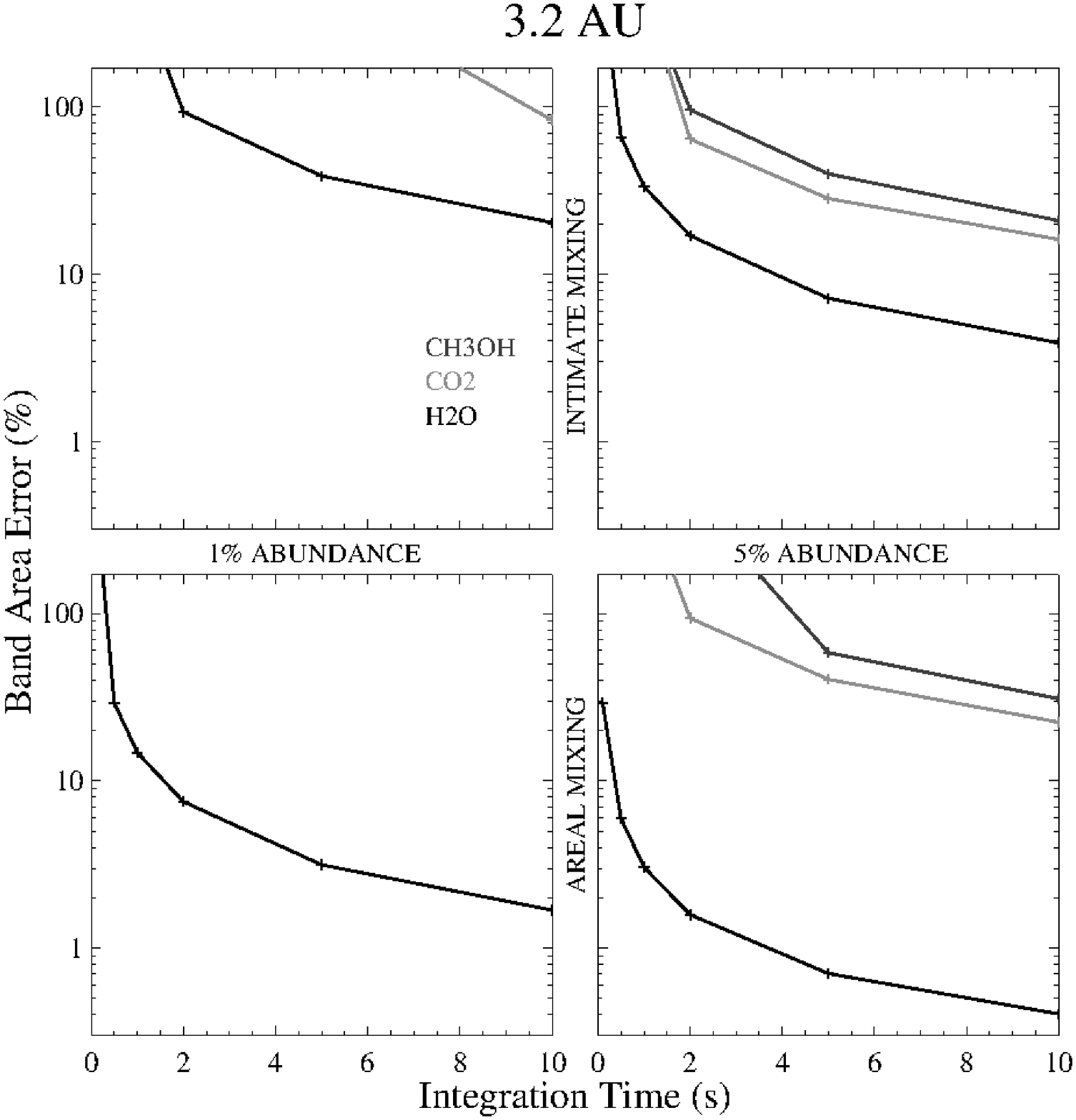}}
\caption{Relative error on the retrieved band area versus different integration times, for heliocentric distance of 3.2 AU. The four panels represent different mixing modalities: intimate (up), areal (down), 1\% abundances (left), 5\% abundances (right).}
\label{fig:banderrors32}
\end{figure}

\begin{figure}[!h]
\centerline{\includegraphics[width=1.\textwidth,clip=]{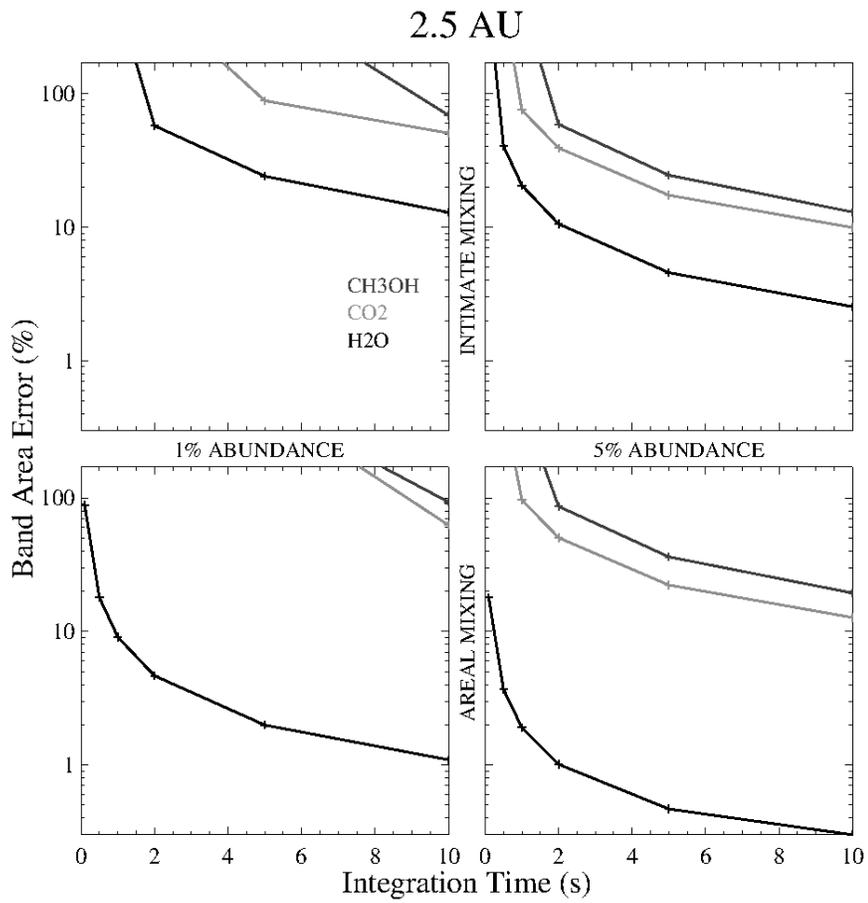}}
\caption{Same of Figure \ref{fig:banderrors32}, for heliocentric distance of 2.5 AU.}
\label{fig:banderrors25}
\end{figure}

\begin{figure}[!h]
\centerline{\includegraphics[width=1.\textwidth,clip=]{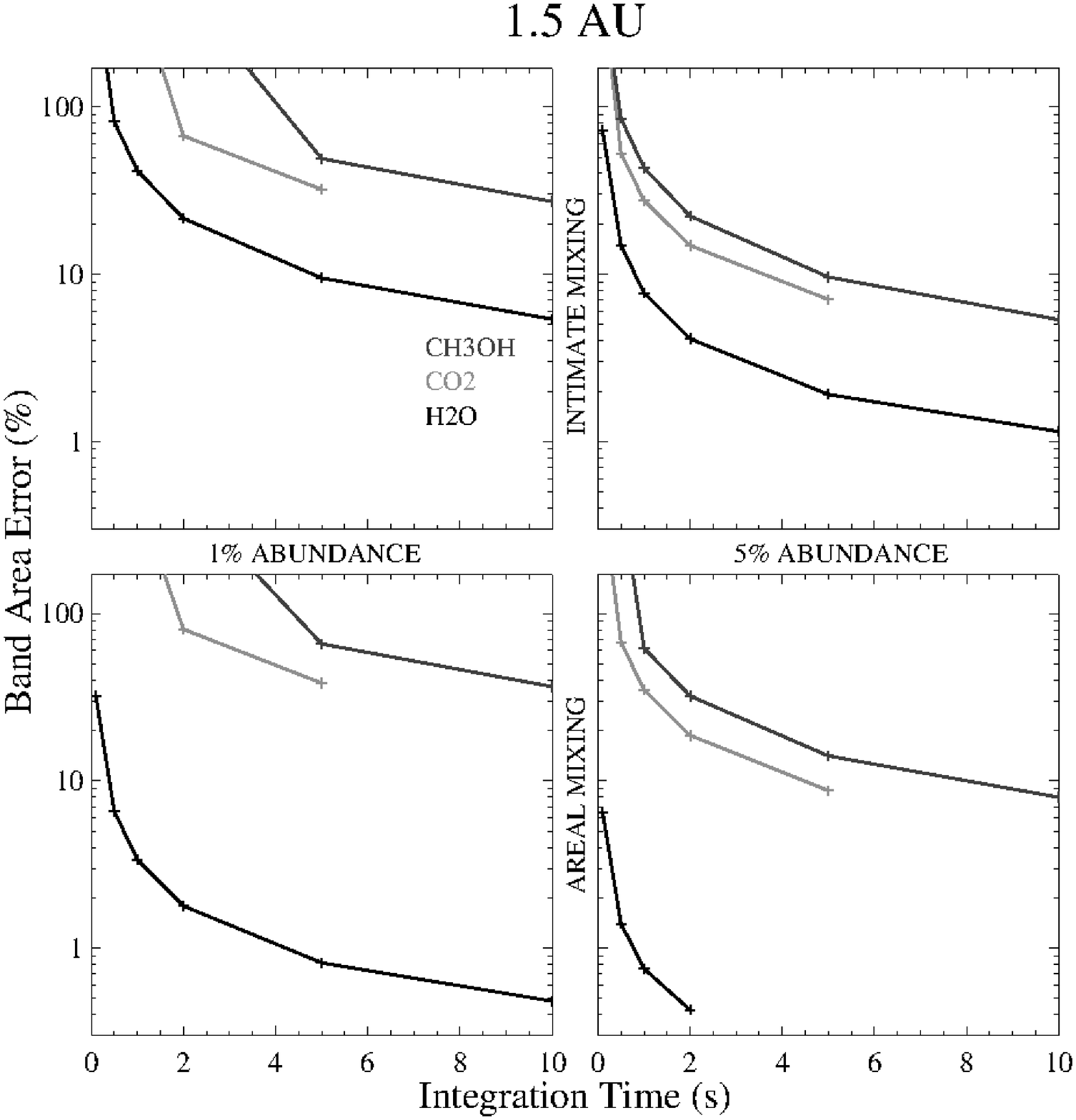}}
\caption{Same of Figure \ref{fig:banderrors32}, for heliocentric distance of 1.5 AU. The missing points represents saturated pixels in correspondence of the absorption bands of the end-members.}
\label{fig:banderrors15}
\end{figure}

The plots in Fig. \ref{fig:banderrors32}, \ref{fig:banderrors25}, \ref{fig:banderrors15} show the error associated to the band area as a function of the integration time for low (left) and high (right) abundances, and for intimate (up) and areal (down) mixing. Each figure shows a different heliocentric distances. The end-members are mixed with the Dark Terrain.

As expected, factors that increase the detectability of a spectral feature for a given end-member are: 1) longer integration time; 2) higher end-member abundance; 3) shorter heliocentric distance. On the other hand, if the signal is too high there is the possibility to reach saturation. Identifying an optimal integration time is therefore necessary for balancing the various factors. An integration time is considered optimal if it is far from the time at which saturation begins and if it permits to correctly detect the end-members.

In the first phase of the mission the comet is far from the Sun. The signal of both reflected sunlight and thermal emission is low. The error associated to each band area is high for short integration times. In general an integration time $> 1\: s$ is needed for the detection of water ice, but a time $> 5\: s$ is required to properly detect carbon dioxide and methanol. However, close to the sun (1.5 AU) 5 s is also a limit beyond which saturation is possible, due to the high signal of thermal emission and reflected sunlight, which prevents the detection of the bands respectively of carbon dioxide and water ice. On the other hand the error on the band area is decreasing because of the higher signal from the nucleus’ surface. 

Although the S/N Simulator is capable to handle variable spectrometer's temperatures, in the study shown in this section we have assumed a nominal temperature of 140 K which is in fact the mean temperature recorded during the mission. However, during some limited phases of the mission, the spectrometer's temperature has risen up to $\sim\:150\: K$. For this reason we have verified how the results are affected by an increment of the spectrometer's temperature. We repeated the same calculation with $T_{S} = 145\: K$ and $T_{S} = 150 \:K$. In general the errors on the band area slightly worsen. However, the main issue is represented by the saturation of pixels that begins at wavelengths gradually smaller with increasing spectrometer's temperature. In some cases, the $CO_{2}$ band is no longer detectable for $T_{S} > 145\:K$ and integration time $> 5\: s$, even at 3.2 AU where the thermal emission from the nucleus surface is lower.

As stated in previous section the variation of the focal plane temperature in the IR channel produces negligible variations in the total signal, at least for the range of temperatures recorded during the mission. 

\chapter{Model-Parameters extractor}
\label{chap:modelextractor}
\section{Algorithms for spectrophotometric analysis}
\label{sec:parametersextractor}
In general the observed data need to be compared with a model to retrieve the surface's properties. 
Referring to the Hapke equations (\ref{eq:hapke} and \ref{eq:rad}) we can distinguish three sets of parameters, which can be retrieved separately because they concern different physical processes: 
\begin{itemize}
\item{parameters describing the thermal emission ($T$ and $\epsilon$ );}
\item{parameters describing the position and the shape of the absorption bands (end-members abundances and grain size) to define the SSA;} 
\item{parameters describing the surface physical properties and structure:

\noindent
 ($p(g), \overline{\theta}, K, B_{S0}, B_{S}(g), B_{C0}, B_{C}(g)$). To retrieve them we need several spectra taken at various phase angles.} 
\end{itemize}

The parameters in each set should be retrieved simultaneously to avoid fictitious constrains. The retrieval procedure has to search for the minimum of the $\chi^{2}_{R}$ variable, namely the best fit between the model ($r^{m}$) and the observed ($r^{o}$) radiance or reflectance. 

\begin{equation}
\chi^{2}_{R} = \left\{\Sigma_{\lambda}[(r^{o}_{i} - r^{m}_{i}) / \sigma_{i}]^{2}\right\}/DOF	
\label{eq:chi}
\end{equation}

Where $\sigma$ is the calculated noise relative to the observed data. $DOF$ are the degrees of freedom, equal to the bands of the spectrum involved in the retrieval process, reduced by the numbers of the parameters to retrieve.

The $\chi^{2}_{R}$ variable is defined in a N-dimensional space, where N is the number of the free parameters to retrieve. So basically the three sets of parameters define three different spaces. We could need many parameters (many dimensions of the space) to fit the data. For example many end-members could be required to fit the absorption bands of a spectrum, and up to 9 parameters are required to fit the phase function with the Hapke model. To perform this task we can take advantage of some curve-fitting algorithm generally used to solve non-linear least squares problems, in particular the gradient-descent algorithm \citep{Snyman} and the Levenberg-Marquardt algorithm \citep{Markwardt}, being respectively a first-order and a second-order optimization algorithm. 

Like other numeric minimization algorithms, these algorithms are iterative procedures. To start a minimization, the user has to provide an initial guess for the parameter vector. In cases with only one minimum, an uninformed standard guess like (0,0,...,0) works fine; in cases with multiple minima, the algorithm converges to the local minimum. For this reason more than one first guess are adopted, trying to explore all the range of possible values for each parameter. The absolute minimum is considered as the best fit if the $\chi^{2}_{R}$ value has a significant lower value than other local minima. 

The best fit gives back a more reliable result if the information of the noise coming from the S/N simulator is used in the definition of the $\sigma_{i}$ in Eq. \ref{eq:chi} (see Fig. \ref{fig:H2Odistribution}). This can be done for every real spectrum acquired by the instrument. This happens because each band is better weighted during the best fit process. Moreover, the resulting $\chi^{2}_{R}$ function coming from the best fit could be compared with the theoretical $\chi^{2}_{R}$ distribution, whose average is 1. This can give a better evaluation on the goodness of the model retrieved. 

The spectral analysis procedure takes as input the observed data, the geometry of the shape model, the calculated S/N of the signal, and an optical constant database. Information on geometry and S/N ratio are not mandatory: the absence of geometry information means we can model only the shape and the position of the absorption bands. In this case the tool can calculate a spectral fit, but a photometric analysis is not possible. In absence of information on the S/N the tool can use another reasonable weight ($\sigma$) in Eq. \ref{eq:chi}, like the measured signal. Without these information the retrieval, even still possible, loses reliability.

\begin{figure}[!h]
\centerline{\includegraphics[width=0.7\textwidth,clip=]{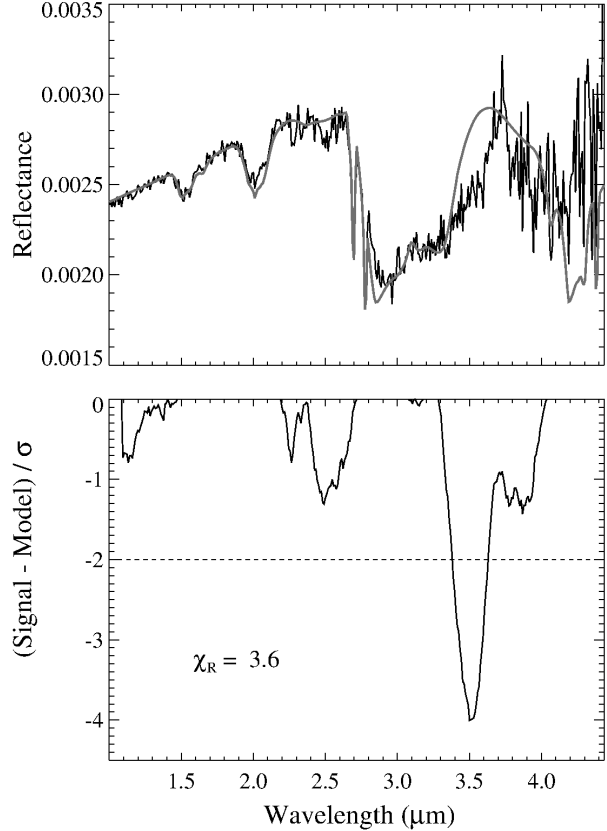}}
\caption{As a first guess the reflectance spectrum is modeled taking into account the dark terrain and the optical constants of the water ice and the carbon dioxide. Up: the synthetic spectrum and the resulting best fit (clear line). The reduced chi-square results to be 3.6. Down: the residuals spectrum smoothed with a 10 bands running box. Only the residuals having the signal overestimated by the model are shown to enlighten absorption bands we are missing. We consider significant a residual if it overcome 2 $\sigma$. We can notice a significant residual at $\sim\:3.5\;\mu m$. We can argue we are missing the absorption band of  methanol.}
\label{fig:step1}
\end{figure}

\begin{figure}[!h]
\centerline{\includegraphics[width=0.7\textwidth,clip=]{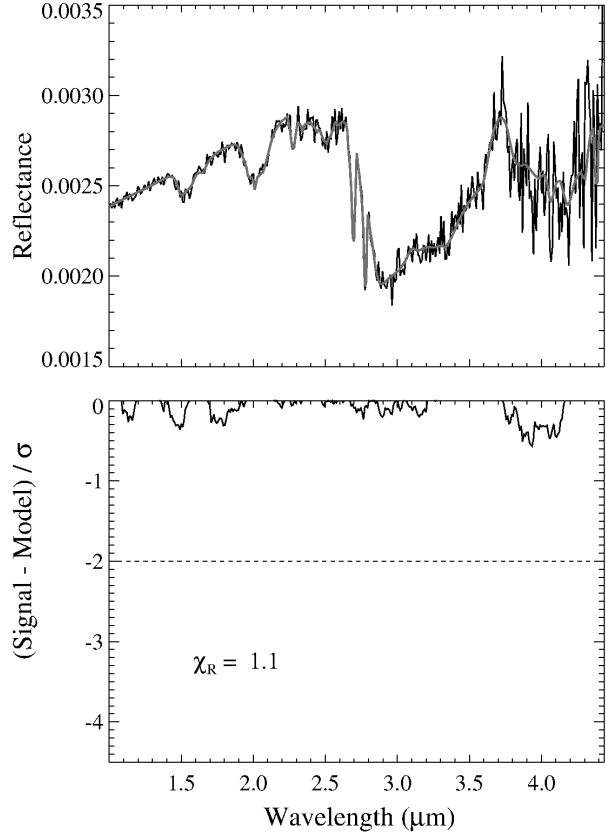}}
\caption{We repeat the procedure of Figure \ref{fig:step1} including methanol. The resulting reduced chi-square is 1.1. The bottom panel show the residuals spectrum keeping the same range of Figure \ref{fig:step1} for the sake of clarity. There are no significant residuals. We can argue we have achieved a good model.}
\label{fig:step2}
\end{figure}

A simple example of the spectral analysis procedure, with simulated spectra, is shown in Figure \ref{fig:step1} - \ref{fig:step2}. The specific composition and viewing geometry is the same shown in figure \ref{fig:radrefl}, being the relative instrumental, geometric and compositional parameters as stated in section \ref{sec:minor}. The respective noise is also simulated.

The resulting reflectance spectrum is analyzed as discussed. Thanks to the analysis of the residuals ($[r^{o}_{i} - r^{m}_{i}] / \sigma_{i})$ a final model is obtained. The fitting procedure have given as output the same abundance and the grain size used to simulate the spectra, showing the functionality of the method.

With the availability of an adequate optical constant database, and the capability of the fast algorithms to handle many end-member, we can perform an automatic procedure to select the end-members as a function of the position of the significant residuals.

Results from the spectral fit allow us to determine the mixture and the grain size. Their knowledge enables us to compute and fix the single scattering albedo $\omega$. Once we have a collection of spectra of the same surface region observed at different phase angles we can model the phase function with a best fit with the Hapke model. The possibility to fix $\omega$ reduces the complexity of phase function fit procedure and allows the decoupling of spectral effect from photometric ones. A fit of the phase function could be performed at each wavelength. However, with the knowledge of $\omega$ we can perform a phase function fit taking into account the whole spectrum, or part of it, allowing a better retrieval of the parameters thanks to the larger statistics. In this way the fit of the phase function is performed after the spectral fit. However, the procedure could be inverted. A phase function fit could be performed at each wavelength before the spectral fit. In this case the single scattering albedo $\omega$ is a free parameter to retrieve. This approach is followed for Lutetia analysis showed in Chapter \ref{sec:lutetia}. 

Once we have a spectrum of the retrieved single scattering albedo it can be modeled by a spectral fit, or it can be assumed as end-member for spectral unmixing of nearby regions which present similar spectral features. 

\section{Parameter errors evaluation}
\label{sec:errorsextractor}
\subsection{Errors from signal uncertainties}
\label{sub:errsignal}
\begin{figure}[!h]
\centerline{\includegraphics[width=0.8\textwidth,clip=]{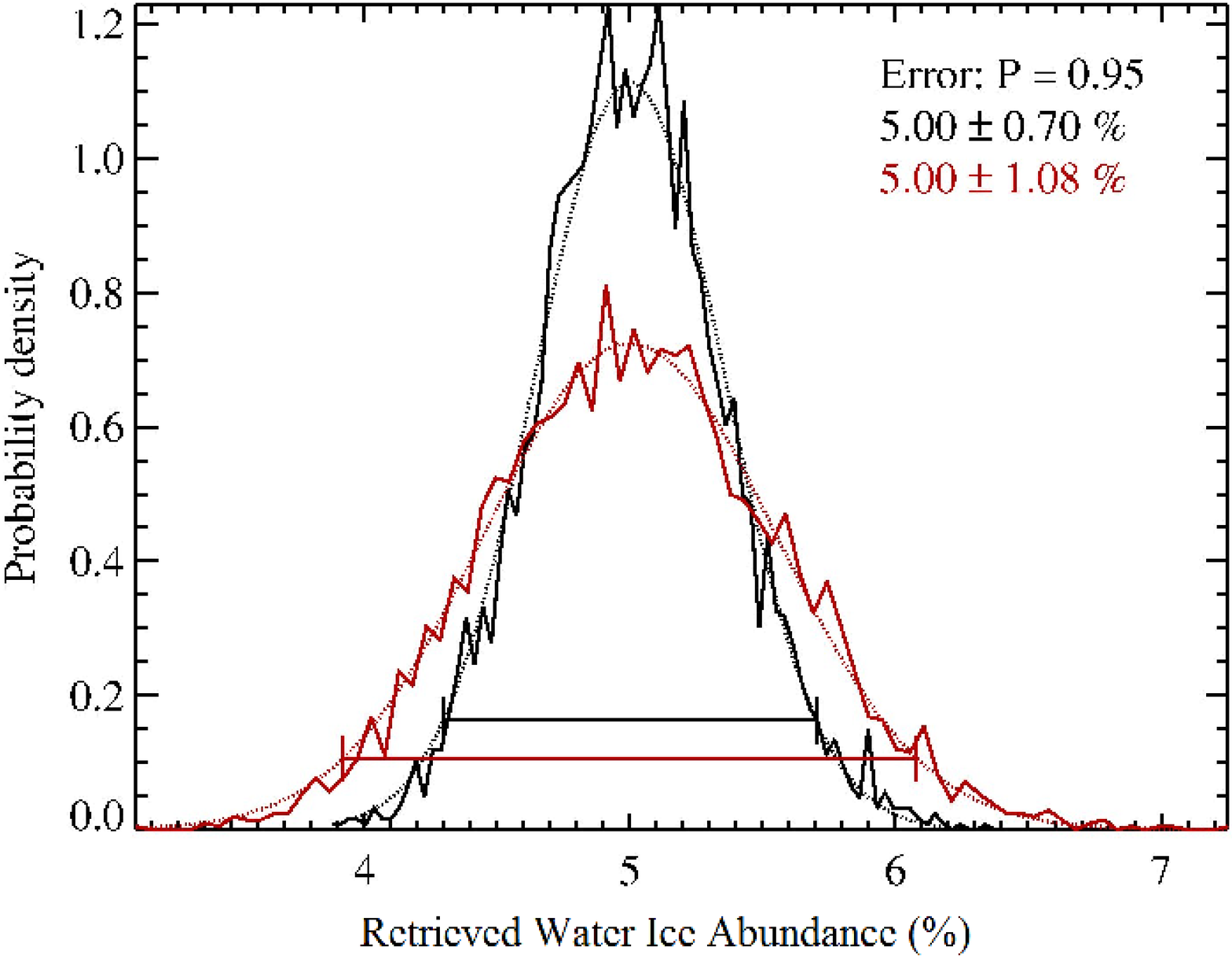}}
\caption{Four thousand noisy spectra are simulated, and the plots show the probability density function (80 bins) of the retrieved abundance of water ice, being the value of the model in input equal to 5\%. The black line shows the resulting distribution in the case $\chi^{2}$ is defined according to Eq. \ref{eq:chi}. The clear line shows the resulting distribution in the case the $\sigma$ in Eq. \ref{eq:chi} is not coming from the S/N simulator but the observed data is used as weight. The theoretical gaussian distributions over plotted (dashed line) have the same mean and dispersion of the resulting distributions. The position of the bars are in the points of the theoretical distributions such that they contain the 95\% of the probability distribution. The distance of the bars from the mean value is indicated in the figure.}
\label{fig:H2Odistribution}
\end{figure}

We can be interested to link the error bars of the spectra to the error of the retrieved parameters. An intuitive method to estimate the total error on the retrieved parameters due to the uncertainties on the input data is given here. After the retrieval procedure we obtain values of spectrophotometric parameters by which we have a model of the spectrum. The model could be used to generate many synthetic spectra adding random poissonian noise, thanks to the information of the $\sigma$ as estimated by the S/N simulator. For each synthetic spectrum generated in this way, the retrieval procedure is repeated, giving as a result different values than those obtained originally. At the end we have a gaussian-like distribution of values for each parameter. Taking the 95\% of the distribution would be a conservative estimation for the error bars of the parameters. An example is shown in Figure \ref{fig:H2Odistribution}.

For real data we could have the situation in which the σ of the S/N calculator underestimates the real noise of the spectrum. This is enlighten by an high value ($>>1$) of the $\chi^{2}_{R}$ coming from the retrieval procedure. In this case to correctly evaluate the error bars of the retrieved parameters, we should scale the calculated error bars by the square root of the $\chi^{2}_{R}$. 

Other sources of error could add uncertainty on the results, like errors on the calibration. Those typically add a fictitious offset to the whole spectrum or in some parts of this. Any systematic error should be removed before starting the retrieval procedure. Other unpredictable errors should be included in this procedure. As an example if we have an estimated uncertainty of 5\% on the absolute calibration, each synthetic spectrum should include a random offset having a $\sigma$ of this amount. 

\subsection{Errors from geometry uncertainties}
The determination of the angles (incidence and emission) of an area subtended by a spatial pixel can presents uncertainty and this should not be neglected because they are required by the model (Eq. \ref{eq:hapke}). The shape of the surface is modeled by many facets that have their own orientations. The area subtended by a pixel is always larger than the area of the facets, and therefore, there are several facets for each pixel.

There are two sources of error that contribute to the uncertainty:
\begin{itemize}
\item{error in the estimation of the average angle. This is usually performed taking into account the four facets at the corners and at the center of the pixel;}
\item{effective angles could be different with respect to the average angles because of the effect of the topographic roughness. This effect is predicted by the Hapke model and should not be included in this evaluation.}
\end{itemize}

To predict the effect of the error on geometry we have to calculate the probability density plot as described in section \ref{sub:errsignal} including a random variation in the orientation of the surface. 

\begin{figure}[!h]
\centerline{\includegraphics[width=0.7\textwidth,clip=]{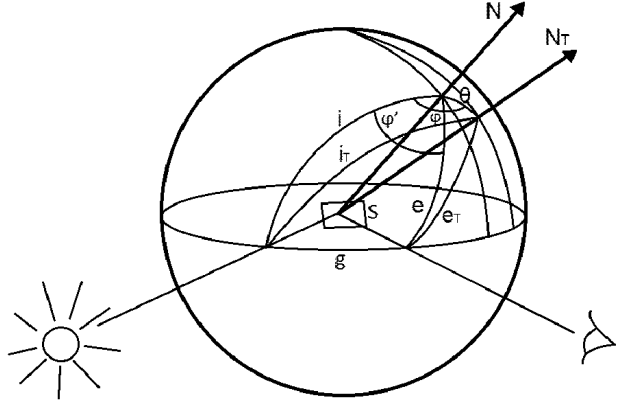}}
\caption{Schematic diagram of tilt coordinates on an element of surface $S$. Being $N$ the normal to the surface, $N_{T}$ the normal tilted, $i$ and $e$ the incidence and emission angle, $i_{T}$ and $e_{T}$ the incidence and emission angle after the tilt of the normal, $\varphi$ and $\varphi'$ respectively the azimuth of the tilt and the azimuth of the emission direction from the direction of the incidence angle; $\theta$ is the tilt angle, and $g$ is the phase angle.}
\label{fig:tilt}
\end{figure}

For each simulation, the normal to the surface is tilted by a random angle $\theta$, that is normally distributed with a fixed standard deviation $\sigma_{\theta}$. The direction of the tilt is determined by the $\varphi$ angle that is uniformly distributed. The cosines of the angles required by the Hapke reflectance model are calculated following the spherical geometry as follows:

\begin{align}
\mu_{0\:TILT}&=\mu_{0} \times cos(\theta)+sin(i) \times sin(\theta) \times cos(\varphi_{i})\\
\mu_{TILT}&=\mu \times cos(\theta)+sin(e) \times sin(\theta) \times cos(\varphi_{e})
\end{align}

where:
\begin{align*}
\varphi_{i}&=\varphi, & \varphi_{e}&= \varphi' - \varphi, & for\:\varphi&<\varphi'\\
\varphi_{i}&=2\pi-\varphi, & \varphi_{e}&= \varphi - \varphi', & for\:\varphi&>\varphi'
\end{align*}

\begin{equation*}
\varphi=arcos\left\{[cos(g) - \mu_{0} \times \mu] / [sin(i) \times sin(e)]\right\}
\end{equation*}

being $\varphi$ and $\varphi'$ respectively the azimuth of the tilt and the azimuth of the emission direction in the reference frame whose z axis is the normal to the surface, that are counted counterclockwise with the azimuth of the incidence direction equal to $0^{o}$ (see Figure \ref{fig:tilt}).

The procedure discussed in section \ref{sub:errsignal} is performed including a random tilt of the surface with a fixed $\sigma_{\theta}$, in a way to modify the value of the incidence and emission angles, and therefore the cosines required by the Hapke reflectance model. 

The dispersion of the retrieved parameters is thus a measure for errors that can be committed in the determination of these parameters, due to both the signal error and the observation geometry uncertainties. 

We can include the evaluation of the error due to geometry information if all the spectra are coming from the same surface region. In case we consider spectra from different surface regions, in order to perform a phase function fit (assuming the same compositional properties), the uncertainties on the geometry have different effects for each spectrum. In this case this last tool is no longer applicable and the errors on geometry can be considered as a further source of noise that increases the $\sigma$ in Eq. \ref{eq:chi}. We should then apply the procedure discussed in section \ref{sub:errsignal} and consider the value of the $\chi^{2}_{R}$ coming from the best fit to correctly evaluate the error bars of the retrieved parameters. This is the case of our Lutetia analysis, discussed in chapter \ref{sec:lutetia}.

\chapter{Deep Impact-HRI data analysis}
\label{DI}
On 2005 July 4, the Deep Impact mission carried out an active planetary experiment by delivering a high-speed impactor against the nucleus of Comet 9P/Tempel1. The mission excavated material from well below the surface, making it available for study both by the Deep Impact flyby spacecraft and by remote sensing from Earth, both on the ground and in Earth orbit, and from elsewhere in the Solar System (Rosetta, Spitzer Space Observatory, Solar and Heliospheric Observatory)  \citep{DI}.

The primary goal of Deep Impact was to understand the difference between the surface of a cometary nucleus and its interior and thus to understand how the material released spontaneously by the nucleus is related to the primitive volatiles that were present in the protoplanetary disk at the age of formation. The other important goal was to understand
the physical properties of the outer layers of the comet  \citep{DI}.

9P/Tempel 1 is a Jupiter-family comet discovered in 1867 by Ernst Wilhelm Leberecht Tempel. With an orbital period at the time of 5.68 years, it was observed on two subsequent apparitions. However, in 1881 owing to a 0.55 AU pass by Jupiter, its orbital period was increased to 6.5 years and its perihelion raised from 1.77 to 2.07 AU. The comet remained ``lost" until \citet{Marsden} calculated that close approaches of the comet to Jupiter in 1941 and 1953 should have decreased the perihelion distance and orbital period. As a result the comet was recovered by Roemer in 1967 and has been observed all of its subsequent perihelion apparitions. Tempel 1 now has an orbital period of 5.5 years and a perihelion distance of 1.5 AU  \citep{DI}.

Deep Impact was launched from Cape Canaveral Air Force Station on 2005 January 12 and it took a short, six-month trajectory to position itself to be overtaken by Comet Tempel 1 on the other side of the Solar System on 2005 July 4. The encounter occurred one day before the comet reached perihelion and 3 days before the comet reached its descending node, thus optimizing the energy delivered in the encounter  \citep{DI}.

The spacecraft consists of two main sections, the 370-kg copper-core ``Smart Impactor" that impacted the comet, and the ``Flyby" section, which imaged the comet from a safe distance during the encounter with Tempel 1.

\begin{figure}[!h]
\centerline{\includegraphics[width=0.4\textwidth,clip=]{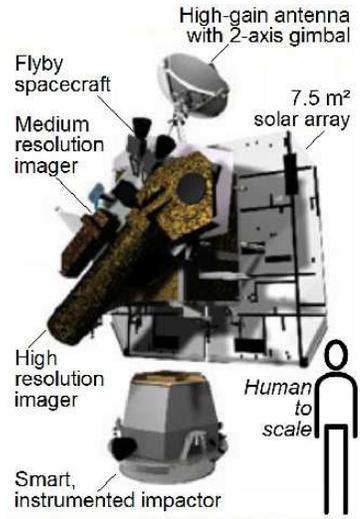}}
\caption{Deep Impact spacecraft.}
\label{fig:spcrft}
\end{figure}

The Flyby spacecraft (see Figure \ref{fig:spcrft}) is about 3.2 meters long, 1.7 meters wide and 2.3 meters high. It includes two solar panels, a debris shield, and several science instruments for imaging, infrared spectroscopy, and optical navigation to its destination near the comet. The spacecraft also carried two cameras, the High Resolution Imager (HRI), and the Medium Resolution Imager (MRI). The HRI is an imaging device that combines a visible-light camera with a filter wheel, and an imaging infrared spectrometer that operates on a spectral range from 1.05 to 4.8 $\mu m$. It has been optimized for observing the comet's nucleus. 
It consisted of a long-focal-length telescope followed by a dichroic beam splitter that reflected (0.3 to 1.0 microns) light through a filter wheel to a CCD for direct, optical imaging (HRIV). The beam splitter transmits the near infrared light to a 2-prisms spectrometer (HRII) \citep{Hampton}.

\begin{table}[!h]
\caption{HRII Spatial characteristics}
\begin{center}
\begin{tabular}{||c|c||}
\hline
\hline
Physical Pixel Size & 36 micrometers\\
\hline
Effective Pixel FOV & 10.0 microradians\\
\hline
Effective FOV & 2.5 milliradians or $0.15^{o}$\\
\hline
\end{tabular}
\end{center}
\end{table}

\begin{table}[!h]
\caption{HRII Spectral characteristics}
\begin{center}
\begin{tabular}{||c|c||}
\hline
\hline
Effective FOV & 10.0 microradians (slitwidth)\\
\hline
Spectral range & 1.05 to 4.8 $\mu m$\\
\hline
Resolution & R = 200 - 800\\
\hline
\end{tabular}
\end{center}
\end{table}

\begin{table}[!h]
\caption{HRIV}
\begin{center}
\begin{tabular}{||c|c||}
\hline
\hline
Pixel Size & 21 micrometers\\
\hline
Pixel FOV & 2.0 microradians\\
\hline
Instrument FOV & 2.0 milliradians or $0.118^{o}$\\
\hline
Surface Scale & 1.4 meters/pixel at 700 km\\
\hline
\end{tabular}
\end{center}
\end{table}

The MRI is the backup device, and was used primarily for navigation during the final 10-day approach. It also has a filter wheel, with a slightly different set of filters.

The Impactor section of the spacecraft contains an instrument that is optically identical to the MRI, called the Impactor Targeting Sensor (ITS), but without the filter wheel. Its dual purpose was to sense the Impactor's trajectory, which could then be adjusted up to four times between release and impact, and to image the comet from close range. 

As the Impactor neared the comet's surface, this camera took high-resolution pictures of the nucleus (as good as 0.2 meters per pixel) that were transmitted in real-time to the Flyby spacecraft before it and the Impactor were destroyed 

(http://en.wikipedia.org/wiki/Deep\_Impact\_(spacecraft)). 

The observations of the impact show distinct signs of layering within the excavated material and provide an upper limit to the strength in the surface layer showing that the material is much weaker than ice and that the nucleus is highly porous. 

The appearance of ice in the coma immediately after the impact, requires that ice must be present in the first outer layers of the surface at the impact site, and it is important to note that fairly pure ice grains were present \citep{DI}.

Key results unrelated to the impact include the determination of a very low thermal inertia, a very high frequency of spontaneous outbursts, apparently correlated with rotational phase in the several weeks preceding the impact, and the existence of small patches of ice on the surface \citep{Sunshine06} that cannot be the primary source of the water observed in the coma \citep{DI}.

Although the impactor spacecraft was destroyed, the flyby spacecraft and its instruments remained healthy in its 3-year, heliocentric orbit after completion
of the mission. After that the Deep Impact flyby spacecraft was retargeted to comet 103P/Hartley 2 as part of an extended mission named EPOXI (Extrasolar Planet Observation and Deep Impact Extended Investigation).

The closest approach to Hartley 2 at 694 km happened on 2010 November 4th, 1 week after perihelion passage and at 1.064 astronomical units (AU) from the Sun. Observations of the comet were carried out for 2 months on approach (5 September to 4 November) and for 3 weeks on departure (4 to 26 November), during which more than 105 images and spectra were obtained \citep{EPOXI}.

\begin{figure}[!h]
\centerline{\includegraphics[width=0.8\textwidth,clip=]{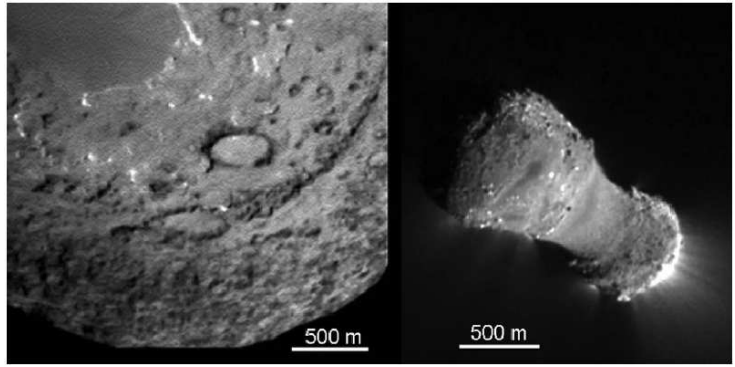}}
\caption{Comparison of a small part of (left) Tempel 1 with (right) Hartley 2 at approximately the same image scale and with nearly identical instruments. Left: Impactor Targeting Sensor (ITS) image iv9000675, 9.1 m/pixel. Right: MRI image mv5004032, 8.5 m/pixel. Sun is to the right  \citep{EPOXI}.}
\label{fig:TempelHartley}
\end{figure}

Prior remote sensing showed that Hartley 2's nucleus has an average radius 1/5 that of comet Tempel 1's nucleus, yet it releases more gas per unit time at perihelion, even when allowing for the smaller perihelion distance of Hartley 2 (1.059 versus 1.506 AU). This puts Hartley 2 in a different class of activity than Tempel 1 or any of the other comets visited by spacecraft. The two comets have very different surface topography (Fig. \ref{fig:TempelHartley}), but whether this is related to the hyperactivity or other processes is still being investigated \citep{EPOXI}.

\section{Tempel 1}

\subsection{Spectral analysis}
\label{sub:spctralTempel}
The calibrated radiance spectra are available through the Planetary Data System website (http://pds.jpl.nasa.gov/). During the flyby of the comet many hyperspectral cubes were taken by the HRI instrument with increasing resolution. We have analyzed the most resolved cube which present the whole nucleus: Exposure ID = 9000036, with a resolution of 164 - 152 m/px (geometry information on this acquisition are courtesy of Tony Farnham). After the impact other cubes with a maximum resolution of 16 m/px were acquired but they show only a little portion of the surface, contaminated by the impact ejecta and without geometry information available. So they are discarded for the following spectrophotometric analysis.

Only data of the spectrometer in the range 1.2 - 4.5 $\mu m$ are used, because of an higher radiometric accuracy \citep{Klaasen1, Klaasen2}.

The first step to perform the analysis is the modeling of the radiance in terms of thermal emission and reflected sunlight (see Eq. \ref{eq:rad}). This is useful for both isolating reflectance spectrum and thermal emission. In this section the former contribution is analyzed, in next section the latter.

The procedure performs the following steps:
\begin{enumerate}
\item{It divides the measured radiance by the solar irradiance corrected for the heliocentric distance ($J/D^{2}$). The result is the reflectance added to the thermal emission.}
\item{At shorter wavelengths (1.2 - 2.5 $\mu m$) the spectrum is not contaminated by thermal emission. A linear fit is performed in this range to extend the reflectance spectrum at longer wavelength. This is reasonable for Tempel 1 spectra that present a quite flat trend. Thus for each pixel a spectral reddening is calculated as the slope of the reflectance spectrum (see Fig. \ref{fig:Tempel4}).}
\item{The contribution of the reflected sunlight in term of radiance is calculated by multiplying the solar irradiance for the interpolated reflectance spectrum.}
\item{The thermal emission is isolated by the total radiance by subtracting the calculated reflected sunlight contribution (previous point) to the measured radiance.}
\end{enumerate}

\begin{figure}[!h]
\centerline{\includegraphics[width=0.8\textwidth,clip=]{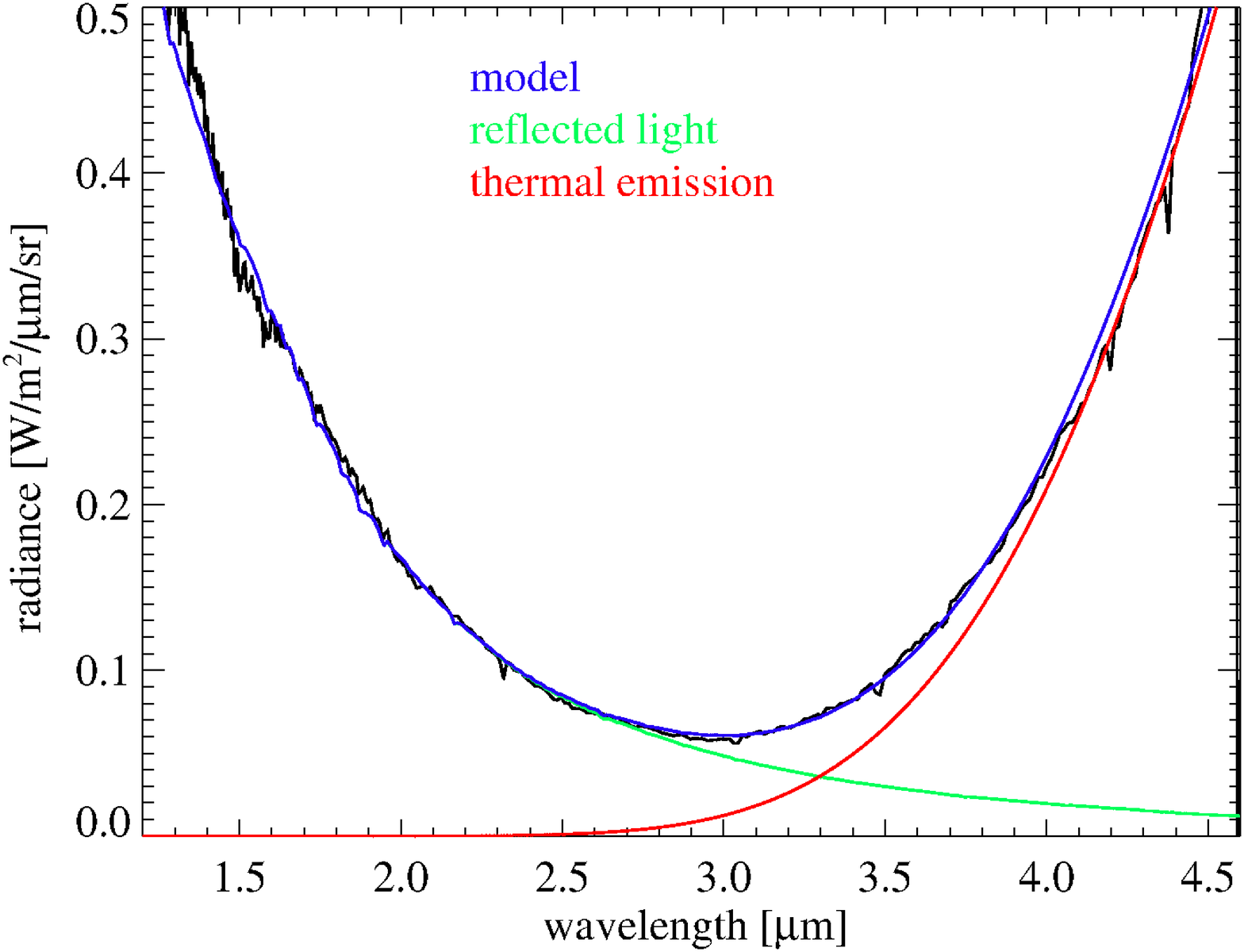}}
\caption{The measured radiance (black line) is modeled in terms of reflected sunlight and thermal emission.}
\label{fig:modelrad}
\end{figure}

In this way the two contributions are isolated and are available for further analysis (see Fig. \ref{fig:modelrad}).

Isolating the thermal emission is also useful to extend the reflectance spectra to longer wavelengths: the thermal emission is modeled as discussed in next section and is subtracted to the measured radiance spectra before dividing it by the solar irradiance. In this way the resulting reflectance spectra are free from thermal emission signal. 

The data of the camera (HRIV) are available for seven different bands from 0.35 to 0.95 $\mu m$ (see Tab \ref{tab:HRIV}). These images are rescaled to the same spatial resolution of the hyper-spectral cubes obtained by the spectrometer. In this way we are able to extend the spectra of each pixel to the visible wavelengths. This task is straightforward because all the used data were acquired in a short time interval compared with the comet's rotation period. Therefore, the dataset analyzed contains only observations taken on the same nucleus hemisphere. Since the radiometric accuracy of the VIS channel is higher than the IR, the latter has been scaled to the former before bridging the spectra of each pixel. This turned to be useful to correct by a factor 2 the calibrated radiance of the HRII instrument \citep{Klaasen1, Klaasen2}. All the spectra used in this work are corrected by this factor.

\begin{figure}[!h]
\centerline{\includegraphics[width=1.\textwidth,clip=]{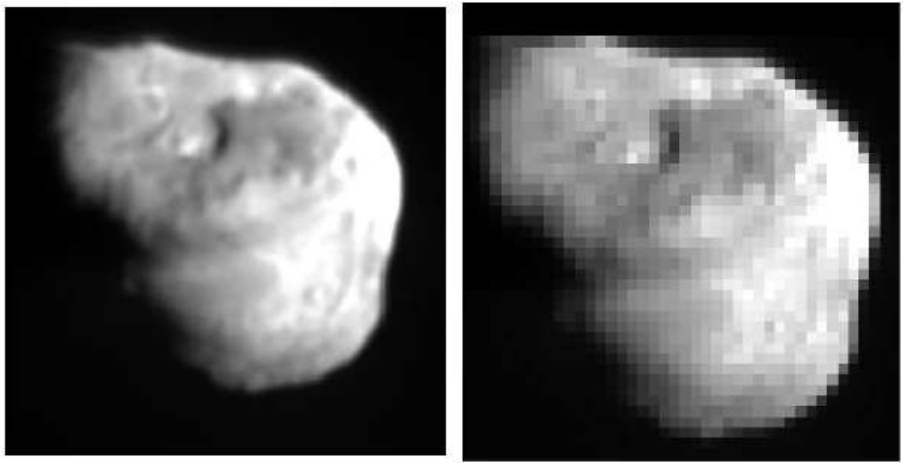}}
\caption{Left panel: HRIV acquisition, Exposure ID = 9000901, right panel: same image after lowering the spatial resolution to match that of the HRII channel.}
\label{fig:Tempel2}
\end{figure}

All of the HRI images are out of focus due to a pre-flight calibration mirror's failure \citep{Klaasen1}. This effect do not affect the HRII acquisition because of the low spatial resolution of the HRII channel, while can be clearly seen for HRIV acquisition (see Fig. \ref{fig:Tempel2}). However, as stated, the HRIV images are rescaled to the same spatial resolution of the HRII channel. Thus this effect do not affect the analysis.

\begin{table}[!h]
\caption{The selected images taken from the HRIV instrument.}
\begin{center}
\begin{tabular}{||c|c|c|c||}
\hline
\hline
Exposure ID	& Filter (nm)	& Pixel scale (m/pixel)	& Distance (km) \\
\hline
\hline
9000906 & 350 & 17.95 & 8977\\
\hline
9000908 & 450 & 17.72 & 8861\\
\hline
9000907 & 550 & 17.84 & 8923\\
\hline
9000901 & 650 & 18.67 & 9337\\
\hline
9000903 & 750 & 18.39 & 9196\\
\hline
9000902 & 850 & 18.55 & 9274\\
\hline
9000905 & 950 & 18.14 & 9071\\
\hline
\end{tabular}
\end{center}
\label{tab:HRIV}
\end{table}

\begin{figure}[!h]
\centerline{\includegraphics[width=0.7\textwidth,clip=]{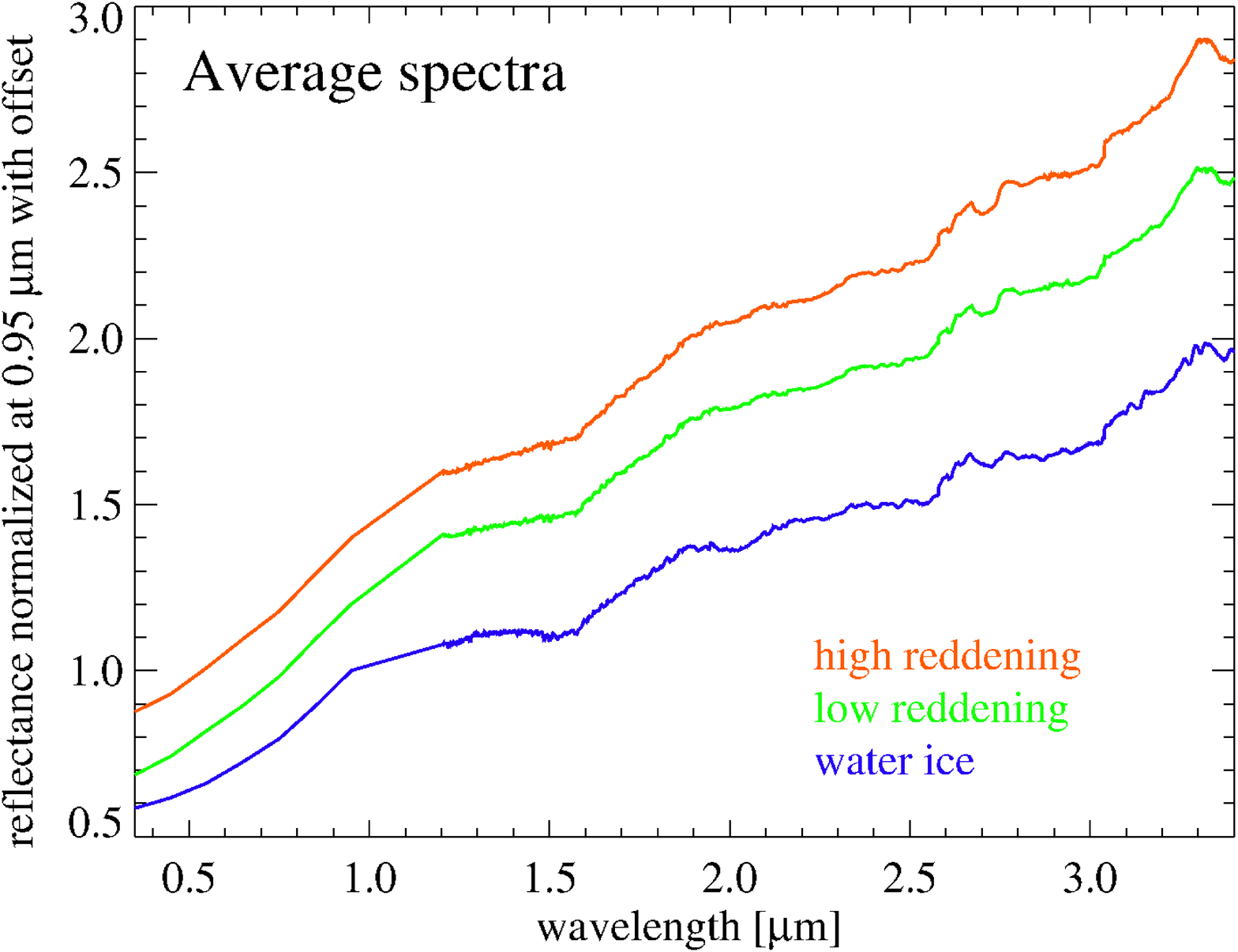}}
\caption{Average spectra of nucleus regions shown in Fig. \ref{fig:icyTempel}}
\label{fig:Tempelspectra}
\end{figure}

\begin{figure}[!h]
\centerline{\includegraphics[width=0.8\textwidth,clip=]{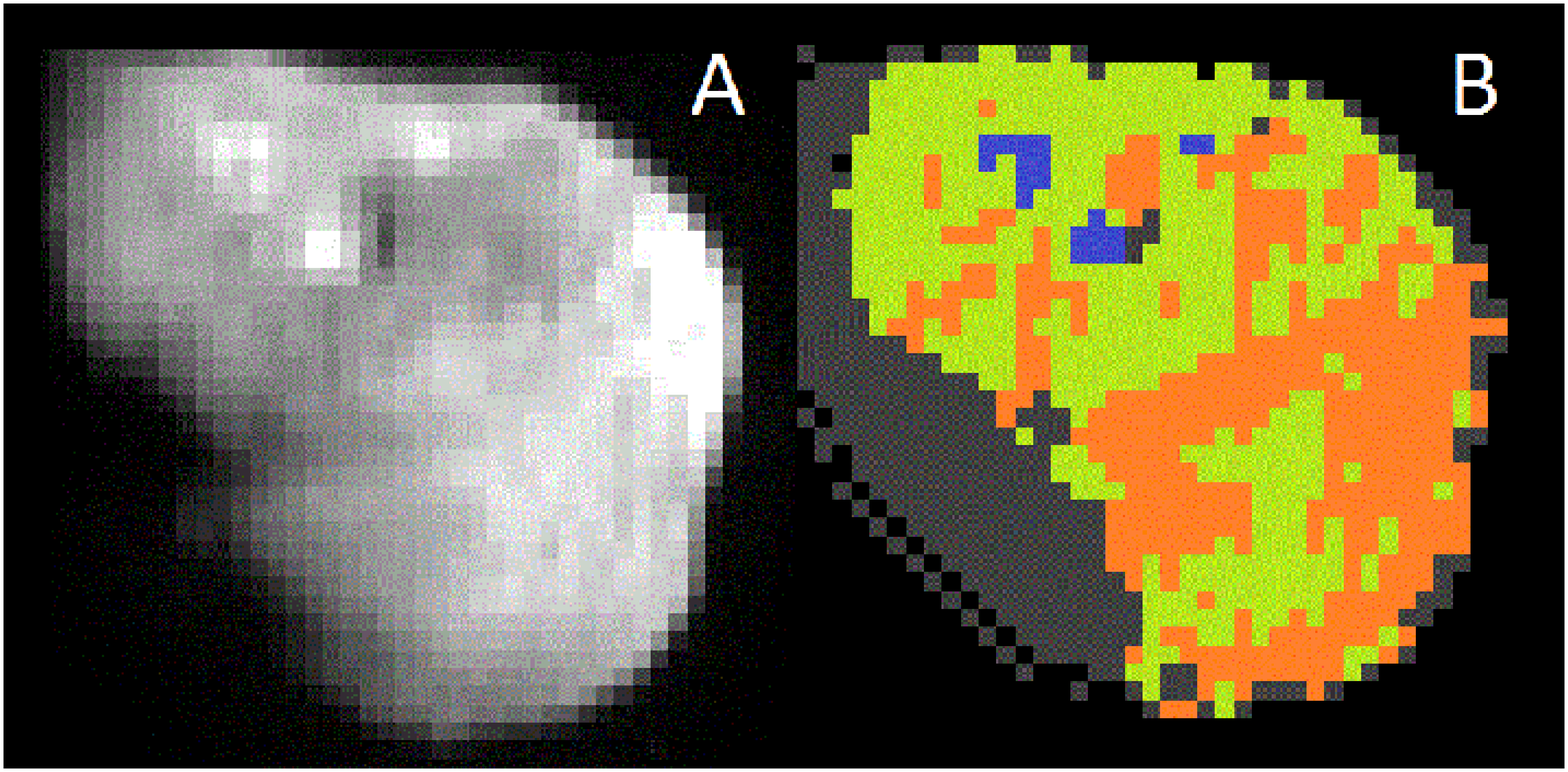}}
\caption{Panel A: image of the nucleus of the comet at $\lambda = 0.35 \: \mu m$ (exposure ID 9000906), rescaled to the same resolution of the hyperspectral cubes. Panel B: spectral classification of the comet's nucleus: in blue the icy regions, in green the low reddening regions, in orange the high reddening regions, in grey the regions which presents unfavourable angles for the analysis.}
\label{fig:icyTempel}
\end{figure}

The VIS-IR hyperspectral cubes are powerful tools to identify icy regions (see Fig. \ref{fig:Tempelspectra}). After having assigned thresholds for the reflectance at visible wavelengths and for the 2.0 $\mu m$ band depth is possible to identify the ice-rich units. In this way we have obtained a similar map to that obtained by \citet{Sunshine06} (see Fig. \ref{fig:icyTempel}).

To define a SSA of the Dark Terrain we need to exclude the icy units from the analysis. Taking into account the Hapke model (Eq. \ref{eq:hapke}), we can derive a SSA from the average spectrum of reflectance of the non-icy units. However, because the available observations are all taken at a single phase angle ($63^{o}$) we have to assume the value of the parameters in the model. At this specific phase angle we assumed that the opposition effect (significant at small phase angles) and the shadowing caused by the roughness (significant at high phase angles) are negligible. Moreover, we fix the porosity parameter and the single particle phase function to 1. In this way we have derived the SSA by inverting the Eq. \ref{eq:hapke}, and we have performed a linear interpolation of it (see Fig. \ref{fig:ssavari})

\begin{figure}[!h]
\centerline{\includegraphics[width=1.\textwidth,clip=]{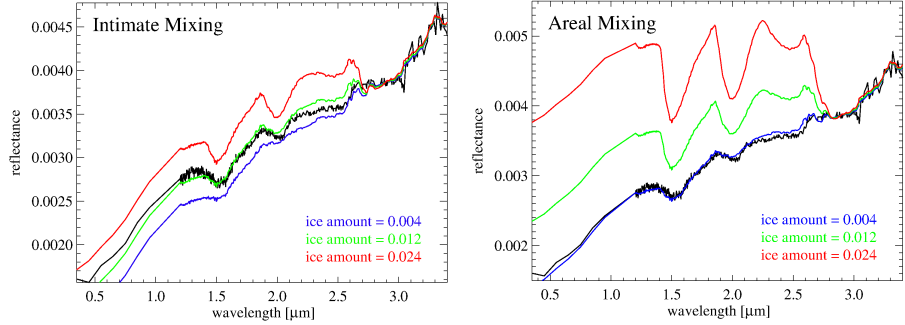}}
\caption{Intimate (left) and areal (right) mixing results: icy region (corresponding to pixels marked in blue in Fig. \ref{fig:icyTempel}) average observed spectrum (black curve); areal mixing spectra with 2.4, 1.2 and 0.4\% of water ice are indicated by red, green and blue curves, respectively. The best-fit are obtained respectively with 1.2\% and 0.4\% water ice-rich spectrum.}
\label{fig:Tempelmixing}
\end{figure}

The extrapolation of the SSA ($\omega$) is useful in modeling the reflectance spectra as a mixture of Dark Terrain and pure water ice. The SSA of pure water ice is obtained from optical constants by \citet{ClarkRN}, \citet{Mastrapa08} and \citet{Warren}, covering a large range of temperature (40 - 260 K), and different states (crystalline and amorphous). In the final fit there are not remarkable differences in the results using different optical constants. As shown in Fig. \ref{fig:Tempelmixing} the low amount of surface ice (and the instrumental sensitivity) don't allow us to infer the water ice state. In particular the 1.65 $\mu m$ secondary absorption feature, characteristic of the crystalline form is not evident on Tempel 1 spectra.

Adopting the areal mixing we obtain from the spectral fits an average grain size of $30 \pm 20\: \mu m$ in diameter with a percentage of water ice of $0.4 \pm 0.2\%$. In the case of intimate mixing we obtain a grain size of $70 \pm 40\: \mu m$ with a percentage of water ice of $1.0 \pm 0.5\%$. The spectra corresponding to the icy regions are well-fitted with both mixing methods.

\begin{figure}[!h]
\centerline{\includegraphics[width=0.6\textwidth,clip=]{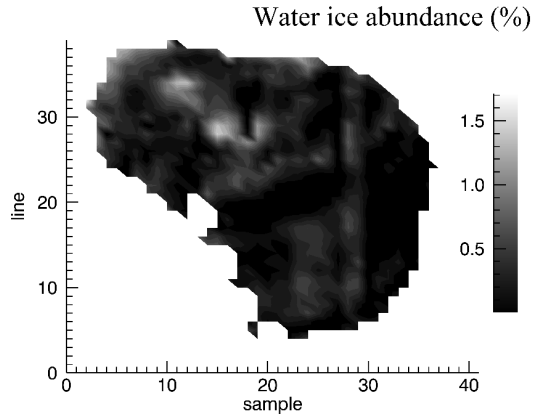}}
\caption{Water ice abundance map, in case we model the mixing between Dark Terrain and pure water ice to be intimate.}
\label{fig:Tempelicemap}
\end{figure}

As a result of the mixing modeling we obtain a map of water ice abundance (see Fig. \ref{fig:Tempelicemap}), whose abundance slightly differs if we take into account an areal or intimate mixing (see Fig. \ref{fig:Tempelmixing}). 
An intimate mixing means that the ice and non-icy components are thermally coupled. This has implications on ice sublimation as discussed in section \ref{sub:alternative}.

\subsection{Thermal emission analysis}
\label{sec:ThermalTempel}
The definition of the single scattering albedo ($\omega$) discussed in previous section is also useful to constrain the thermal properties of the material making up the comet surface. In particular here we are interested to derive the integrated emissivity ($\overline{\epsilon}_{h}$). Following \citet{Davidsson} it can be approximated with:

\begin{equation}
	\overline{\epsilon}_{h} \approx 1 - A_{b}
\end{equation}

\noindent
where $A_{b}$ is the bolometric albedo defined as the average of the spectral spherical albedo $A_{s}(\lambda)$ weighted by the spectral irradiance of the Sun $J_{s}(\lambda)$:

\begin{equation}
	A_{b} = \frac{\int_{0}^{\infty}A_{s}(\lambda)J_{s}(\lambda)d\lambda}{\int_{0}^{\infty}J_{s}(\lambda)d\lambda}
\end{equation}

\noindent
and the spherical albedo ($A_{s}$) is the total fraction of incident irradiance scattered by a body into all directions. It can be approximated by (Eq. 11.45b in \citet{Hapke}):

\begin{equation}
	A_{s} \approx \frac{1-\gamma}{1+\gamma}\biggl(1-\frac{1}{3}\frac{\gamma}{1+\gamma}\biggl)
\end{equation}

\noindent
where $\gamma = \sqrt{1-\omega}$.

Although the integral to derive $A_{b}$ should be calculated from 0 to infinity, most of the solar radiation comes from the wavelengths covered by the channels HRIV and HRII (from $\sim 0.2$ to $\sim 5 \mu m$ ), hence an integration on this range is a good approximation.

The overall calculation brings to the value of $\overline{\epsilon}_{h} = 0.987$

To calculate the Temperature and the emissivity from the data it is followed the work of \citet{Davidsson}. With respect to the previous work of \citet{Groussin2} he admitted the possibility of a lower emissivity than that resulting from the very dark surface of the comet, which should be close to 1. The lower emissivity is justified by the effect of roughness which produces self heating effects (see Eq. \ref{eq:selfheating}) which are directly linked to the roughness of the surface. 

Following the method discussed in section \ref{sec:parametersextractor} a best fit is performed with the model of thermal emission (Planck's law) to retrieve effective emissivity ($\epsilon_{eff}$) and Temperature ($T$):

\begin{equation}
B = \epsilon_{eff} \times \frac{2\:h\:c^{2}}{(\lambda\cdot10^{-6})^{5}}\:\frac{1}{e^{\frac{h\:c}{\lambda\cdot10^{-6}\:k_{B}\:T)}-1}} \times 10^{-6}	
\label{eq:planck}
\end{equation}

\noindent
where:
\begin{itemize}
\item[]{the factor $10^{-6}$ is needed for a $\lambda$ in input in [$\mu m$] and the dimensions in output to be: [$W/sr/\mu m/m^{2}$]}
\item[]{$h = 6.62 \cdot 10^{-34} \:J\cdot s$ is the Planck constant} 
\item[]{$K_{B} = 1.38 \cdot 10^{-23} \:J/K$ is the Boltzmann constant}
\item[]{$c = 2.998 \cdot 10^{8}\:m/s$ is the speed of light}
\end{itemize}

The effective emissivity $\epsilon_{eff}$ is linked to the self heating ($\xi$) by \citep{Davidsson}:

\begin{equation}
\epsilon_{eff} = (1-\xi\overline{\epsilon}_{h} )\overline{\epsilon}_{h} 
\end{equation}

Thanks to the retrieval for each spatial pixel, maps of temperature, effective emissivity and self heating are obtained. 

\begin{figure}[!h]
\centerline{\includegraphics[width=1.\textwidth,clip=]{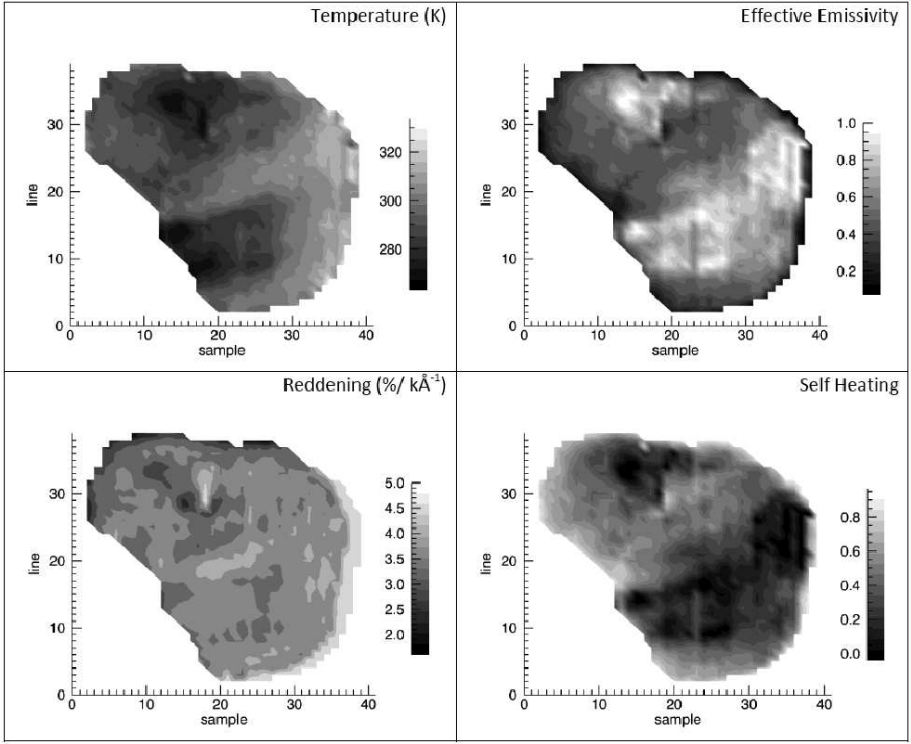}}
\caption{Upper left panel: temperature $T(K)$; upper right panel: effective emissivity $\epsilon_{eff}$; lower right panel: self heating; lower left panel: reddening (\%/k\AA), it is calculated with respect the value of the reflectance at $\lambda = 1.7\:\mu m$ for a direct comparison with the work of \citet{Davidsson}.}
\label{fig:Tempel4}
\end{figure}

The maps obtained (see Fig. \ref{fig:Tempel4}) are very similar to those in \citet{Davidsson}.

The position of the ice-rich units (see Fig. \ref{fig:icyTempel} and Fig. \ref{fig:Tempelicemap}) is coincident with that of the lowest reddening and lowest Temperature regions (see Fig. \ref{fig:Tempel4}), confirming the consistency of the maps.

\subsection{Thermal emission analysis: alternative approach}
\label{sub:alternative}
The interpretation of the results of thermal analysis exposed in the previous section poses some difficulties. 

\citet{Sunshine06} pointed out that the presence of water ice would imply temperature lower than the free sublimation temperature ($\sim 200 K$), while resulting temperatures are close to 300 K. A possible explanation is that water ice is physically and thermally decoupled from the rest of the surface at higher temperature, and the water ice involves only a little portion of the surface subtended by the pixels. 

Another difficulty rises when we try to link the self heating parameter ($\xi$) to the mean slope ($\overline{\theta}$), being both related to the roughness of the surface. The two parameters can be linked univocally as proposed by \citet{Lagerros}. However, as stressed by \citet{Davidsson}, from the resulting mean $\xi$, the mean $\overline{\theta}$ should be $55^{o}$, significantly above the value obtained by \citet{Li06}: $\overline{\theta} = 16^{o}$, and well beyond the assumption of little angles for the derivation of the mean slope parameter by \citet{Hapke}. Possible explanation of the discrepancy between the two parameters is that they measure the roughness at fundamentally different size scales: several works indicate that the mean slope parameter ($\overline{\theta}$) measures roughness at the smallest size scales where the shadows still exist, namely sub-millimeter scale, while the self heating parameter ($\xi$) could measure roughness where this effect is important, that is decimeter scale or larger.

Another possible explanation is that the model neglects the effect of the shadowing caused by the roughness of the surface. To enlighten this point we can make a meaningful example: let's take the case in which the half of the surface with an emissivity $\epsilon \sim 1$ subtended by a pixel is at temperature $< 160$ K because it is in shadow. Below this temperature, a spectrometer working at wavelengths up to 5 $\mu m$ loses sensitivity to the thermal emission. Let's suppose the other half of the surface subtended by the pixel is at a temperature of 300 K. Let's also suppose the self heating effects are negligible. The radiance emitting from such a pixel, in the range 1 - 5 $\mu m$ would be equal to the half of the radiance of a black body at 300 K. The temperature retrieved by the model exposed in previous section would be 300 K, with an effective emissivity of 0.5, even if self heating effects are negligible. This effect could be enhanced by observing at higher phase angle, where shadow effects prevail. In this case we could overestimate the mean temperature and the mean roughness of the surface region subtended by a pixel.

\begin{figure}[!h]
\centerline{\includegraphics[width=1.\textwidth,clip=]{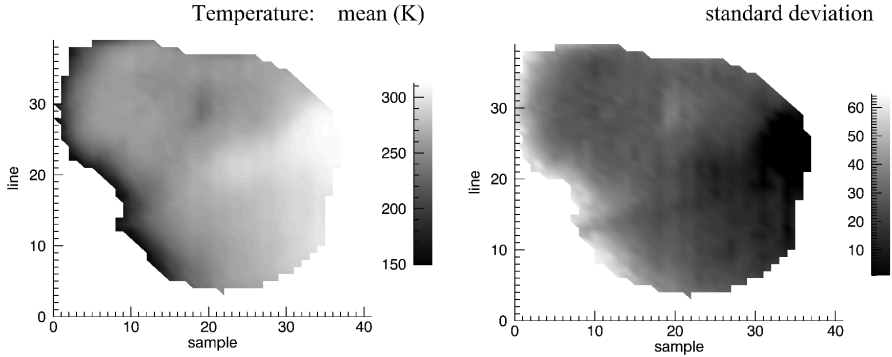}}
\caption{Mean (left) and standard deviation (right) map of the temperature on the surface, according to the Gaussian distribution of Planck functions model.}
\label{fig:Tempelgauss}
\end{figure}

Here a different approach in the interpretation of the thermal emission is presented. It can overcome most of the problems just discussed. The intent of this approach is to admit the surface subtended by a pixel to be made by different portions emitting at different temperatures because of the roughness effects. The thermal emission is modeled as a sum of Planck functions at different temperatures in the range T = 0 - 400 K (with an integrated emissivity $\overline{\epsilon}_{h}$ as calculated in section \ref{sec:ThermalTempel}), weighted by a Gaussian function $G(T)$, whose mean and standard deviation are the free parameters to retrieve. The mean value represents the most frequent value of temperature on the surface subtended by the pixel, while the standard deviation is the measure of the dispersion of the temperature values caused by roughness effects. Results are mapped in figure \ref{fig:Tempelgauss}.

The map in figure \ref{fig:Tempelgauss} (left panel) shows temperatures significantly lower that those in figure \ref{fig:Tempel4} (upper left panel). However, it should be stressed that the results coming from the two models are not contradictory because are based on different assumptions. If we use the model presented in previous section we should keep in mind that the radiance is more affected by thermal emission coming from higher temperature regions.
This last method is recommended if the risk of contamination by shadows is high, and we are interested in modeling a distribution of temperatures on scale smaller than that of the pixel.

The icy regions mapped in figure \ref{fig:Tempelicemap} present a mean temperature  $\overline{T} \sim 250$, with a standard deviation of $\sigma_{T }\sim 30$. According to a Gaussian distribution the probability to have temperature $< 200 K$ with this parameters is $\sim 5 \%$. Thus, such a portion of the surface is compatible with the presence of the exposed water ice without risk of sublimation. This result is consistent with the abundance of water ice retrieved in those regions ($< 2 \%$).

\section{Hartley 2}

\subsection{Spectral analysis}
\label{sub:spctralHartley}

The calibrated radiance spectra are available through the Planetary Data System website (http://pds.jpl.nasa.gov/). During the flyby of the comet many hyperspectral cubes were taken by the HRI instrument. As in the case of Tempel 1 analysis, we take into account the most resolved cube which present the whole nucleus: Exposure ID = 5005001, with a resolution of 28.5 m/px (geometry information on this acquisition are courtesy of Tony Farnham).  

The radiance is modeled in terms of thermal emission and reflected sunlight (see Eq. \ref{eq:rad}) following the steps discussed in section \ref{sub:spctralTempel}.
The spectra of Hartley 2 appear to be more differentiated across the surface than those of Tempel 1. For this reason a different approach for the spectral analysis is followed: the spectra are modeled as binary intimate mixtures of amorphous carbon \citep{Zubko} and pure water ice \citep{ClarkRN, Mastrapa08, Warren}, thus avoiding the definition of a dark terrain, as done for Tempel 1.

Following the method discussed in section \ref{sec:parametersextractor}, the abundance and grain size of the two endmembers are retrieved for each pixel of the comet's nucleus.

\begin{figure}[!h]
\centerline{\includegraphics[width=1.\textwidth,clip=]{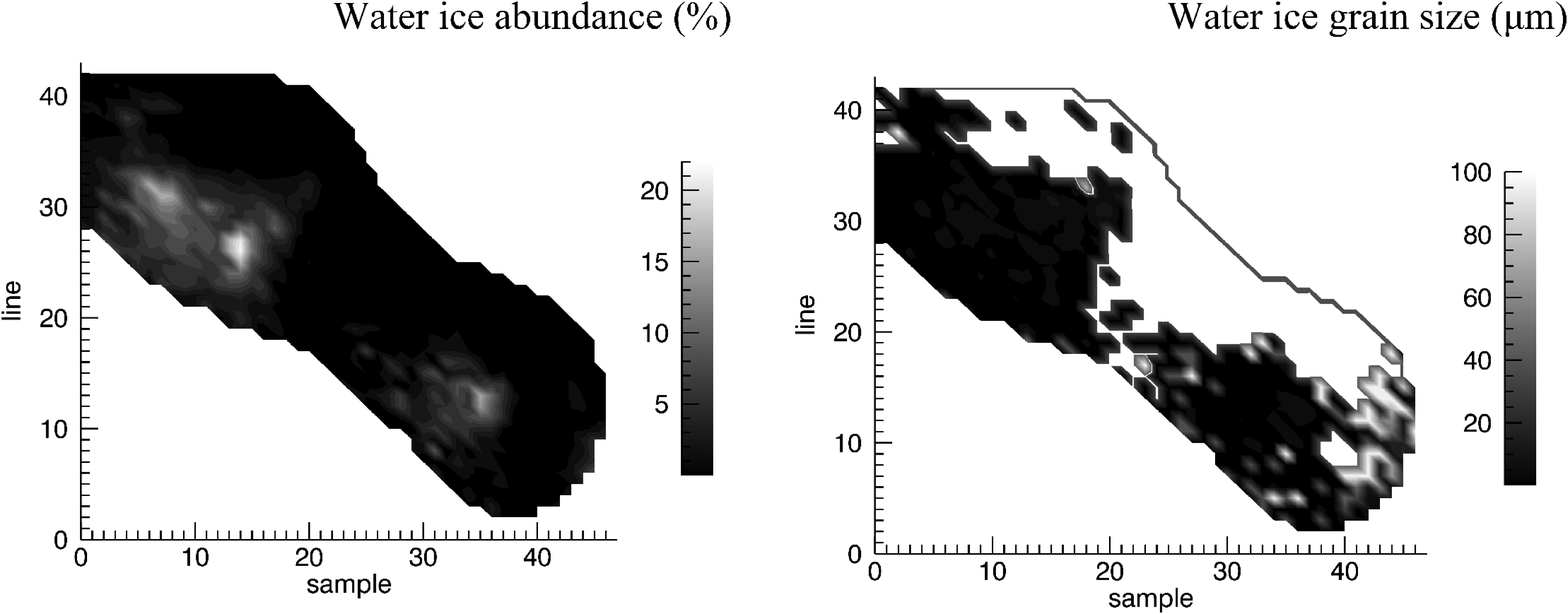}}
\caption{Water ice abundance and grain size map of Hartley 2, in case we model the mixing between amorphous carbon and pure water ice to be intimate.}
\label{fig:Hartleyice}
\end{figure}

\begin{figure}[!h]
\centerline{\includegraphics[width=0.8\textwidth,clip=]{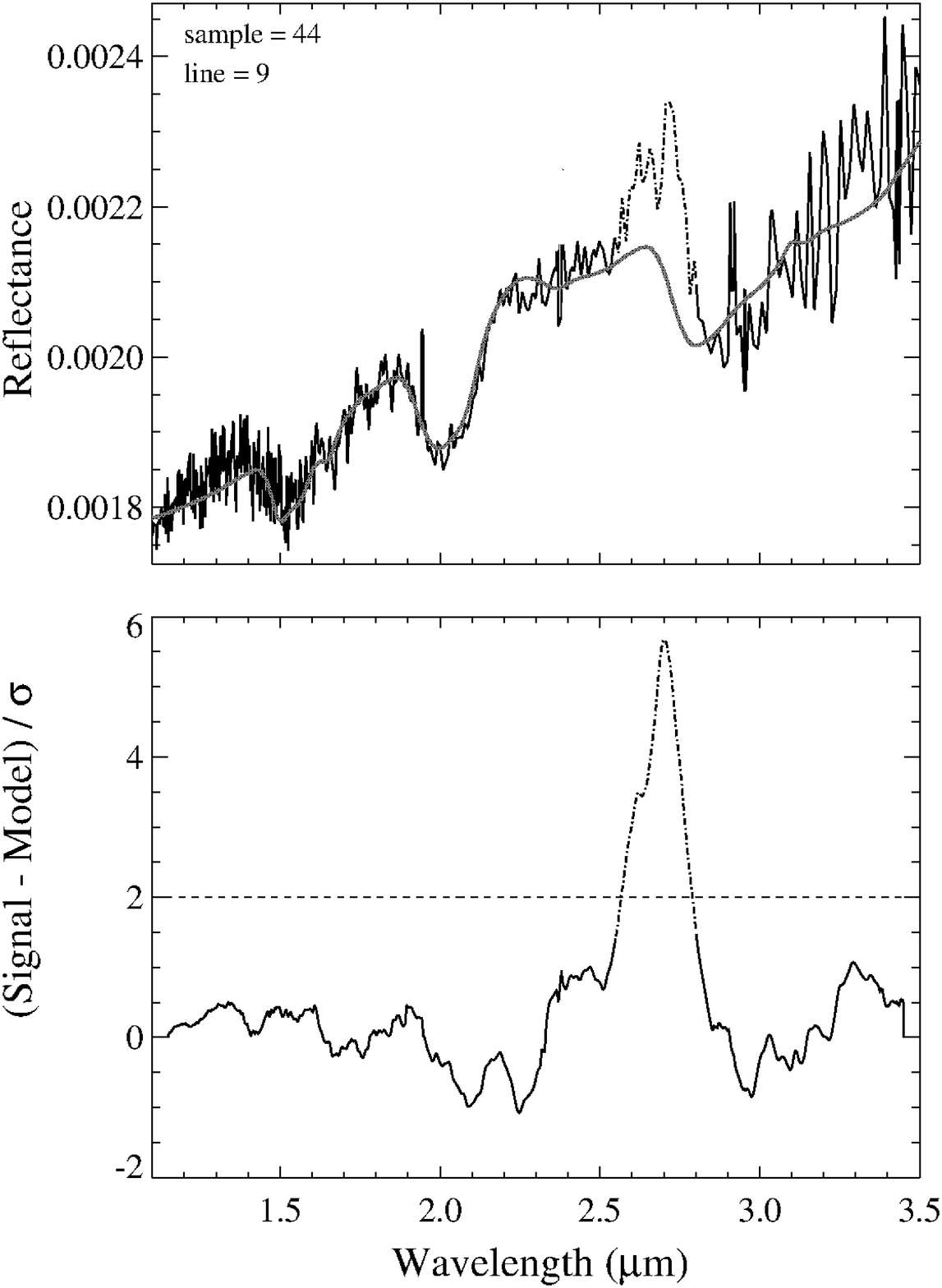}}
\caption{Upper panel: the measured spectrum and the retrieved model. Lower panel: residuals analysis. The spectrum of residual is smoothed by a running box. A significant residual ($> 2 \sigma$) at $\lambda \sim 2.7\: \mu m$ is related with the water vapour emission \citep{EPOXI} which is not taken into account in the model. The best fit is performed skipping the spectral range 2.55 - 2.8 $\mu m$ (dashed spectrum), related to this water vapour emission. Retrieved water ice abundance: 1 \%; retrieved grain sizes: water ice = $100\: \mu m$, amorphous carbon = $4\: \mu m$.}
\label{fig:Hartleyhigs}
\end{figure}

\begin{figure}[!h]
\centerline{\includegraphics[width=0.8\textwidth,clip=]{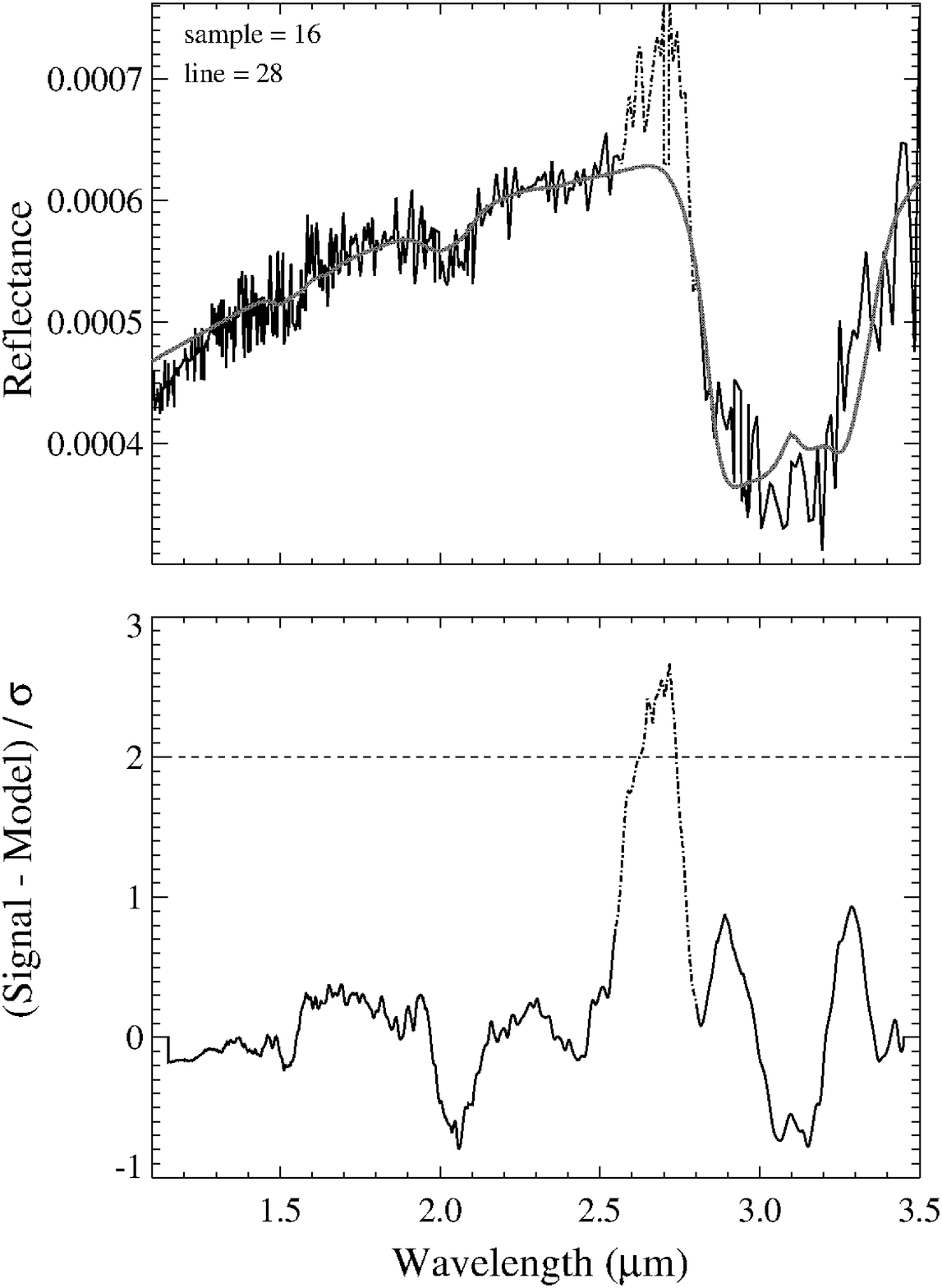}}
\caption{Upper panel: the measured spectrum and the retrieved model. Lower panel: residuals analysis. The spectrum of residual is smoothed by a running box. A significant residual ($> 2 \sigma$) at $\lambda \sim 2.7\: \mu m$ is related with the water vapour emission \citep{EPOXI} which is not taken into account in the model. The best fit is performed skipping the spectral range 2.55 - 2.8 $\mu m$ (dashed spectrum), related to this water vapour emission. Retrieved water ice abundance: 8 \%; retrieved grain sizes: water ice = $7\: \mu m$, amorphous carbon = $2\: \mu m$.}
\label{fig:Hartleylogs}
\end{figure}

The abundances retrieved can be intended as the area covered by that material as usual, but because the grain size of the two end-members are independent each other, it cannot be also intended as a numerical density anymore.

Figure \ref{fig:Hartleyice} shows large regions on the nucleus surface where the water ice is relatively abundant (up to $\sim 20 \%$). The position of these regions on the nucleus surface match with the distribution of water ice in the inner coma as a product of activity of the comet \citep{EPOXI}. Moreover, the region where exposed ice present larger grain size seems to be correlated with the most abundant and heterogeneous ($H_{2}O, CO_{2}, CN$) ambient outgassing \citep{EPOXI}. 

Figure \ref{fig:Hartleyhigs} (upper panel) shows the measured spectrum and the best fit model, from which abundance and grain size of water ice are retrieved. The spectrum is related to the clearest region of right panel of figure \ref{fig:Hartleyice}, which present larger grain size of the water ice particles. The analysis of the goodness of the model is given in the lower panel.

Figure \ref{fig:Hartleylogs} shows the same analysis of figure \ref{fig:Hartleyhigs} for a pixel in a dark region of right panel of figure \ref{fig:Hartleyice} (small grain size region). The difference of the spectra coming from the two regions (Fig. \ref{fig:Hartleyhigs} and \ref{fig:Hartleylogs}) is clearly visible: the former presents the three absorption bands of water ice ($\lambda \sim$ 1.5, 2.0 and 3.0 $\mu m$), the latter present a large absorption band at $\lambda \sim$ 3.0 $\mu m$, being the other two weak or absent.

\subsection{Thermal emission analysis}

\begin{figure}[!h]
\centerline{\includegraphics[width=0.9\textwidth,clip=]{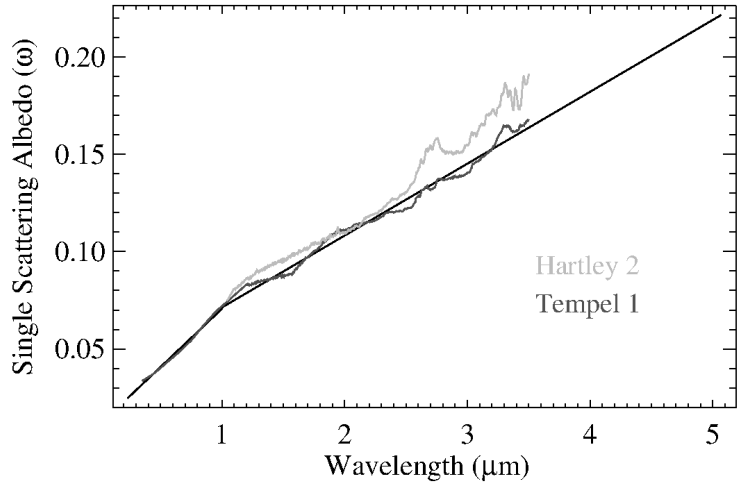}}
\caption{The single scattering albedo ($\omega$) calculated for the two comets. The high level of uncertainty coming from the thermal emission removal prevents the calculation of the SSA beyond 3.5 $\mu m$. The SSA derived from Hartley 2 data presents contamination of water vapour emission ($\lambda \sim 2.7$) not removed here. Superimposed is the linear interpolation of the Tempel 1 SSA, which has been used for the spectra simulation in Chapter \ref{chap:spectrasimulation}. The two slopes are related to the two different channels of the instrument: HRIV and HRII.}
\label{fig:SSA2comets}
\end{figure}

As discussed in section \ref{sec:ThermalTempel}, the thermal analysis is performed following the work of \citet{Davidsson}. The spectrum of single scattering albedo for comet Harley 2 is extracted from reflectance spectra in the same way of Tempel 1 (see section \ref{sub:spctralTempel}). The slope and the level of the albedo ($\omega$) of the two comets are very similar as shown in Fig. \ref{fig:SSA2comets}. For this reason we assume for Hartley 2 the same integrated emissivity $\overline{\epsilon}_{h}$ already calculated for Tempel 1 in section \ref{sec:ThermalTempel}.

\begin{figure}[!h]
\centerline{\includegraphics[width=1.\textwidth,clip=]{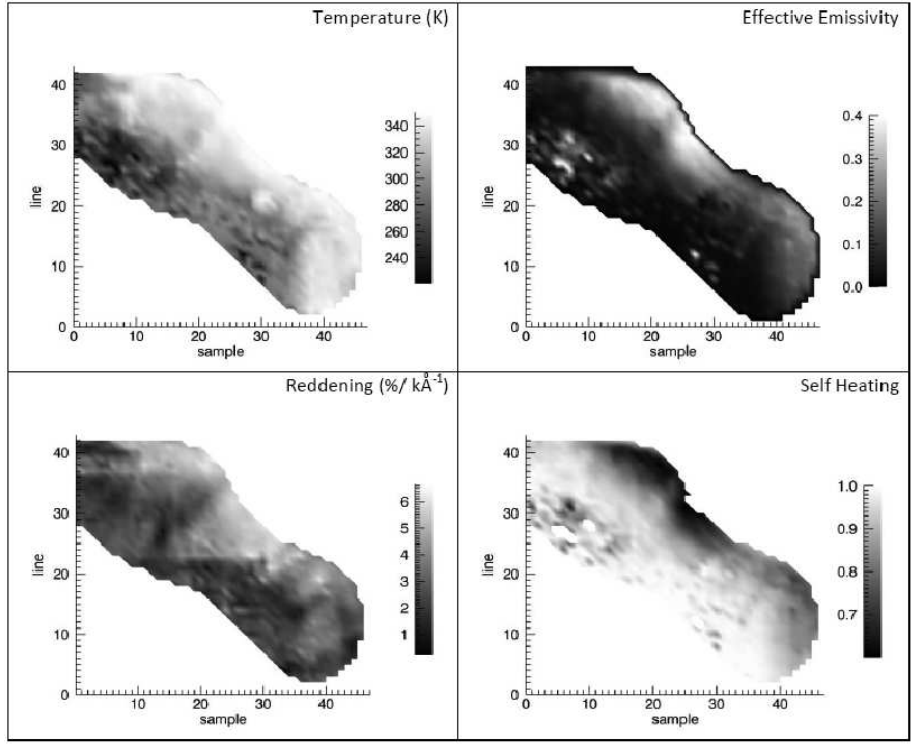}}
\caption{Upper left panel: temperature $T(K)$; upper right panel: effective emissivity $\epsilon_{eff}$; lower right panel: self heating; lower left panel: reddening (\%/k\AA), it is calculated with respect the value of the reflectance at $\lambda = 1.7\:\mu m$ for a direct comparison with the work of \citet{Davidsson}.}
\label{fig:Hartley4}
\end{figure}

Maps of Temperature, effective emissivity, self heating and reddening are calculated as discussed in section \ref{sec:ThermalTempel}, and are shown in Fig. \ref{fig:Hartley4}. 

The reddening map (lower left panel) shows results in the same range than those of Tempel 1 (see Fig. \ref{fig:Hartley4}). The two horizontal stripes are products of artifacts on the spectra.

As in the case of Tempel 1 the map of temperature (upper left panel) is consistent with the direction of the Sun. The map of effective emissivity (upper right panel) show very low values, that are even more problematic in their interpretation than those of Tempel 1. The related map of self heating shows as a consequence very high values. Moreover, higher values are located in regions weakly illuminated by the Sun, showing that shadowing effects are important and produce misleading results.

\begin{figure}[!h]
\centerline{\includegraphics[width=1.\textwidth,clip=]{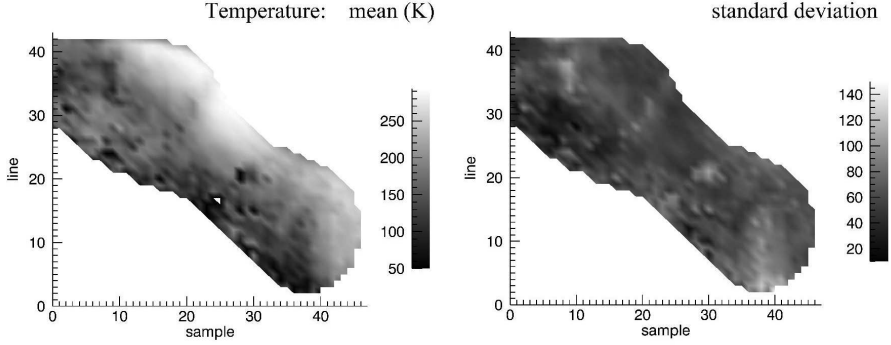}}
\caption{Mean (left) and standard deviation (right) map of the temperature on the surface, according to the Gaussian distribution of Planck functions model (see section \ref{sub:alternative}.)}
\label{fig:Hartleygauss}
\end{figure}

The alternative approach discussed in section \ref{sub:alternative} is applied to Hartley 2 data. Results are shown in Fig. \ref{fig:Hartleygauss}.

As in the case of Tempel 1, temperature map in Fig. \ref{fig:Hartleygauss} (left panel) shows lower values than those in Fig. \ref{fig:Hartley4} (upper left panel), because of the different modeling. However, the alternative approach can better explain the presence of exposed water ice up to 20 \% in abundance (see Fig. \ref{fig:Hartleyice}), having those regions a mean temperature $\leq 200 \:K$. 

The dispersion of Temperature distribution (right panel) shows a mean value $\sigma_{T} \sim 60$. This higher value with respect to that resulting from Tempel 1 ($\sigma_{T} \sim 30$) could be mostly related to the higher phase angle during Hartley 2 data acquisition ($84^{o}$ against $63^{o}$), which produce more shadows on the surface and so more temperature differentiation. 

\chapter{Rosetta-VIRTIS data analysis}

\section{Lutetia photometric analysis} 
\label{sec:lutetia}
\begin{figure}[!h]
\centerline{\includegraphics[width=0.8\textwidth,clip=]{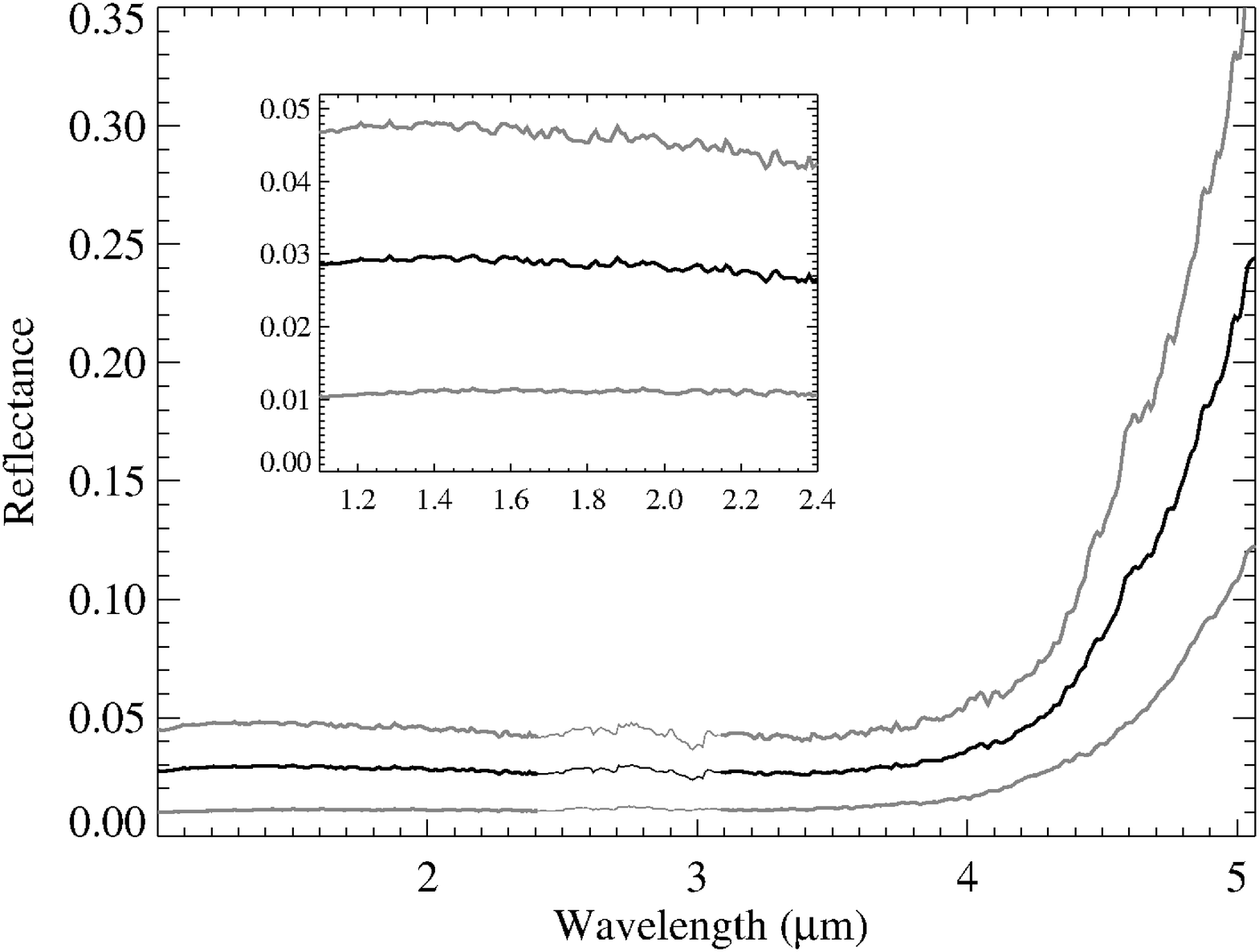}}
\caption{Lutetia mean reflectance spectrum in the 1.0 - 5.0 $\mu m$ spectral range. Where the line is thin there are instrumental artifacts in the spectrum; the rise above 4 $\mu m$ is due to the surface thermal emission. In light color are shown the $1 \sigma$ spectra. The panel inside the plot shows a detail of the reflectance within the wavelength range taken into account for the analysis.}
\label{fig:averagelutetia}
\end{figure}

On 10th of July 2010 the Rosetta spacecraft was directed to a fly-by with the main belt asteroid 21 Lutetia, a large sized asteroid of $\sim 100$ km in diameter. Its ancient surface age (determined from crater counting) coupled with its complex geology and high density suggests that Lutetia is most likely a primordial planetesimal \citep{Sierks}. This was also confirmed by the spectroscopic observations performed by the VIRTIS instrument \citep{CoradiniL}, which have shown no absorption features, of either silicates or hydrated minerals (see Figure \ref{fig:averagelutetia}). 

The availability of accurate geometric information \citep{Keihm} (incidence, emission and phase angles) for each pixel in the VIRTIS acquisitions and in all phases of the fly-by gives us the opportunity to perform a disk-resolved photometric analysis. 

Analysis of the normalized spectral variation across the observed surface at highest resolution shows a remarkable uniformity of the surface spectral properties. For this reason we can assume all the surface having the same photometric characteristics, and we can use the large amount of available spectra in different viewing geometry to improve the retrieval of the photometric parameters.

\subsection{Data and reduction}
During the close encounter with Lutetia, VIRTIS obtained different high resolution disk-resolved hyperspectral cubes from IR and VIS channels, at a wide range of phase angles. To study the photometric properties, we focused on the IR channel from 1.1 to 2.4 $\mu m$ which is a reliable spectral region for the absolute photometric calibration (see Figure \ref{fig:averagelutetia}), and do not present contamination from thermal emission. The images were calibrated through the standard VIRTIS calibration pipeline \citep{Filacchione} to reflectance units. Out of statistics pixels are removed by applying the despiking procedure discussed in section \ref{sec:artifact}.

\begin{figure}[!h]
\centerline{\includegraphics[width=0.5\textwidth,clip=]{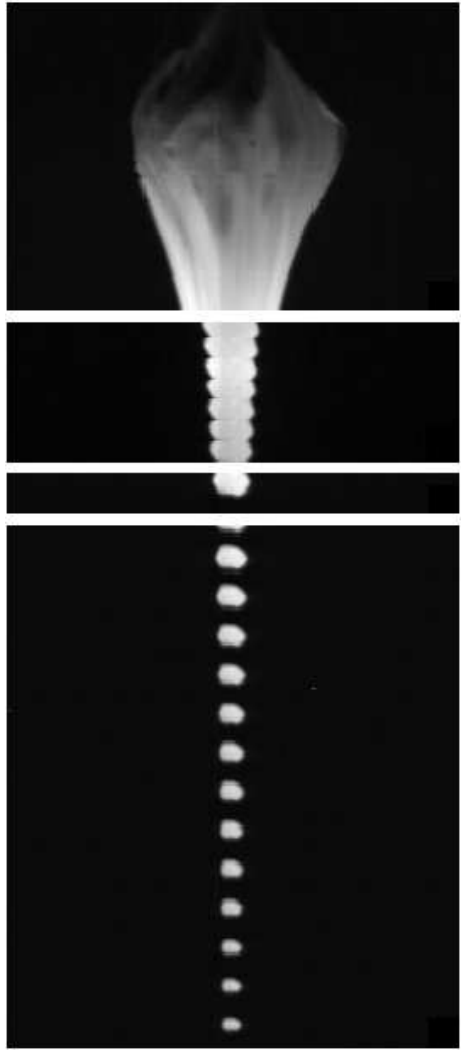}}
\caption{The four hyperspectral cubes analyzed at $\lambda = 2.0 \:\mu m$. They are (from up to down) the four listed in table \ref{tab:lutetiadataset}.}
\label{fig:lutetia4}
\end{figure}

Table \ref{tab:lutetiadataset} summarizes the hyperspectral cubes with high spatial resolution taken into account for the analysis (see also Fig. \ref{fig:lutetia4}). This dataset spans across a wide range of phase angles $0.01^{o} - 136.4^{o}$, however, as the angular radius of the Sun at the Lutetia distance (2.72 AU) is $0.1^{o}$ observations at smaller phases do not improve the knowledge of the photometric properties of the object. Pixels with phase angles smaller than $0.1^{o}$ are then discarded.

\begin{table}[!h]
\caption{The four cubes taken into account for the analysis. They are identified with their SpaceCraft Event Time (SCET). Each line of the cubes corresponds to a specific phase angle. In the last column the number of spatial pixels selected for the analysis.}
\begin{center}
\begin{tabular}{||c|c|c|c|c||}
\hline
\hline
SCET & km/px & Phase angle range & Lines & Selected pixels\\
\hline
\hline
00237396952 & 1.7 & $12.2^{o} - 136.4^{o}$ &	175	& 9177\\
\hline
00237396112 & 3.5 & $0.0^{o} - 2.0^{o}$ & 80 & 1545 \\
\hline
00237395857 & 5.6 & $3.0^{o} - 3.2^{o}$ & 23 & 155 \\
\hline
00237394252 & 6.2 & $4.0^{o} - 7.5^{o}$ & 296 & 620 \\
\hline
\end{tabular}
\end{center}
\label{tab:lutetiadataset}
\end{table}

The information on the geometry is used for a first selection of the spatial pixels to be taken into account. Pixels near the limb (within 2 pixels) are excluded to avoid pixels not completely filled by the projected area. Moreover, pixels whose incidence or emission angle overcome $70^{o}$ are excluded to avoid the non-linear effect of the topographic roughness which causes large uncertainties for the model. 

Pixels are further analyzed line by line: for each pixel the average of the reflectance over the considered spectral range is taken into account. This quantity is studied as a function of the geometry according to the Lommel-Seeliger law. A linear correlation is expected, because the phase angle is fixed within the lines: $<r> \sim [\mu_{0} / (\mu+\mu_{0})]$ (see Eq. \ref{eq:hapke}). The scatter of the pixels from a linear trend is due mostly to geometry information errors. The pixels whose average signal overcome $2 \sigma$ of the scatter are excluded from the analysis.

\subsection{Method and models}
Given the flatness of the spectra in the considered range ($1.1-2.4\:\mu m$) and the need to mitigate the signal uncertainties due to the noise, best fits of the phase function are performed taking the average over each 10 bands, that is each $0.1\: \mu m$.

The Hapke model (Eq. \ref{eq:hapke}) is used for the first three hyperspectral cubes in Table \ref{tab:lutetiadataset}. The least resolved of the four (00237394252) presents 14 scans of the full nucleus, permitting a full-disk analysis. Although the geometries of this cube are available it is preferred to treat it with an integrated formula that takes into account the whole disk, losing the information on the shape model. This permits to avoid large uncertainties of geometry information, and to test the consistency of the Hapke formula in the full disk version:

\begin{equation}
\begin{split}
<FDR> \:=\: &\frac{\int_{A} r(i,e,g) S(i,e,g) \mu dA}{\int_{A} \mu dA} = \\
&\textbf{(}\; K/8 \left\{ 1-sin(\frac{g}{2}) \times tan(\frac{g}{2}) \times ln[cot(\frac{g}{4})] \right\}\times\\
&\left\{[1+B_{S0}B_{S}(g)p(g)-1]\omega + 4r_{0}(1-r_{0}) \right\}+\\
& \frac{2}{3}r_{0}^{2} \times [sin(g) + (\pi-g) cos(g)]\; \textbf{)} \times (1+B_{C0}B_{C}(g)) \times S(g,\overline{\theta})\: / \:\pi
\label{eq:hapkefull}
\end{split}
\end{equation}

\noindent
Where:
\begin{itemize}
\item[]{$r_{0} = (1 - \gamma) / (1 + \gamma)$}
\item[]{$\gamma = \sqrt{(1- \omega)}$}
\item[]{$S(g,θ)$ is the correction for the shadowing function depending on the roughness parameter $\overline{\theta}$ (Eq 12.61 pag. 330 in \citet{Hapke}).}
\end{itemize}

Eq. \ref{eq:hapkefull} is straightforwardly derived by Eq. 11.35 - 11.42, p. 299 in \citet{Hapke} and assumes a spherical surface. The integration gives as a result the reflectance of the full disk averaged over the projected surface. This quantity is directly comparable to the measured reflectance averaged over all the pixels of the disk.

\begin{figure}[!h]
\centerline{\includegraphics[width=1.\textwidth,clip=]{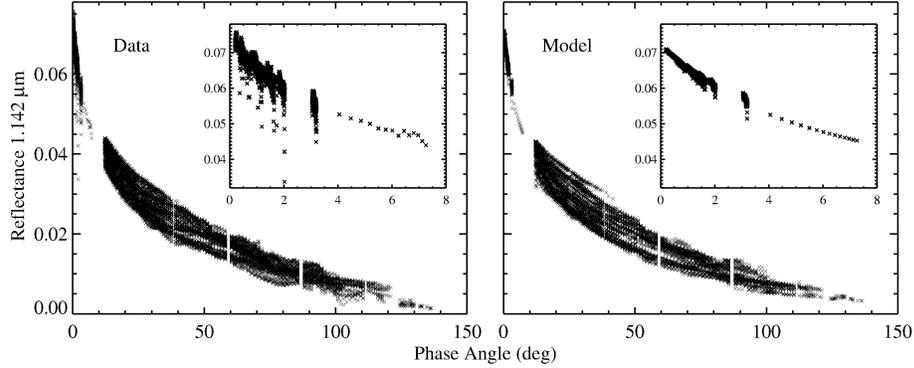}}
\caption{Left panel: reflectance centered in 1.142 $\mu m$, filter 0.1 $\mu m$ wide, plotted against phase angle for all selected spectra. Scatter of the points is due to different observation geometry (incidence and emission angle). 
In the zoom is shown a detail for small phase angles, three of the four hyperspectral cubes cover this range. From $4^{o}$ to $7.5^{o}$ the points are the result of the average over the whole disk (14 scans). 
Right panel: the best fit of the model. It is able to account for different observation geometry. Most of the discrepancy between data and model are probably due to errors in the geometry information.}
\label{fig:lutetiaresults}
\end{figure}

The phase function fit is performed taking into account all the 10877 spectra coming from the first three cubes of Table \ref{tab:lutetiadataset}, applying Eq. \ref{eq:hapke}, and the 14 spectra resulting from the average full disk reflectance coming from the last cube of Table \ref{tab:lutetiadataset}, using Eq. \ref{eq:hapkefull} (see Figure \ref{fig:lutetiaresults}).

The 9 parameters of the Hapke model hardly fit a unique solution. This problem is recurrent in this kind of analysis as enlighten by \citet{HelfShe}. In particular three couple of parameters cannot be constrained separately: the porosity parameter $K$ and the single scattering albedo $\omega$, and the four parameters concerning the opposition effect ($B_{S0}$, $h_{S}$ and $B_{C0}$, $h_{C}$).

To constrain $K$ it is followed the suggestion of \citet{HelfShe}, linking the porosity parameter $K$, to the angular amplitude of the SHOE $h_{S}$, which  also depends on the porosity. It is used the Eq. 9.26 pag. 234 in \citet{Hapke}, that is derived for equant particles, larger than the wavelength, and with a narrow size distribution.

To constrain the opposition effect parameters it is followed \citet{Ciarniello11} by eliminating $B_{C0}$ and $h_{C}$ from the fit but still taking into account a possible contribution of the CBOE by permitting $B_{S0}$ to overcome 1. It is preferred to model SHOE instead of CBOE because the typical predicted angular width of the SHOE is larger than that of the CBOE \citep{Hapke}. Moreover, the relatively dark surface make the opposition effect more affected by SHOE as it is linked to single scattering properties, while the CBOE is linked to multiple scattering.

\subsection{Results and discussion}
\begin{figure}[!h]
\centerline{\includegraphics[width=1.\textwidth,clip=]{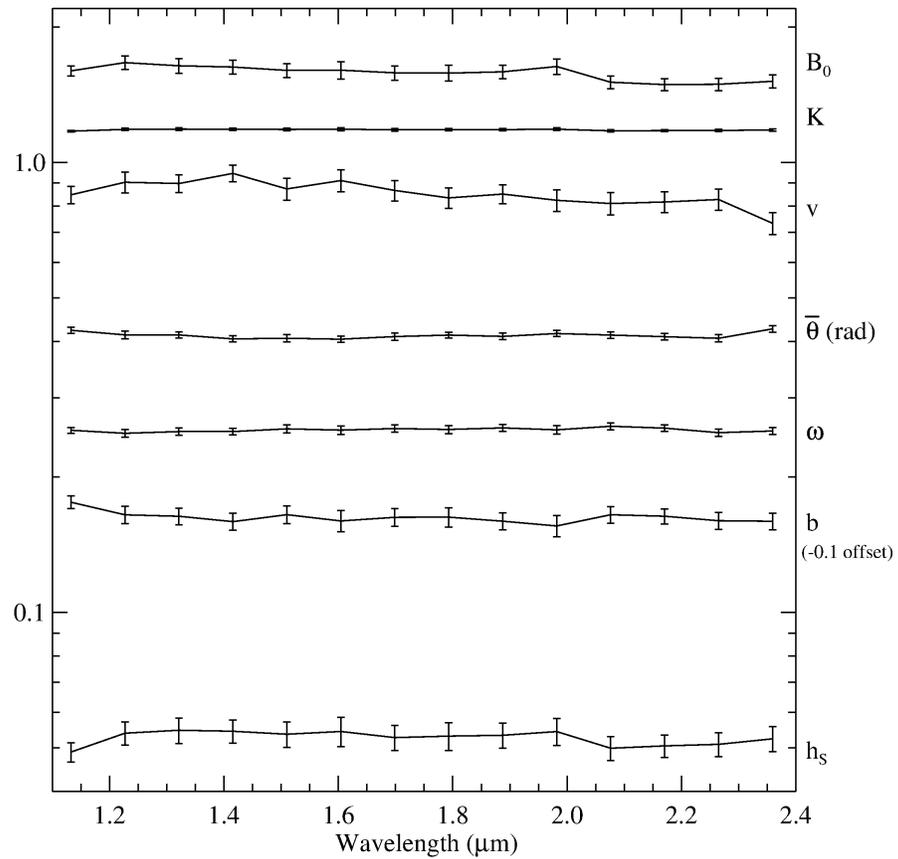}}
\caption{The retrieved parameters are plotted as a function of the wavelength. The mean slope $\overline{\theta}$ is expressed in radians, like the $h_{S}$ parameter. The $b$ parameter is plotted with an offset of -0.1 for the sake of clarity. Error bars are calculated as discussed in section \ref{sec:errorsextractor}.}
\label{fig:lutetiaparameters}
\end{figure}

The best fit is performed with the method explained in section \ref{sec:parametersextractor}. Figure \ref{fig:lutetiaparameters} shows the best fit for one wavelength. The parameters retrieved as a function of the wavelength present a flat trend, as we should expect considering the flatness of the spectrum (see Table \ref{tab:lutetiaparameters}). Scatter of the points from this trend are well explained by the error bars, that are calculated as explained in section \ref{sec:errorsextractor} (see Figure \ref{fig:lutetiaparameters}).

\begin{table}[!h]
\caption{Retrieved parameters of the Hapke model as a function of the wavelength for photometric analysis of Lutetia. In the first line the weighted average over the wavelengths range is considered.}
\begin{center}
\begin{tabular}{||c|c|c|c|c|c|c|c||}
\hline
\hline
$\lambda (\mu m)$ & $\omega$ & $K$ & $b$ & $v$ & $\overline{\theta}\; (^{o})$ & $B_{0}$ & $h_{S}$\\
\hline
Average & 0.2544 & 1.181 & 0.263 & 0.852 & 23.64 & 1.575 & 0.0525\\
\hline
\hline
1.142 & 0.2540 & 1.174 & 0.276 & 0.847 & 24.28 & 1.598 & 0.0490\\
\hline
1.236 & 0.2501 & 1.184 & 0.265 & 0.904 & 23.71 & 1.667 & 0.0539\\
\hline
1.331 & 0.2523 & 1.186 & 0.264 & 0.898 & 23.71 & 1.640 & 0.0547\\
\hline
1.425 & 0.2523 & 1.186 & 0.259 & 0.946 & 23.23 & 1.629 & 0.0544\\
\hline
1.519 & 0.2558 & 1.184 & 0.265 & 0.873 & 23.31 & 1.602 & 0.0536\\
\hline
1.614 & 0.2542 & 1.185 & 0.260 & 0.912 & 23.20 & 1.604 & 0.0544\\
\hline
1.708 & 0.2563 & 1.182 & 0.263 & 0.867 & 23.51 & 1.582 & 0.0527\\
\hline
1.803 & 0.2551 & 1.183 & 0.263 & 0.834 & 23.69 & 1.580 & 0.0531\\
\hline
1.897 & 0.2571 & 1.183 & 0.260 & 0.851 & 23.53 & 1.590 & 0.0533\\
\hline
1.991 & 0.2546 & 1.185 & 0.256 & 0.824 & 23.88 & 1.634 & 0.0544\\
\hline
2.086 & 0.2593 & 1.176 & 0.265 & 0.810 & 23.69 & 1.508 & 0.0499\\
\hline
2.180 & 0.2568 & 1.177 & 0.264 & 0.817 & 23.49 & 1.490 & 0.0505\\
\hline
2.274 & 0.2509 & 1.178 & 0.260 & 0.827 & 23.30 & 1.492 & 0.0509\\
\hline
2.369 & 0.2531 & 1.181 & 0.259 & 0.732 & 24.45 & 1.515 & 0.0524\\
\hline
\end{tabular}
\end{center}
\label{tab:lutetiaparameters}
\end{table}

Results (Table \ref{tab:lutetiaparameters}) are similar to those coming from other asteroids: Steins \citep{Spjuth}, Annefrank \citep{Hillier}, Itokawa \citep{Lederer, Kitazato}, Eros \citep{ClarkB, Domingue, Li04}, Toutatis \citep{Hudson}, Castalia \citep{Hudson97}, Ida \citep{Helfenstein96}, Gaspra \citep{Helfenstein94}. However, comparisons must be made with care because of different channels used for observation, and because variations in the models. 

The filling factor resulting from the mean retrieved $K$ parameter is: $\phi = 0.125$ (Eq. 7.45b, pag. 167 in \citet{Hapke}).

The two parameters of the Heyney-Greenstein function ($b$, $v$) model the single particle phase function to be mostly backscattering. The retrieved value for $v$ (0.852 in average) is always within the limits given in Eq. \ref{eq:limitv} (1.50 in average). 

The retrieved $B_{0}$ overcomes 1, indicating that both SHOE and CBOE are present. 

$K$, $\overline{\theta}$ and $h_{S}$ should not depend on wavelength, at least in the little range considered, because they are linked to the structure of the surface. The next step of the analysis would be to run the retrieval by fixing these parameters to their average value, and obtaining the other parameters as a function of the wavelength. In this case all the values obtained are constant within the errors, because the flatness of the spectra, thus this further step does not add any significant information.

\section{Comet 67P/Churyumov-Gerasimenko}

\subsection{Despiking and artifacts removal}
\label{sec:artifact}
The acquired spectra are always affected by spikes caused by cosmic rays and other solar particles hitting the detectors. Their removal is specially needed to correctly fit the measured data with models. A routine has been developed to address this task. It is devoted to deal with spectra showing features at scales much larger than the width of the bands of the instrument ($\sim 10$ nm). The procedure works as follows:
\begin{enumerate}
\item{The users have to set the number of parts in which the spectrum has to be divided. Each part will be interpolated with a polynomial fit of an order settable by the user. The number of parts and the degree of the polynomial fit depends on the complexity of the spectrum in terms of spectral features.}
\item{The standard deviation of the spectrum with respect to the polynomial fit is calculated for each part. The signals overcoming an adjustable number of standard deviations is set equal to the polynomial fit. Normally for the spikes removal a number of 4 $\sigma$ is enough.}
\item{The previous points are repeated until there are not anymore spikes to cut.}
\end{enumerate}

Some parts of the spectra are further affected by some artifacts like the crack of the filter in the IR channel \citep{Filacchione} and a not proper characterization of the Instrument Transfer Function. Many artifacts are proportional to the absolute level of the signal. 
The despiking procedure turned to be useful for artifact removal: thanks to the settable level of intensity of despiking, the artifact can also be removed. 

However, an intense despiking procedure cannot be applied in blind way to all the spectra: the risk to cut real features is high, especially if the spectral feature involves few instrumental bands. 

For this reason a map of artifacts is developed for both VIS and IR channels. To produce this map the despiking procedure is applied to spectral cubes which are assumed to present not spectral features at small scale. The map shows the relative difference of the signal before and after despiking:

\begin{equation}
Artifact\:Map\:(band,sample) = \frac{signal - signal\:despiked}{signal\:despiked}
\label{eq:artifact}
\end{equation}

\begin{figure}[!h]
\centerline{\includegraphics[width=0.7\textwidth,clip=]{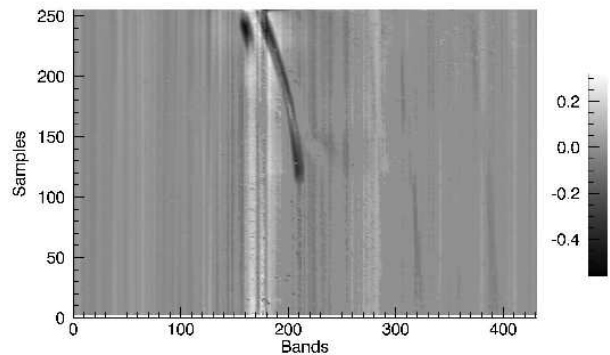}}
\caption{The map of artifact in the IR channel. It is the median of the Eq. \ref{eq:artifact} for a large numbers of acquisitions. The crack of the filter is clearly visible.}
\label{fig:artifactmapir}
\end{figure}

\begin{figure}[!h]
\centerline{\includegraphics[width=0.7\textwidth,clip=]{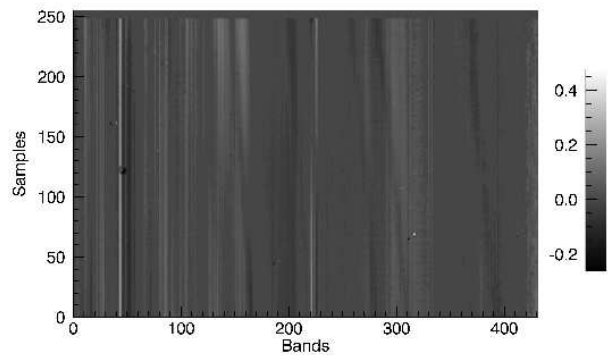}}
\caption{The map of artifact in the VIS channel. It is the median of the Eq. \ref{eq:artifact} for a large numbers of acquisitions.}
\label{fig:artifactmapvis}
\end{figure}

The calculation is performed for each acquisition frame, i.e. each line of the cubes, and the median is performed. As a result the maps show the artifacts that are signal dependent, thus they can be used for their removal:

\begin{equation}
Signal\:(band,sample) = signal/(1+Artifact\:Map)
\label{eq:removal}
\end{equation}

\begin{figure}[!h]
\centerline{\includegraphics[width=0.7\textwidth,clip=]{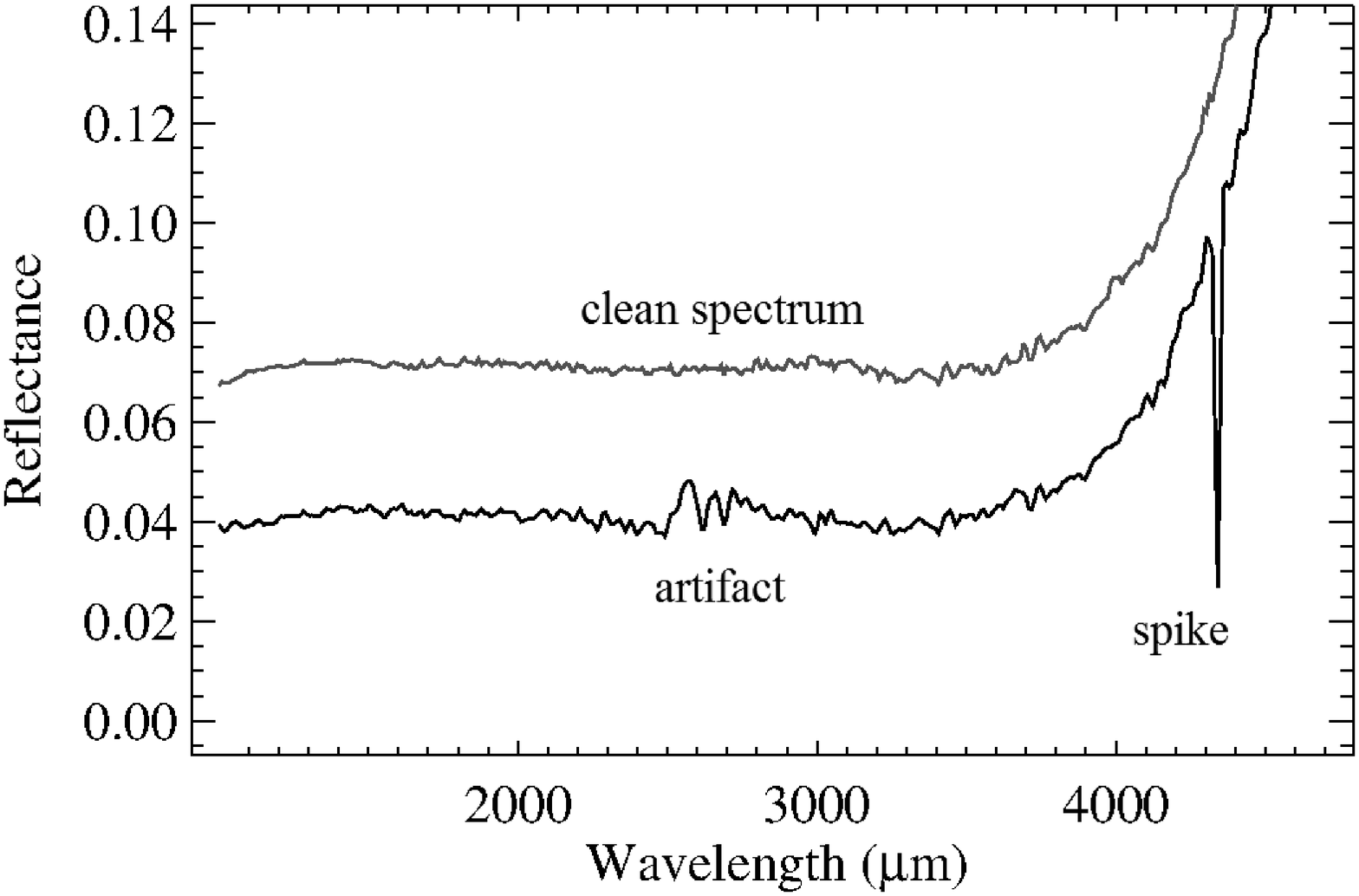}}
\caption{An example of cleaned spectrum of Lutetia after despiking and artifact removal.}
\label{fig:lutetiadespiked}
\end{figure}

Figure \ref{fig:lutetiadespiked} shows the result of the cleaning procedure for a spectrum acquired during Lutetia flyby.

The fact that the maps calculated with Comet 67P CG data work well if applied to Lutetia data makes us sure we are removing artifacts and not spectral features.

\subsection{Spectral analysis}
A complete spectrophotometric analysis of the nucleus of comet 67P/C-G is outside the scope of the present work. 

The aim of this section is to show the very first results and, as a consequence, the first problems we have to face.

The spectra of the resolved nucleus, collected in the first phase of the mission, are cleaned as discussed in previous section. Subsequently the thermal emission is removed as explained in section \ref{sub:spctralTempel}. This allows to isolate the contribution of the reflected solar light, namely the reflectance spectra.

The bridging of the two channels of VIRTIS-M (VIS and IR) allows to have a unique view of the spectra from 0.2 to 5 $\mu m$, being the two spectral channels slightly overlapping at $\lambda = 1.0 \: \mu m$.

Despite of the high spatial resolution, the comet's spectra appear to be very uniform along all the surface and basically featureless, except for a broad band at 2.9 - 3.7 $\mu m$ and a tentative band at 0.9 $\mu m$.

Others distinctive characteristics of the spectra are the very low reflectance and two different slopes separated by a knee at about $1.2 \: \mu m$.

Increases of the band depth and width of this broad absorption, correlated also to changes in the spectral slopes, occur in active areas (neck regions in Fig. \ref{fig:navcam}).

\begin{figure}[!h]
\centerline{\includegraphics[width=0.8\textwidth,clip=]{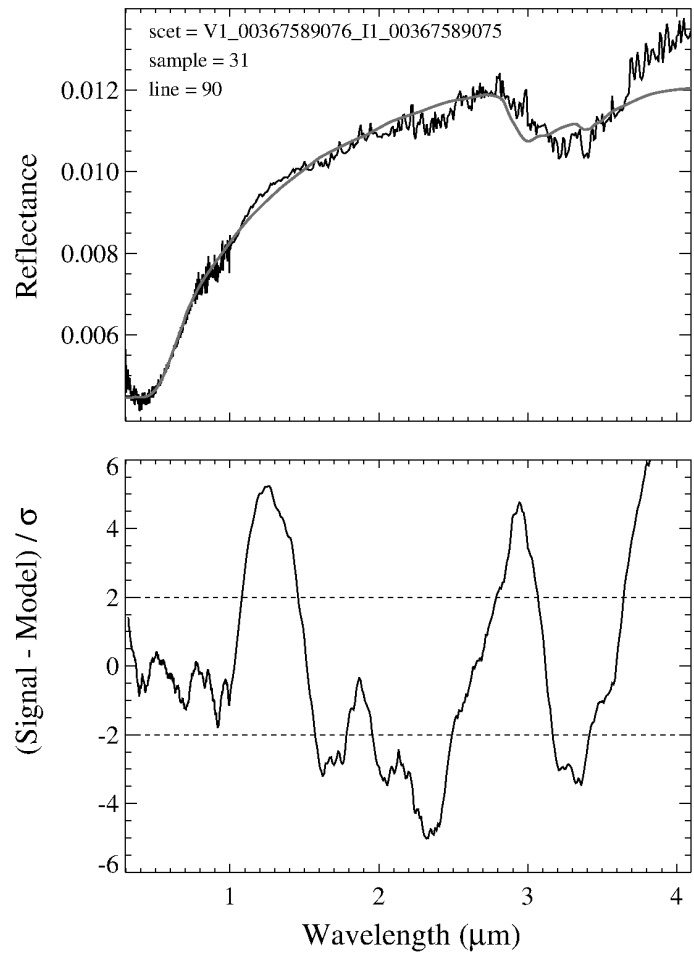}}
\caption{Upper panel: the measured spectrum and the retrieved model. Lower panel: residuals analysis. The spectrum of residual is smoothed by a running box. Significant residuals ($> 2 \sigma$) affect many parts of the spectrum, which is in fact poorly fitted by the model. For a complete discussion refer to the text. Retrieved abundances: tholin = 7.5 \%, amorphous carbon = 92.5 \%; retrieved grain sizes: tholin = $2.5\: \mu m$, amorphous carbon = $1.1\: \mu m$.}
\label{fig:CG}
\end{figure}

The method for spectral analysis discussed in section \ref{chap:modelextractor} and already applied for Hartley 2 (see section \ref{sub:spctralHartley}) has been applied for some regions of the comet 67P/C-G, and a typical result is showed in Figure \ref{fig:CG}.  

The spectra are modeled as intimate mixtures taking into account the optical constants of amorphous carbon \citep{Zubko}, pure water ice \citep{ClarkRN, Mastrapa08, Warren}, and Titan tholis \citep{Imanaka}. 

The best fit procedure apparently does not need the presence of water ice. However, in active areas, where the depth of the absorption band at 2.9 - 3.7 $\mu m$ slightly increase, presence of small amounts of water ice (1 - 2 \%) seems to improve the best fit.

The retrieved abundance of tholins is $\sim$ 8 \%. Even if the slopes overall the spectrum are well fitted, many features are poorly fitted as highlighted by the residuals analysis: the model is not able to fit the knee at 1.2 $\mu m$. In general the measured spectrum is more flat than the model in the range 1.2 - 2.5 $\mu m$. 

The main issue is represented by the broad band at 2.9 - 3.7 $\mu m$: it seems the optical constants available to us are not able to correctly produce the right position of this band, and its smooth profile. 

Moreover, from $\lambda \sim 4 \: \mu m$, the signal diverges from the model. This could be due again to the not correct optical constants we are using to produce the model or, in this case to a not complete removal of thermal emission that at these wavelengths provides the same contribution of the reflected sunlight to the radiance, and longward becomes predominant.

Finally, at the very little grain sizes retrieved, the approximation of geometrical optics is not fully satisfied, and consequently the Hapke model could be no longer valid.

\chapter{Conclusions}
This work has started with a careful analysis of the performance of the VIRTIS instrument, based on data acquired during the overall cruise phase of Rosetta spacecraft. This analysis is finalized to a better comprehension of the behaviour of the instrument with respect to the signal coming from the comet 67P/CG - main target of the mission - under all the observing conditions planned during the journey of the comet toward the Sun. 

This has brought to the modeling of the Signal to Noise ratio (Chapter 2) which has useful applications for the following analysis.

To understand how the VIRTIS instrument handles the spectra of the comet, ``possible" spectra of a comet have been simulated. The base for their simulation is the Hapke radiative transfer model described in Chapter 3. The modeling of the signal to noise ratio has further allowed to add simulated noise to the spectra. 

The tool to produce syntethic noisy spectra is useful expecially in the planning phase, to understand under which condition a signal is correctly detected by the instrument and therefore to optimize the instrument operational parameters depending on the observing conditions. Moreover, the analysis of the error on the diagnostic absorption bands allow us useful comparisons with real spectra. If a certain end-member is not observed we have the opportunity to fix an upper limit to its concentration thanks to the method proposed in Chapter 3: for a given viewing geometry and instrumental parameters, the upper limit of its abundance corresponds to the 100\% error on the band area.

The modeling of the signal to noise ratio of the instruments is the input of a new method developed to perform spectrophotometric analysis. This method, described in Chapter 4 allows an accurate and fast retrieval of the model-parameters we are interested in, and an estimate of their errors.

This model has been applied to data obtained from Deep Impact mission during the encounter of the comet Tempel 1 and its extended investigation during a second fly-by with the comet Hartley 2. The DI spacecraft carries a spectrometer similar to VIRTIS. Results obtained are similar to those obtained by previous works on the same targets, showing the reliability of the method. The investigation of Tempel 1 and Hartley 2 has been fruitful for at least two other reasons: new results on the abundances and grains size of water ice detected are obtained, and the development of an alternative method to perform thermal emission analysis. The latter in particular allows a better explanation of the presence of exposed water ice on the surface of the comets thanks to a modeling of distribution of temperatures in the surface subtended by a pixel.

We have then applied the same methods to the VIRTIS data.

The photometric model of the Lutetia asteroid have been derived applying the developed tools. Results of the photometric parameters retrieved are very similar to recent works on other asteroids, using the Hapke's model. However, the present study differs from the previous by the introduction of the $K$ porosity parameter. We must emphasize that the reliability of this parameter is still under investigation. \citet{Ciarni14} have shown that the updated Hapke's model including $K$ \citep{Hapke} improves the output at medium-large phase angles but also pointed out an inconsistency between analytical solution and numerical simulations when opposition effect and $K$ parameter are modeled at the same time, and the latter is $>> 1$. However, this is not our case because the retrieval of porosity $K$ parameter gives back a value close to 1. The porosity modelling deserves further investigations, especially because it can constrain thermal inertia of the asteroid surface.

The final step of this work coincides with the first step of the analysis of comet 67P/CG, the main target of the Rosetta mission. The first result obtained by applying the developed methods to the spectra of the comet are very encouraging although still preliminary. 

The very low reflectance of the nucleus as observed by VIRTIS coupled with distinct spectral slopes in VIS and IR ranges suggests the presence of macromolecular carbon-bearing compounds. The very broad absorption in the 2.9 - 3.7 $\mu m$ is compatible with nonvolatile organic macromolecular materials, indicating a complex mixture of various types of C-H and/or O-H chemical groups.  

From the observations performed at heliocentric distance in the range 3.6 - 3.3 AU there is no evidence even at the highest spatial resolution of 15-25 m/pixel, of ice-rich patches, indicating a generally dehydrated nature for the entire surface layer currently illuminated by the Sun.

\begin{figure}[!h]
\centerline{\includegraphics[width=0.8\textwidth,clip=]{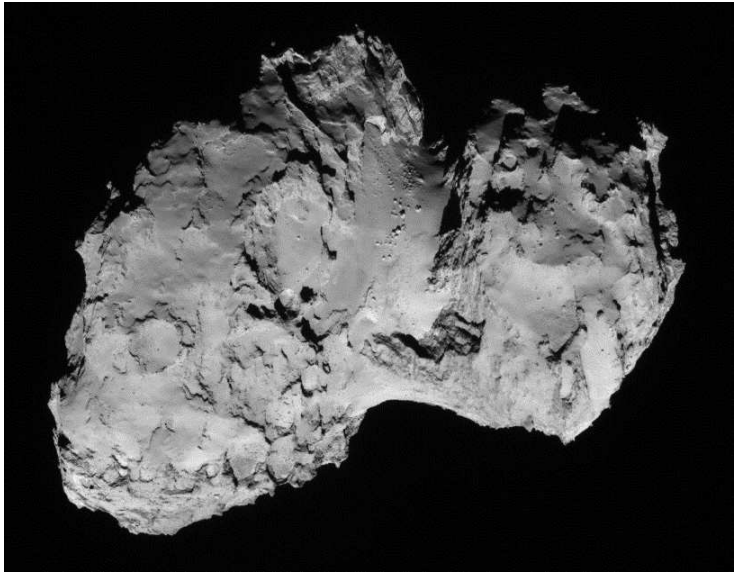}}
\caption{Rosetta navigation camera image taken on 19 August 2014 at about 79 km from comet 67P/Churyumov-Gerasimenko. The comet nucleus is about 4 km across. (\it{http://www.esa.int/spaceinimages/Images/2014/08/Comet\_on\_19\_August\_2014\_-\_NavCam})}
\label{fig:navcam}
\end{figure}

Further investigations are needed to asses the physical and compositional properties of the surface. Attempts should be made to understand the origin of its shape in relation to the spectrophotometric analysis of the three parts in which it can be divided: the head, the neck and the body (see Figure \ref{fig:navcam}).

The time is the fourth dimension in which our analysis will move, as well as the comet which will change its heliocentric distance. Although this further complicates the investigation, it could represent the key to unveil the secrets of the comet.

\chapter*{Ringraziamenti}
Questo lavoro non sarebbe stato possibile senza Fabrizio Capaccioni, Gianrico Filacchione e Mauro Ciarniello, indispensabili per la mia formazione e compagni di viaggio in questa avventura.

\noindent
Un ringraziamento va inoltre a:

tutto il Team di VIRTIS, 

tutti i colleghi dell'IAPS,

la mia famiglia.

\mbox{}~\\ 
\bibliographystyle{elsarticle-harv}
\bibliography{tesi}

\end{document}